\documentclass[12pt]{article}
\usepackage{graphicx}

\usepackage{scicite}

\usepackage{times}
\usepackage{longtable}
\usepackage{caption}
\usepackage{pdflscape}
\usepackage{rotating}
\usepackage{txfonts}

\def\psr{PSR\,J0348+0432}

\newcommand\fd{\hbox{$.\!\!^{\reset@font\romn d}$}}
\newcommand\fh{\hbox{$.\!\!^{\reset@font\romn h}$}}
\newcommand\fm{\hbox{$.\!\!^{\reset@font\romn m}$}}
\newcommand\fs{\hbox{$.\!\!^{\reset@font\romn s}$}}
\newcommand\fdg{\hbox{$.\!\!^\circ$}}

\newcommand\farcs{\hbox{$.\!\!^{\prime\prime}$}}
\newcommand\fp{\hbox{$.\!\!^{\reset@font\reset@font\scriptscriptstyle\romn p}$}}
\newcommand\arcmin{\hbox{$^\prime$}}
\newcommand\arcsec{\hbox{$^{\prime\prime}$}}

\def\aj{AJ}%
\def\araa{ARA\&A}%
\def\apj{ApJ}%
\def\apjl{ApJ}%
\def\apjs{ApJS}%
\def\aap{A\&A}%
\def\mnras{MNRAS}%
\def\prd{Phys.~Rev.~D}%
\def\prl{Phys.~Rev.~Lett.}%
\def\pasp{PASP}%
\def\sci{Science}%
\def\sovast{Soviet~Ast.}%
\def\nat{Nature}%
\def\physrep{Phys.~Rep.}%
\newenvironment{sciabstract}{%
\begin{quote} \bf}
{\end{quote}}

\newcounter{lastnote}

\title{A Massive Pulsar in a Compact Relativistic Binary\footnote{This is the authors' version of the work. The definite version is published in {\it Science} Online, 26 April 2013,  Vol: 340, Issue: 6131  doi: 10.1126/science.1233232 }} 
\author{John Antoniadis,$^{1}$ \footnote{Member of the International Max Planck Research School for Astronomy and Astrophysics at the Universities of Bonn and Cologne; e-mail: jantoniadis@mpifr-bonn.mpg.de}~ 
Paulo C. C. Freire,$^{1}$ 
Norbert Wex,$^{1}$ 
Thomas M. Tauris,$^{2,1}$\\
Ryan S. Lynch,$^{3}$ 
Marten H. van Kerkwijk,$^{4}$ 
Michael Kramer,$^{1,5}$
Cees Bassa,$^5$ \\
Vik S. Dhillon,$^6$ 
Thomas Driebe,$^7$ 
Jason W. T. Hessels,$^{8,9}$ 
Victoria M. Kaspi,$^3$ \\
Vladislav I. Kondratiev,$^{8,10}$ 
Norbert Langer,$^2$ 
Thomas R. Marsh,$^{11}$ \\
Maura A. McLaughlin,$^{12}$ 
Timothy T. Pennucci,$^{13}$
Scott M. Ransom,$^{14}$ \\
Ingrid H. Stairs,$^{15}$ 
Joeri van Leeuwen,$^{8,9}$ \\
Joris P. W. Verbiest,$^{1}$
David G. Whelan,$^{13}$
}
\date{}
\begin{document}
\maketitle
\begin{center}
\small{$^{1}$Max-Planck-Institut f{\"u}r Radioastronomie, 
Auf dem H{\"u}gel 69, 53121 Bonn, Germany}\\
\small{$^{2}$Argelander Institut f{\"u}r Astronomie, 
Auf dem H{\"u}gel 71, 53121 Bonn, Germany}\\ 
\small{$^3$Department of Physics, McGill University,}
\small{3600 University Street, Montreal, QC H3A 2T8, Canada}\\
\small{$^{4}$Department of Astronomy and Astrophysics, 
University of Toronto,}\\
\small{50 St. George Street, Toronto, ON M5S 3H4, Canada}\\
\small{$^{5}$Jodrell Bank Centre for Astrophysics, 
The University of Manchester,}\\
\small{Alan Turing Building, Manchester, M13 9PL, UK}\\
\small{$^{6}$Department of Physics \& Astronomy, University of Sheffield, 
Sheffield, S3~7RH, UK}\\
\small{$^{7}$Deutsches Zentrum f{\"u}r Luft- und Raumfahrt e.V. (DLR),}\\
\small{Raumfahrtmanagement, K{\"o}nigswinterer Str.\ 522--524, 53227 Bonn,
Germany}\\
\small{$^8$ASTRON, the Netherlands Institute for Radio Astronomy,}\\
\small{Postbus 2, 7990 AA Dwingeloo, The Netherlands}\\
\small{$^9$Astronomical Institute ``Anton Pannekoek'', 
University of Amsterdam,} \\
\small{Science Park 904, 1098 XH Amsterdam, The Netherlands}\\
\small{$^{10}$Astro Space Center of the Lebedev Physical Institute,}\\ \small{Profsoyuznaya str. 84/32, Moscow 117997, Russia}\\
\small{$^{11}$Department of Physics, University of Warwick, Coventry, CV4 7AL, UK}\\
\small{$^{12}$Department of Physics, West Virginia University,}
\small{111 White Hall, Morgantown, WV 26506, USA}\\
\small{$^{13}$Department of Astronomy, University of Virginia, P.O. Box 400325, 
Charlottesville, VA 22904, USA}\\
\small{$^{14}$National Radio Astronomy Observatory, 520 Edgemont Rd., Charlottesville VA 22903, USA}\\
\small{$^{15}$Department of Physics and Astronomy, 
University of British Columbia}\\
\small{6224 Agricultural Road, Vancouver, BC V6T 1Z1, Canada}\\

\end{center}

\begin{sciabstract}
Many physically motivated extensions to general relativity (GR) 
predict significant deviations in the properties 
of spacetime surrounding  massive neutron stars. 
We report the measurement of a 2.01$\pm$0.04 solar mass (M$_\odot$) 
pulsar in a 2.46-hr orbit with a 0.172$\pm$0.003 M$_\odot$ white dwarf.
The high pulsar mass and the compact orbit make this system a sensitive
laboratory of a previously untested strong-field gravity regime.
Thus far, the observed orbital decay agrees with GR, supporting its
validity even for the extreme conditions present in the system. The resulting 
constraints on deviations support the use of GR-based templates for 
ground-based gravitational wave detectors. Additionally,
the system strengthens recent constraints on the properties of dense
matter and provides insight to binary stellar astrophysics and pulsar recycling.

\end{sciabstract}

Neutron stars (NSs) with masses above $1.8$\,M$_\odot$  
manifested as radio pulsars are valuable probes of  fundamental physics in 
extreme conditions unique in the observable Universe and inaccessible to
 terrestrial experiments. 
Their high masses are directly linked to the equation-of-state (EOS) of matter at 
supra-nuclear densities \cite{lp04,dpr+10} and constrain the 
lower mass limit for production of astrophysical black holes (BHs).
Furthermore, they possess extreme internal gravitational fields which result in 
gravitational binding energies substantially higher than 
those found in more common, 1.4\,M$_\odot$ NSs.
Modifications to GR, often motivated by the desire for a unified model
of the four fundamental forces, can generally imprint 
measurable signatures in gravitational waves (GWs) radiated
by systems containing such objects, even if deviations from GR 
vanish in the Solar System and in less massive NSs \cite{de93,de96,will93}. 

However, the most massive NSs known today reside in long-period
binaries or other systems unsuitable for GW radiation tests.  
Identifying a massive NS in a compact, relativistic binary is 
thus of key importance for  understanding 
gravity-matter coupling under  extreme conditions. 
Furthermore, the existence of a massive NS in a relativistic orbit can
also be used to test current knowledge of close binary evolution. 

\subsection*{Results}
\paragraph*{\psr\ \& optical observations of its companion}
\psr, a pulsar spinning at 39\,ms in a 2.46-hr orbit with 
a low-mass companion, was detected by a 
recent survey \cite{drift,discovery_paper}
conducted with the Robert C. Byrd Green Bank Telescope (GBT). 
Initial timing observations of the binary yielded
an accurate astrometric  position, which allowed us to identify
its optical counterpart in the Sloan Digital Sky Survey (SDSS) archive \cite{som}.
The colors and flux of the counterpart
are consistent  with a low-mass white dwarf (WD) with a helium core
at a distance of $d \sim 2.1$\,kpc.
Its relatively high apparent brightness ($g' =20.71 \pm 0.03$\,mag) allowed us 
to resolve its spectrum using the  Apache Point Optical Telescope.
These observations revealed  deep Hydrogen lines, typical of low-mass 
WDs, confirming our preliminary identification.
The radial velocities of the WD mirrored that of \psr,
also verifying that the two stars are gravitationally bound. 

In December 2011 we obtained 
phase-resolved spectra of the optical counterpart using 
the FORS2 spectrograph of the Very Large Telescope (VLT). 
For each spectrum, we measured the radial 
velocity which we then  folded modulo the system's orbital period. 
Our orbital fit to the velocities constrains the semi-amplitude of their
modulation to be $K_{\rm WD}=351\pm4$\,km\,s$^{-1}$  
(Fig.~1; see also Materials \& Methods). 
Similarly, the orbital solution from radio-pulsar timing yields
$K_{\rm PSR} = 30.008235\pm0.000016$\,km\,s$^{-1}$ for the pulsar. 
Combined, these constraints imply a mass ratio, 
$q = M_{\rm PSR} / M_{\rm WD} = K_{\rm WD}/K_{\rm PSR} =11.70 \pm 0.13$.

Modeling of the Balmer-series lines in a high signal-to-noise average 
spectrum formed by the coherent addition of individual spectra (Fig.~1b) shows 
that the WD has an effective temperature  of 
$T_{\rm eff}=(10120 \pm 47_{\rm stat} \pm 90_{\rm sys})$\,K and
 a surface gravity of  $\log_{10} (g\,[{\rm cm\,s^{-2}}]) = ( 6.035 \pm 
 0.032_{\rm stat} \pm 0.060_{\rm sys})$\,dex. 
 Here the systematic 
 error is an overall estimate of uncertainties 
 due to our fitting technique and flux calibration \cite{som}.
We found no correlation of this measurement with orbital phase and no
signs of rotationally-induced broadening in the spectral lines \cite{som}.
Furthermore, we searched for variability using the 
ULTRACAM instrument \cite{ultracam} on the 4.2-m William-Herschel 
Telescope at La Palma, Spain. 
The lightcurves, spanning 3\,hours in total,  
have a root-mean-square scatter of $\sim$ 0.53, 0.07 and 0.08\,mag in 
$u'$, $g'$ and $r'$ respectively and show no evidence for variability 
over the course of the observations. 
The phase-folded light-curve shows no variability either. 
Additionally, our calibrated magnitudes are consistent with the SDSS catalogue 
magnitudes, implying that the WD shone at a constant flux over   
this $\sim 5$\,yr timescale \cite{som}.

\paragraph*{Mass of the white dwarf}
The surface gravity of the WD scales with its mass and the inverse 
square of its radius ($g\equiv GM_{\rm WD}/R_{\rm WD}^2$). 
Thus, the  observational constraints combined with a theoretical 
finite-temperature mass-radius relation for low-mass WDs 
yield a unique solution for the mass of the 
companion \cite{vbjj05}.  Numerous such models exist in the literature, 
the most detailed of which are in good agreement for very low mass WDs 
($<0.17-0.18$\,M$_\odot$), but differ substantially for higher masses [e.g. \cite{sar+01,pac07, kba+10}]. 
The main reason for this  is the difference in the predicted size of the 
hydrogen envelope, which determines whether the main energy source of the star 
is residual hydrogen burning (for ``thick'' envelopes) or the latent heat of the core (for ``thin'' envelopes).

In the most widely accepted scenario, WDs lose their thick hydrogen 
envelope  only if their mass exceeds a threshold. The exact location of the latter is still uncertain 
but estimated to be around $\sim$ 0.17 -- 0.22\,M$_\odot$ [e.g. \cite{sar+01,pac07, kba+10}]. 
Two other pulsars with WD companions, studied in the literature, strongly suggest that 
this transition threshold 
is indeed most likely close to 0.2\,M$_\odot$ \cite{vbjj05,avk+12}. 
In particular, the WD companion of PSR\,J1909$-$3744 has a well-determined mass of 0.20\,M$_\odot$ \cite{jhb+05}, a large characteristic age 
of a several Gyr and a WD companion that appears to be hot \cite{vbjj05}, suggesting that its envelope is thick. 
For this reason we base the WD mass estimate on cooling tracks with thick hydrogen atmospheres for masses up to 0.2\,M$_\odot$, which we constructed using the 
``MESA'' stellar evolution code \cite{pbd+11,som}.
Initial models were built  for masses identical to the ones in 
\cite{sar+01} --- for which previous comparisons have 
yielded good agreement with observations \cite{avk+12} --- 
with the addition of tracks with 0.175 and 0.185\,M$_\odot$ for finer coverage (Fig.~2).
For masses up to 0.169\,M$_\odot$ our models show excellent agreement with \cite{sar+01};  
our 0.196\,M$_\odot$ model though is quite different,  
because it has a thick envelope instead of a think one.  
Being closer to the constraints for the WD companion to \psr, it yields 
a more conservative mass constraint: 
$M_{\rm WD}$ = 0.165 -- 0.185 at 99.73\% confidence (Fig.~3 \& Table~1), which we adopt.
The corresponding radius is $R_{\rm WD}$ = 0.046 -- 0.092\,R$_\odot$ 
at 99.73\% confidence.
Our models yield a cooling age of $\tau_{\rm cool}\sim 2$\,Gyr.

\paragraph*{Pulsar mass}
The derived WD mass and the observed mass ratio $q$ 
imply a NS mass in the range 
1.97 -- 2.05\,M$_{\odot}$ at 68.27\%  or
1.90 -- 2.18\,M$_{\odot}$ at 99.73\% confidence.  Hence, 
\psr\  is only the second NS with a precisely determined mass around 
 $2$\,M$_{\odot}$, after PSR\,J1614$-$2230 \cite{dpr+10}. 
It has a 3-$\sigma$ lower mass limit 0.05\,M$_\odot$ higher than the latter,  
and therefore provides a verification, using a different method, of the 
constraints on the EOS of super-dense matter present in NS interiors 
\cite{opr+10,dpr+10}. 
For these masses and the known orbital period, 
GR predicts that the orbital period should decrease at the rate of
$\dot{P}_{\rm b}^{\rm \, GR} = (-2.58_{-0.11}^{+0.07}) \times 
10^{-13}$\,s\,s$^{-1}$ (68.27\% confidence) due to energy loss 
through GW emission. 

\paragraph*{Radio observations}
Since April 2011 we have been observing \psr\/
with the 1.4\,GHz receiver of
the 305-m radio telescope at the Arecibo Observatory,
using its four Wide-band Pulsar Processors \cite{dsh00}. 
In order to verify the Arecibo data, we have been
independently timing \psr\ at 1.4\,GHz using
the 100-m radio telescope in Effelsberg, Germany. The two timing data sets
produce consistent rotational models, providing added confidence in both.
Combining the Arecibo and Effelsberg data with the initial
GBT observations \cite{discovery_paper},
we derive the timing solution presented in Table~1.
To match the arrival times, the solution requires a 
significant measurement of orbital decay,
$\dot{P}_{\rm b} = (-2.73 \pm 0.45) \times 10^{-13}$\,s\,s$^{-1}$
(68.27\% confidence).
 
The total proper motion and
distance estimate (Table~1) allows us to calculate 
the kinematic corrections to $\dot{P}_{\rm b}$ from
its motion in the Galaxy, plus any contribution due
to possible variations of Newton's gravitational 
constant $G$: $\delta \dot{P}_{\rm b} = 
0.016\pm0.003\times 10^{-13} \,\rm s\,s^{-1}$. This is negligible compared to the measurement uncertainty.
Similarly, the small rate of rotational energy loss of the pulsar (Table~1) excludes 
any substantial contamination due to mass loss from the
system; furthermore we can exclude  substantial contributions to
$\dot{P}_{\rm b}$ from tidal effects 
(see \cite{som} for details).
Therefore, the
observed $\dot{P}_{\rm b}$ is caused by GW emission 
and its magnitude is entirely consistent with the one predicted by GR:
$\dot{P}_{\rm b} / \dot{P}_{\rm b}^{\rm \, GR} = 1.05 \pm 0.18$ (Fig.~3). 

If we {\em assume} that GR is the correct theory of gravity, we can then
derive the component masses from the intersection of
the regions allowed by $q$ and $\dot{P}_{\rm b}$ (Fig.~3):
$M_{\rm WD} = 0.177_{-0.018}^{+0.017} \, \rm M_{\odot}$ and
$M_{\rm PSR} = 2.07_{-0.21}^{+0.20} \, \rm M_{\odot}$ (68.27\% confidence).
These values are not too constraining yet. However, the
uncertainty of the measurement of $\dot{P}_{\rm b}$
decreases with $T^{-5/2}$ (where $T$ is the timing baseline);
therefore this method will yield very precise mass measurements
within a couple of years.

\subsection*{Discussion}

\paragraph*{\psr\ as a testbed for gravity} 
There are strong arguments for GR not to be valid beyond a (yet unknown) critical 
point, like its incompatibility with quantum theory and its prediction of the 
formation of spacetime singularities under generic conditions. Therefore, it 
remains an open question if GR is the final description of macroscopic gravity. 
This strongly motivates testing  gravity regimes that have not been tested 
before, in particular regimes where gravity is strong and highly non-linear. 
Presently, binary pulsars provide the best high-precision experiments to probe 
strong-field deviations from GR and the best tests of the radiative 
properties of gravity \cite{dt92_nw,sta03,ksm+06,dam09_nw,fwe+12}. Among these 
systems \psr\ has a special role: it is the first massive ($\sim$ 2\,M$_\odot$) 
NS in a relativistic binary orbit. The orbital period of \psr\ is only 15 
seconds longer than that of the double pulsar system, but it has $\sim$ 2 times 
more fractional gravitational binding energy than each of the double pulsar NSs. 
This places it far outside the presently tested binding energy range [see 
Fig.~4a \& \cite{som}]. Because the magnitude of strong-field effects generally  
depends non-linearly on the binding energy, the measurement of orbital decay  
transforms the system into a gravitational laboratory for a previously untested 
regime, qualitatively very different from what was accessible in the past. 

In physically consistent and extensively studied alternatives, gravity is
generally mediated by extra fields (e.g.\ scalar) in addition to the tensor
field of GR\cite{will93}. A dynamical coupling between matter and these extra
fields can lead to prominent deviations from GR that only occur at the high
gravitational binding energies of massive NSs. One of the prime examples is the
strong-field scalarization discovered in \cite{de93}. 
If GR is not valid, in the \psr\ system where
such an object is closely orbited by a weakly self-gravitating body, one
generally expects a violation of the strong equivalence principle that in turn
leads to a modification in the emission of gravitational waves. While in GR the
lowest source multipole that generates gravitational radiation is the
quadrupole, alternative gravity theories generally predict the presence of
monopole and dipole radiation, on top of a modification of the other multipoles
\cite{will93}. For a binary system, the leading change in the orbital period is
then given by the dipole contribution, which for a (nearly) circular orbit reads
\cite{som}:
\begin{equation}
  \dot{P}_{\rm b}^{\rm dipolar} \simeq -\frac{4\pi^2 G}{c^3 P_{\rm b}}\,
  \frac{M_{\rm PSR}M_{\rm WD}}{M_{\rm PSR} + M_{\rm WD}} \,
  (\alpha_{\rm PSR} - \alpha_{\rm WD})^2 \;,
\end{equation}
where $\alpha_{\rm PSR}$ is the effective coupling strength between the NS and 
the ambient fields responsible for the dipole moment [e.g.\ scalar fields in 
scalar-tensor gravity], and $\alpha_{\rm WD}$ is the same parameter for the 
WD companion. The WD companion to \psr\ has a fractional gravitational binding 
energy ($E_{\rm grav}/M_{\rm WD}c^2$) of just 
$-1.2 \times 10^{-5}$, and is therefore a weakly self-gravitating object. 
Consequently, $\alpha_{\rm WD}$ is practically identical to the linear 
field-matter coupling $\alpha_0$, which is well constrained ($|\alpha_0| < 
0.004$) in Solar System experiments \cite{will06_nw,dam09_nw}. 

For $\alpha_{\rm PSR}$, the situation is very different. Even if $\alpha_0$ is
vanishingly small, $\alpha_{\rm PSR}$ can have values close to unity, due to a 
non-linear behavior of gravity in the interaction between matter and the 
gravitational fields in the strong-gravity regime inside NSs 
\cite{de93,de96}. A significant $\alpha_{\rm PSR}$ for NSs up to 
$1.47$\,M$_\odot$ has been excluded by various binary pulsar experiments 
\cite{fwe+12,som}. The consistency of the observed GW damping 
($\dot{P}_{\rm b}$) with the GR predictions for \psr\ (Table~1) implies 
$|\alpha_{\rm PSR} - \alpha_0 | < 0.005$ (95\% confidence) and consequently 
excludes significant strong-field deviations, even for massive NSs of 
$\sim 2$\,M$_\odot$. 
 
To demonstrate in some detail the implications of our results for possible 
strong-field deviations of gravity from Einstein's theory, we confront our 
limits on dipolar radiation with a specific class of scalar-tensor  
theories, in which gravity is mediated by a 
symmetric second-rank tensor field $g_{\mu\nu}^\ast$ and 
by a long-range (massless) scalar field $\varphi$. Scalar-tensor theories  are 
well motivated and  consistent theories of gravity, extensively studied in the 
literature [e.g.\ \cite{fm03,goe12}]. For this reason, they are the most natural 
framework for us to illustrate the gravitational phenomena that can be probed 
with \psr.

Concerning the EOS of NS matter, in our calculations we use 
the rather stiff EOS ``.20''of \cite{hkp81} that supports (in GR) NSs of up to 
$2.6$\,M$_\odot$. We make this choice for two reasons: i) a stiffer EOS generally 
leads to more conservative limits when constraining alternative gravity 
theories, and ii) it is able to support even more massive NSs than \psr, which 
are likely to exist \cite{frb+08,vbk11,rfs+12}. Furthermore, in most of our conclusions a specific EOS is used only for illustrative purposes, and the 
obtained generic results are EOS independent. 

Fig.~4b illustrates how \psr\ probes a non-linear regime of gravity that has not 
been tested before. A change in EOS and gravity theory would lead to a modified 
functional shape for the effective coupling strength, $\alpha_{\rm PSR}$.
However, this would not change the general picture: even in the strong 
gravitational 
field of a 2\,M$_\odot$ NS gravity seems to be well described by GR and there 
is little space for any deviations, at least in the form of long-range fields,
which influence the binary dynamics. Short range interactions, like massive 
Brans-Dicke gravity \cite{abw+12} with a sufficiently large scalar mass (heavier 
than $\sim 10^{-19}$\,eV/$c^2$), cannot be excluded by \psr. Nevertheless, as we 
will argue below, in combination with the upcoming ground-based GW detectors, 
this could lead  to  particularly illuminating insights into the properties of gravitational 
interaction.
 

\paragraph*{Constraints on the phase evolution of neutron-star mergers}
The first likely direct GW detection from astrophysical sources by 
ground-based laser interferometers, like the LIGO (Laser Interferometer 
Gravitational Wave Observatory, \cite{ligo}) and the VIRGO 
\cite{virgo} projects, will mark the beginning of a new era of GW astronomy 
\cite{ss09}. One of the most promising sources for these detectors are 
in-spiralling compact binaries, consisting of NSs and BHs, whose orbits are 
decaying towards a final coalescence due to GW damping. While the signal 
sweeps in frequency through the detectors' typical sensitive bandwidth between 
about 20 Hz and a few kHz, the GW signal will be deeply buried in 
the broadband noise of the detectors \cite{ss09}. To detect it, one will have to 
apply a matched filtering technique, i.e.~correlate the output of the detector 
with a template wave form. Consequently, it is crucial to know the binary's 
orbital phase with high accuracy for searching and analyzing the signals from 
in-spiraling compact binaries. Typically, one aims to lose less than one 
GW cycle in a signal with $\sim$ $10^4$ cycles. For this reason, 
within GR such calculations have been conducted with great effort by various 
groups up to the 
3.5 post-Newtonian order, i.e.~all (non-vanishing) terms up to order $(v/c)^7$, providing 
sufficient accuracy for a detection \cite{bla06,will11_nw,mag08}. 

If the gravitational interaction between two compact masses is different from 
GR, the phase evolution over the last few thousand cycles, which fall into the 
bandwidth of the detectors, might be too different from the (GR) template in 
order to extract the signal from the noise. In scalar-tensor gravity for 
instance, the evolution of the phase is driven by radiation reaction, which is 
modified because the system loses energy to scalar GWs 
\cite{will94,de98}. Depending on the difference between the effective scalar 
couplings of the two bodies, $\alpha_A$ and $\alpha_B$, the 1.5 post-Newtonian dipolar 
contribution to the phase evolution could drive the GW signal many cycles 
away from the GR template. For this reason, it is desirable that potential 
deviations from GR in the interaction of two compact objects can be tested and 
constrained prior to the start of the advanced GW detectors. For ``canonical'' 
$1.4$\,M$_\odot$ NSs and long-range gravitational fields, this 
has already been achieved to a high degree in binary pulsar experiments, e.g.\ 
\cite{de98}. So far, the best constraints on dipolar gravitational wave damping 
in compact binaries  come from the observations 
of the millisecond pulsar PSR~J1738+0333 \cite{fwe+12}. However, as discussed in 
detail above, these timing experiments are insensitive to strong-field 
deviations that might only become relevant in the strong gravitational fields
associated with high-mass NSs. Consequently, the dynamics of a merger of a 2\,M$_\odot$ NS with a ``canonical'' NS or a BH might have a significant
contribution from dipolar GWs. With our constraints on dipolar radiation 
damping from the timing observations of \psr, given above, we can already 
exclude a deviation of more than $\sim0.5$ cycles from the GR template 
during the observable in-spiral caused by additional long-range gravitational 
fields, for the whole range of NS masses observed in nature (see Fig.~5, and 
\cite{som} for the details of the calculation). 
This compares to the precision of GR templates based on the 3.5 post-Newtonian approximation \cite{bla06,mag08}.
Furthermore, in an 
extension of the arguments in \cite{will94,de98} to massive NSs, our result 
implies that binary pulsar experiments are already more sensitive for testing 
such deviations than the upcoming advanced GW detectors. 

Finally, as mentioned before, our results on \psr\ cannot exclude dipolar 
radiation from short-range fields. Hence, if the range of the 
additional field in the gravitational interaction happens to lie between the 
wavelength of the GWs of \psr\ and the wavelength of the merger signal
($\sim$ $10^9$\,cm; $\sim$ $10^{-13}$\,eV$/c^2$), then the 
considerations concerning the applicability of the GR template given here do 
not apply. On the other hand, in such a case the combination of binary pulsar 
and LIGO/VIRGO experiments can be used to constrain the mass of this 
extra field.

\paragraph*{Formation, past and future evolution of the system}
The measured spin period $P$ and spin-period derivative $\dot{P}$ 
of \psr, combined with the masses 
and orbital period of the system (Table~1), form a peculiar 
set of parameters that gives insight to binary stellar evolution. 
The short 2.46-hr orbital period 
is best understood from evolution via a common envelope where the NS 
is captured in the envelope of the WD progenitor, leading to efficient removal of 
orbital angular momentum on a short timescale of $\sim 10^3$\,yr \cite{il93}. 
This implies that the NS was born with an initial mass close 
to its current mass of 2.01\,M$_{\odot}$, because very little accretion was possible. Whereas the slow 
spin period of $\sim$~$39\,{\rm ms}$ and the unusually strong 
magnetic field \cite{som} of a 
few~$10^9\,{\rm G}$ (Table~1) provide further support for this scenario, 
the low WD mass contradicts the 
standard common-envelope hypothesis by requiring a progenitor star mass 
smaller than
$2.2$\,M$_{\odot}$, because more massive stars would leave behind more massive 
cores  \cite{som,tlk11}. For such low donor star masses, however, the mass ratio of the 
binary components is close to unity, leading to dynamically stable mass transfer 
without forming a common envelope \cite{ts99,prp02}. 
One potential solution to this mass discrepancy for common-envelope evolution 
is to assume that the
original mass of the WD was $\ge 0.4\,{\rm M}_{\odot}$ and
that it was subsequently evaporated by the pulsar wind \cite{fgld88} when
\psr\ was young and energetic, right after its recycling phase \cite{som}.
Such an evolution could also help explain the formation of
another puzzling system, PSR\,J1744$-$3922 \cite{brr+07}.
However, we find that this scenario is quite unlikely given that
the observed spectrum of the WD in \psr\ only displays hydrogen lines, 
which is not expected if the WD was indeed a stripped remnant of a much more 
massive helium or carbon-oxygen WD. Furthermore, it is unclear 
why this evaporation process should have come to a complete stop when the WD reached its  current mass of $0.17\,{\rm M}_{\odot}$. 
A speculative hypothesis to circumvent 
the above-mentioned problems would be a common-envelope evolution with hypercritical accretion, where $\sim$ 0.6\,M$_{\odot}$ of material was efficiently transferred to a 
1.4\,M$_{\odot}$ NS \cite{che93,som}.

An alternative, and more promising, formation scenario is evolution via 
a close-orbit low-mass {X}-ray  binary (LMXB) 
with a $1.0-1.6$\,M$_{\odot}$ donor star that suffered from loss of 
orbital angular momentum due to magnetic braking \cite{ps89,prp02,vvp05}. This requires a 
finely tuned truncation of the mass-transfer process which is not yet understood 
in detail, but is also required for other known recycled pulsars \cite{som} with short 
orbital periods of $P_{\rm b}\le 8\,{\rm hr}$ and low-mass helium WD companions with 
$M_{\rm WD}\approx 0.14 - 0.18\,{\rm M}_\odot$. 
The interplay between magnetic braking,
angular momentum loss from stellar winds (possibly caused by irradiation)
and mass ejected from the vicinity of the NS 
is poorly understood and current stellar evolution models have difficulties
reproducing these binary pulsar systems.
One issue is that the converging LMXBs most often do not detach
but keep evolving with continuous mass transfer to more and more compact systems
with $P_{\rm b} \leq 1\,{\rm hr}$ and ultra-light donor masses smaller than
$0.08\,{\rm M}_{\odot}$. 

Using the Langer stellar evolution code \cite{som}, 
we have attempted 
to model the formation of the \psr\ system via LMXB evolution (Fig.~6).
To achieve this, we forced 
the donor star to detach its Roche~lobe at $P_{\rm b}\sim5\,{\rm hr}$, 
such that the system subsequently shrinks in size to its present
value of $P_{\rm b}\simeq 2.46\,{\rm hr}$ due to GW radiation within 
2~Gyr, the estimated cooling age of the WD. 
An illustration of the past and future evolution of \psr\ from
the two different formation channels is shown in Fig.~7.

An abnormality of \psr\ in view of the LMXB model 
is its slow spin period of 
$P\sim 39\,{\rm ms}$ and, in particular, the high value for the spin period
derivative, $\dot{P} = 2.41 \times 10^{-19}$\,s\,s$^{-1}$. 
These values correspond to an
inferred surface magnetic flux density of $B\sim 2\times 10^9\,{\rm G}$, 
which is high compared to most other recycled pulsars \cite{tlk12}.
However, a high $B$ value naturally explains the slow spin period of \psr\ from a 
combination of spin-down during the Roche-lobe decoupling phase \cite{tau12}
and subsequent magnetic dipole radiation from this high-magnetic-field pulsar \cite{tlk12,som}.
Another intriguing question concerning this evolutionary channel is the spread in NS masses.
In the five currently known NS-WD systems with $P_{\rm b}\le 8\,{\rm hr}$, the NS masses
span a large range of values, ranging from $\sim 1.4$ up to 2.0\,M$_\odot$.
The lower masses imply that the mass transfer during the LMXB phase is extremely inefficient
--- only about $30\%$ of the material leaving the donor is accreted by the NS \cite{jhb+05,avk+12}.
If this is indeed the case, and one assumes that the physical processes that lead to the formation
of these systems are similar, it is likely that \psr\ was born with an initial mass of 
$1.7\pm0.1$\,M$_\odot$, providing further support for a non-negligible fraction
of NSs born massive \cite{tlk11}.

Emission of GWs will continue to shrink the orbit of \psr\ and in 
400\,Myr (when $P_{\rm b}\simeq23\;{\rm min}$) the WD will fill its 
Roche~lobe and possibly leave behind a planet orbiting the pulsar \cite{bbb+11,vnv+12}. Alternatively, if \psr\ is near the upper-mass limit for NSs
then a BH might form via accretion-induced collapse of the massive NS in a 
cataclysmic, {$\gamma$}-ray burst-like event \cite{grbs2}.

\subsection*{Materials \& Methods}

\paragraph*{Radial velocities and atmospheric parameters}

A detailed log 
of the VLT observations can be found in \cite{som} (Figure\,S1 \& Table\,S1).  
We extracted the spectra following closely the method used in \cite{avk+12} 
and compared them with template spectra to measure the radial velocities.
Our best fits for the WD had 
reduced $\chi^2$ minimum values of $\chi^2_{\rm red, min} = 1.0 - 1.5$ \cite{som}. 
Uncertainties were taken to be the difference in velocity over 
which $\chi^2$ increases by $\chi^2_{\rm red, min}$ to account for 
the fact that $\chi^2_{\rm red, min}$ is not equal to unity \cite{avk+12}.
After transforming the measurements to the reference frame of the
Solar System Barycenter (SSB), we folded them using the radio-timing ephemeris 
described below. We then fitted for the semi-amplitude of the radial
velocity modulation, $K_{\rm WD}$, and the systemic radial velocity with respect
to the SSB, 
$\gamma$, assuming a circular orbit and keeping the time of passage through the ascending node, $T_{\rm asc}$, fixed to the best-fit value of the radio-timing 
ephemeris. 
Our solution yields $K_{\rm WD}=351\pm 4$\,km\,s$^{-1}$ and 
$\gamma = -1\pm20$\,km\,s$^{-1}$ \cite{som}.

Details of the Balmer lines in the average spectrum of \psr,
created by the coherent  addition of the individual spectra
shifted to zero velocity, are shown in Fig.\,1b.
We modeled the spectrum using a 
 grid of detailed  hydrogen atmospheres \cite{koe08}. 
These models incorporate the improved treatment
of pressure broadening of the absorption lines presented in \cite{tb09}.
As mentioned above, our fit yields $T_{\rm eff}=(10120 \pm 35_{\rm stat} \pm 90_{\rm sys})$\,K 
for the effective temperature and $\log_{10} g = ( 6.042 \pm 0.032_{\rm stat} \pm 0.060_{\rm sys})$ 
for the surface gravity \cite{som}.
 The $\chi^2$ map shown in Fig.~2a  is inflated to take into account systematic uncertainties. 
The average spectrum was also searched for rotational broadening. Using the 
analytic profile of \cite{gray05} to convolve the model atmospheres, we 
scanned the grid of velocities $0\le v_{\rm r} \sin i \le 2000$\,km\,s$^{-1}$ 
with a step size of 100\,km\,s$^{-1}$. The result is consistent with no rotation 
and our 1-$\sigma$ upper limit is $v_{\rm r} \sin i \le 430$\,km\,s$^{-1}$.

\paragraph*{Modeling of the white-dwarf mass}
Low-mass WDs are thought to form naturally within the age of the 
Universe via mass transfer in a binary, either through  
Roche-lobe overflow or common-envelope evolution. 
In both cases, the WD forms when the envelope mass drops below
a critical limit, which depends primarily on the mass of the stellar core, 
forcing the star to contract and detach from its Roche lobe. 
After the contraction, the mass of the relic envelope  is fixed for a given core mass, but 
further reduction of its size may occur shortly before the star enters the final 
cooling branch due to hydrogen shell flashes which force the star to re-expand to 
giant dimensions. Additional mass removal via Roche-lobe overflow as well as rapid shell 
hydrogen burning through the CNO cycle 
may then lead to a decrease of the envelope size and affect 
the cooling history and atmospheric parameters. 
To investigate the consequence of a reduced envelope size for 
the WD companion to \psr, we constructed WD models 
in which we treat the envelope mass as a free parameter \cite{som}.
For the WD companion to \psr, 
an envelope mass below the critical limit for hydrogen 
fusion is not likely for two main reasons:

First, for a pure helium composition, the observed surface gravity translates 
to a WD mass of $\sim 0.15$\,M$_\odot$ and a cooling age of 
$\sim20$\,Myr, which is anomalously small. 
Such a small age would also imply a large increase in 
the birth and in-spiral rate of similar relativistic NS--WD systems \cite{kkl+04}. 
Furthermore, post-contraction flash episodes on the WD are not sufficient 
to remove the entire envelope. Therefore, creation of a pure helium WD
requires large mass loss rates before the progenitor contracts, 
which is unlikely. For small progenitor masses ($\le 1.5$\,M$_\odot$) large mass
loss prevents contraction and the star evolves to a semi-degenerate companion on
a nuclear timescale that exceeds the age of the Universe. 
For more massive progenitors ($> 1.5$\,M$_\odot$) the core grows beyond
$\sim 0.17$\,M$_\odot$ in a short timescale and ultimately leaves a
too-massive WD.

Second, even for  envelope 
hydrogen fractions as low as $X_{\rm avg} = 10^{-6}$, the observed 
temperature and surface gravity cannot be explained simultaneously: The low 
surface gravity would again require a small mass of $\sim 0.15$\,M$_\odot$. 
However, in this case the surface hydrogen acts like an insulator, preventing the 
heat of the core from reaching the stellar surface. As a result, temperatures as high 
as 10000\,K can only be reached for masses above $\sim 0.162$\,M$_\odot$.

Past a critical envelope mass, the pressure at the bottom of the envelope 
becomes high enough to initiate hydrogen-shell burning. The latter then becomes 
the dominant energy source and the evolutionary time-scale increases; 
the radius of the star grows by $\sim\,50\%$ (depending on the mass),  
expanding further for larger envelopes. 
The dependence of the surface gravity on the radius 
implies that the observed value translates to a 
higher mass as the envelope mass increases.
Therefore, the most conservative lower limit for the WD mass
 (and thus for \psr, given the fixed mass ratio) 
is obtained if one considers models with the absolute 
minimum envelope mass required for hydrogen burning. 
In this scenario, the mass of the WD  is  
in the range 0.162 -- 0.181\,M$_\odot$  at 99.73\% confidence \cite{som}.
Despite this constraint being marginally consistent with our observations \cite{som}, 
it is not likely correct due to the high degree of fine-tuning. 

For these reasons we have adopted the assumption that the WD companion to PSR\,J0348+0432 
has a thick envelope as generally expected for WDs with such low surface gravity and high temperature.

\paragraph*{Radio-timing analysis}
The Arecibo observing setup \cite{som} and data reduction are
similar to the well tested ones described in
\cite{fwe+12}. Special care is taken with saving raw search data,
which allows for iterative improvement of the ephemeris and
eliminates orbital-phase dependent smearing of the pulse profiles,
which might contaminate the measurement of $\dot{P}_{\rm b}$ \cite{nsk+08}.
From this analysis we derive 7773 independent measurements of
pulse times of arrival (TOAs)
with a root-mean-square (rms) uncertainty smaller than 10\,$\mu$s.
Similarly the Effelsberg observations yield a total of 179 TOAs with uncertainties smaller than $20\,\mu$s.

We use the {\sc tempo2} timing package \cite{hem06} to derive
the timing solution presented in Table~1, using 8121 available
TOAs from GBT \cite{discovery_paper},  Arecibo and Effelsberg.
The motion of the radiotelescopes relative to the barycenter of
the Solar System was computed using the DE/LE 421 Solar
System ephemeris \cite{de421}, published by the Jet Propulsion Laboratories.
The orbit of \psr\ has a very low eccentricity, therefore we use the
``ELL1'' orbital model \cite{lcw+01} to describe the motion of the pulsar.

For the best fit, the reduced $\chi^2$ of the timing residuals
(TOA minus model prediction) is 1.66,
a result similar to what is obtained in timing observations of other millisecond
pulsars. The overall
weighted residual rms is 4.6$\,\mu$s.
There are no unmodeled systematic trends in the
residuals; either as a function of orbital phase or as a function of
time. Therefore $\chi^2 > 1$ is most likely
 produced by under-estimated TOA uncertainties.
We increased our estimated TOA uncertainties for each telescope 
 and receiver to produce a reduced $\chi^2$ of unity on short timescales; 
 for our dominant dataset (Arecibo) the errors 
 were multiplied by a factor of 1.3.

 This produces more conservative estimates
of the uncertainties of the timing parameters; these have been
verified using the Monte Carlo statistical method described in
\cite{fwe+12}: when all parameters are fitted, the
Monte Carlo uncertainty ranges are very similar to those
estimated by {\sc tempo2}. As an example,
{\sc tempo2} estimates
$\dot{P}_{\rm b} = (-2.73 \pm 0.45) \times 10^{-13}$\,s\,s$^{-1}$
(68.27\% confidence) and the Monte Carlo method yields
$\dot{P}_{\rm b} = (-2.72 \pm 0.45) \times 10^{-13}$\,s\,s$^{-1}$
(68.27\% confidence),
in excellent agreement.
The observed orbital decay appears to be stable; no higher
derivatives of the orbital period are detected \cite{som}.

\clearpage

\noindent
Table~1. \\{\bf Observed and Derived Parameters for the \psr\ system}\\  
Timing parameters for the \psr\ system, indicated with their
1-$\sigma$ uncertainties as derived by {\sc tempo2}
where appropriate (numbers in parentheses refer to errors on the last digits). 
The timing parameters are calculated for the reference epoch MJD\,$56000$, 
and are derived from TOAs in the range MJD\,$54872-56208$. \newline
*For these timing parameters  
we have adopted the optically derived
parameters (see text for details).

\begin{table}
\begin{footnotesize}
\begin{center}
\begin{tabular}{l l}
\hline
 \multicolumn{2}{c}{{\bf Optical Parameters}} \\
\hline
Effective temperature, $T_{\rm eff}$ (K) \dotfill & $10120\pm47_{\rm stat}\pm 90_{\rm sys}$ \\
Surface gravity, $\log_{10}(g[{\rm cm\,s^{-1}}])$ \dotfill & $6.035\pm0.032_{\rm stat}\pm 0.060_{\rm sys}$ \\
Semi-amplitude of orbital radial velocity, $K_{\rm WD}$\,(km\,s$^{-1}$) \dotfill & $351 \pm 4$\\
Systemic radial velocity relative to the Sun, $\gamma$\,(km\,s$^{-1}$) 
\dotfill & $-1 \pm 20$ \\
\hline
  \multicolumn{2}{c}{{\bf Timing Parameters}} \\
\hline

 Right ascension, $\alpha$ (J2000) \dotfill & $03^{\rm h}\; 
 48^{\rm m}\;43^{\rm s}.639000(4)$ \\
 Declination, $\delta$ (J2000) \dotfill &
$+04^\circ \, 32\arcmin\, 11\farcs4580(2)$ \\
 Proper motion in right ascension, $\mu_{\alpha}$ (mas yr$^{-1}$)
\dotfill & $+4.04(16)$ \\
 Proper motion in declination, $\mu_{\delta}$ (mas yr$^{-1}$)
\dotfill & $+3.5(6)$ \\
 Parallax, $\pi_d$ (mas) \dotfill & $0.47$* \\
 Spin frequency, $\nu$ (Hz)                  \dotfill & $25.5606361937675(4)$ \\
 First derivative of $\nu$, $\dot{\nu}$ ($10^{-15}$ Hz s$^{-1}$)
\dotfill & $-0.15729(3)$ \\
 Dispersion measure, DM (cm$^{-3}$ pc)       \dotfill & $40.46313(11)$ \\
 First derivative of DM, DM1 (cm$^{-3}$ pc yr$^{-1}$) \dotfill &
$-0.00069(14)$  \\
 Orbital period, $P_{\rm b}$ (d) \dotfill & $0.102424062722(7)$  \\
 Time of ascending node, $T_{\rm asc}$ (MJD) \dotfill & $56000.084771047(11)$ \\
Projected semi-major axis of the pulsar orbit,
$x$ (lt-s) \dotfill & $0.14097938(7)$ \\
$\eta \equiv e \sin \omega $ \dotfill & $(+1.9 \pm 1.0) \times 10^{-6}$  \\
$\kappa \equiv e \cos \omega $ \dotfill & $(+1.4 \pm 1.0) \times 10^{-6}$  \\
First derivative of $P_{\rm b}$, $\dot{P}_{\rm b}$ ($10^{-12}$\,s\,s$^{-1}$)
\dotfill & $- 0.273(45)$\\

\hline
  \multicolumn{2}{c}{{\bf Derived Parameters}} \\
\hline

 Galactic longitude, $l$ \dotfill & $183\fdg3368$\\
 Galactic latitude, $b$ \dotfill & $-36\fdg7736$\\
 Distance, $d$ (kpc) \dotfill & $2.1(2)$ \\
 Total proper motion, $\mu$ (mas yr$^{-1}$)  \dotfill & $5.3(4)$\\
 Spin period, $P$ (ms) \dotfill & $39.1226569017806(5)$ \\
 First derivative of $P$, $\dot{P}$ ($10^{-18}$
 s\,s$^{-1}$) \dotfill & $0.24073(4)$ \\
 Characteristic age, $\tau_c$ (Gyr)  \dotfill & $2.6$ \\
 Transverse magnetic field at the poles, $B_0$ ($10^9$\,G) \dotfill & $\sim$ $2$ \\
 Rate or rotational energy loss, $\dot{E}$ ($10^{32}$ erg s$^{-1}$)
\dotfill & $\sim$ $1.6$ \\
 Mass function, $f$ (M$_{\odot}$)   \dotfill & $0.000286778(4)$ \\
 Mass ratio, $q \equiv M_{\rm PSR}/M_{\rm WD}$ \dotfill & $11.70(13)$ \\
 White dwarf mass, $M_{\rm WD}$  (M$_{\odot}$)  \dotfill & $0.172(3)$ \\
 Pulsar mass, $M_{\rm PSR}$ (M$_{\odot}$) \dotfill & $2.01(4)$\\
 ``Range" parameter of Shapiro delay, $r$ ($\mu$s) \dotfill & $0.84718$* \\
 ``Shape" parameter of Shapiro delay, $s \equiv \sin i$ \dotfill &
$0.64546$* \\
 White dwarf radius, $R_{\rm WD}$  (R$_{\odot}$)  \dotfill & $0.065(5)$ \\
 Orbital separation, $a$ ($10^9$\,m)  \dotfill & $0.832$ \\
 Orbital separation, $a$ (R$_{\odot}$)  \dotfill & $1.20$ \\
 Orbital inclination, $i$ \dotfill & $40\fdg2(6)$\\
 $\dot{P}_{\rm b}$ predicted by GR, $\dot{P}_{\rm b}^{\rm GR}$
 ($\rm 10^{-12}$\,s\,s$^{-1}$) \dotfill & $-0.258^{+0.008}_{-0.011}$\\
 $\dot{P}_{\rm b}/\dot{P}_{\rm b}^{\rm GR}$ \dotfill & $1.05 \pm 0.18$ \\
 Time until coalescence, $\tau_m$ (Myr) \dotfill & $\sim$ $400$ \\
 \hline

 \hline
\end{tabular}
\end{center}
\end{footnotesize}
\label{table:timing}
\end{table}

\newpage
Figure~1.\\ {\bf Radial Velocities and Spectrum of the White Dwarf Companion to \psr.}\\
\textbf{Upper:} Radial velocities of the WD companion to \psr\ plotted 
against the orbital phase (shown twice for clarity). Over-plotted  is the 
best-fit orbit of the WD (blue line) and the mirror orbit of the pulsar (green). 
\textbf{Lower:} 
Details of the fit to the Balmer lines (H$\beta$ to H12) 
    in the average spectrum of the WD companion to \psr\ created by the coherent addition 
    of 26 individual spectra shifted to
    zero velocity. Lines from H$\beta$ (bottom) to H12 are shown.
    The red solid lines are the best-fit atmospheric model (see text). 
    Two models with ($T_{\rm eff},\log_{10} g$) = (9900\,K,\,5.70)
    and ($T_{\rm eff},\log_{10} g$) =(10200\,K,\,6.30), each $\sim 3$-$\sigma$ 
    off from the best-fit central value (including systematics) are shown
   for comparison (dashed blue lines).
 $    \begin{array}{cc}

   \includegraphics[scale=0.38]{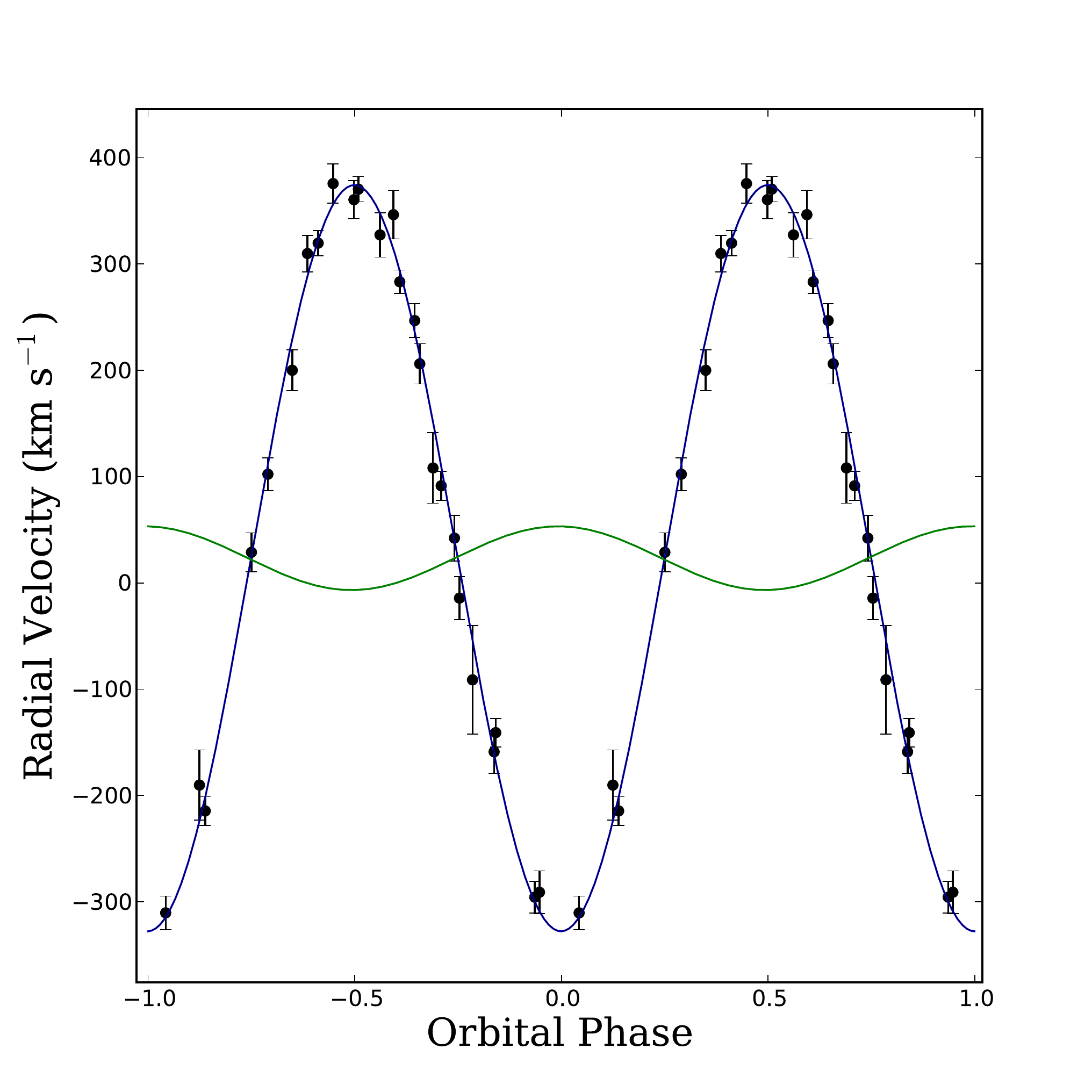} &
   \includegraphics[scale=0.38]{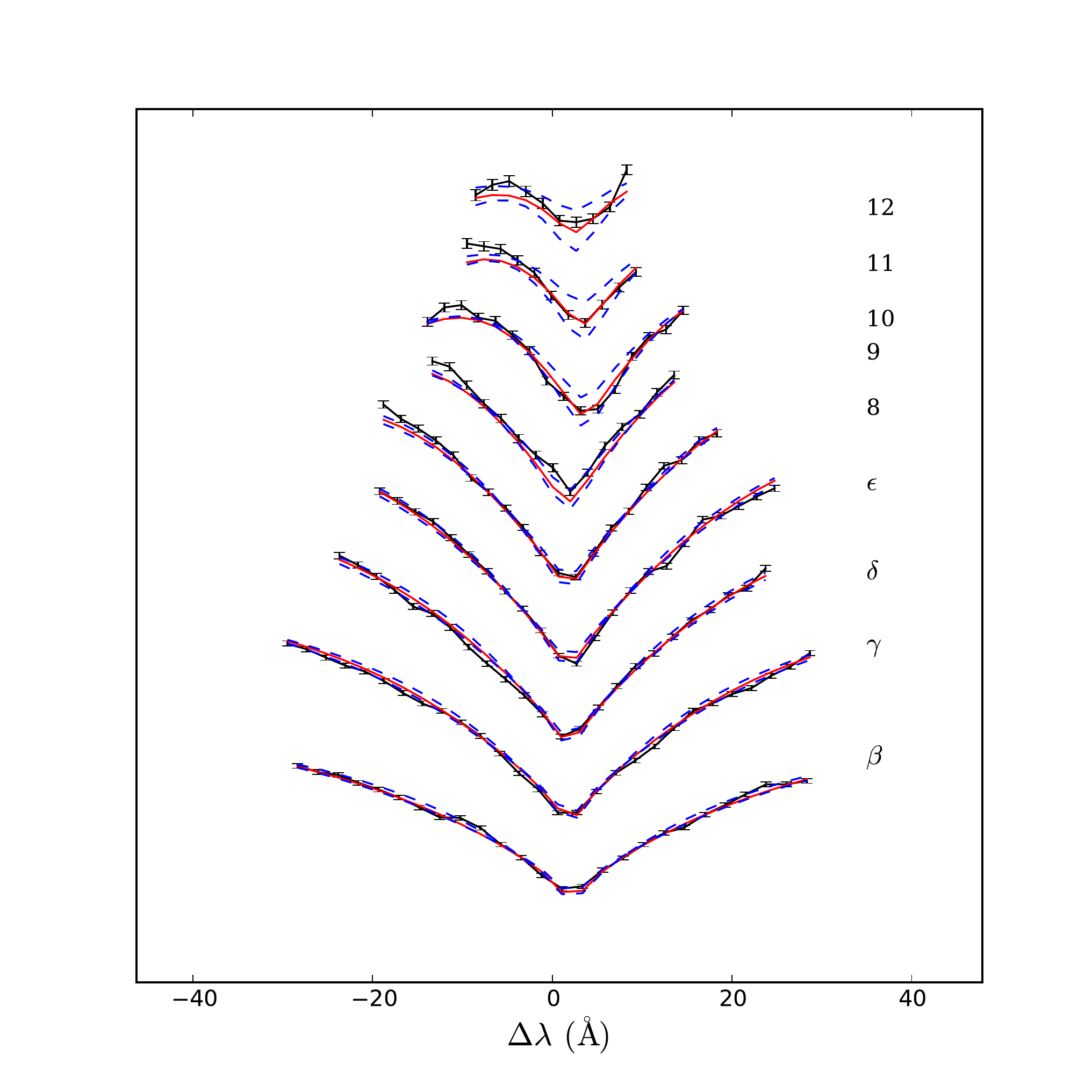}\\

   \end{array}$

\newpage
Figure~2. \\ {\bf Mass Measurement of the White Dwarf Companion to \psr}\\
\textbf{Left:} Constraints on effective temperature, $T_{\rm{eff}}$, and surface gravity, $g$, 
for the WD companion to \psr\ compared with theoretical WD models. The shaded  areas depict 
the $\chi^2 - \chi^2_{\rm min} = 2.3$, 6.2 and 11.8 intervals (equivalent to 1, 2 and 3-$\sigma$) of our fit to the average spectrum. Dashed lines show the 
detailed theoretical cooling models of \cite{sar+01}. Continuous  lines depict 
tracks with thick envelopes for masses up to $\sim$ 0.2\,M$_\odot$ that yield the 
most conservative constraints for the mass of the WD. 
\textbf{Right:} Finite-temperature mass-radius relations for our models 
together with the constraints imposed from modeling of the 
  spectrum  (see text). Low mass -- high 
  temperature points are an
  extrapolation from lower temperatures.
 $    \begin{array}{cc}

   \includegraphics[scale=0.38]{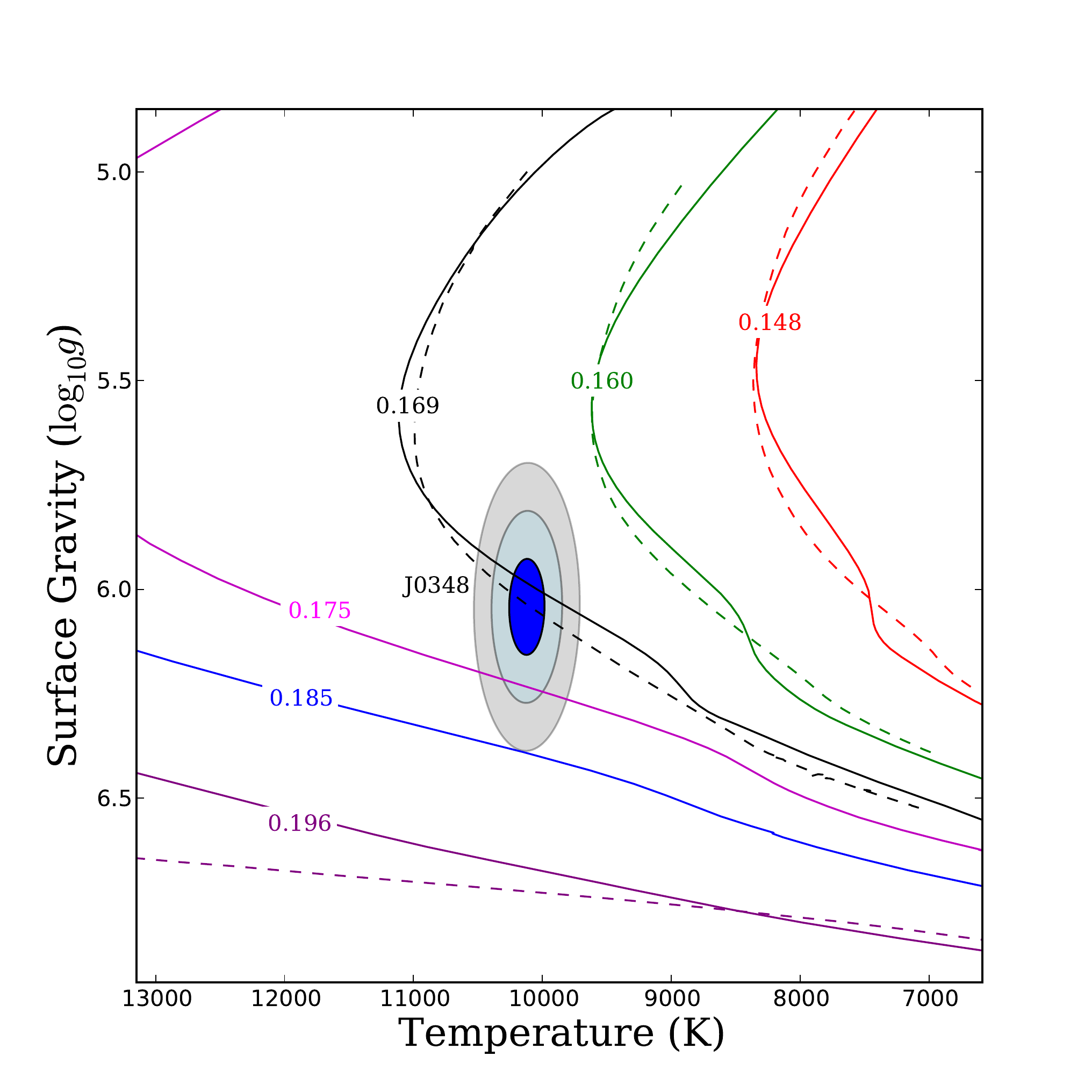} &
   \includegraphics[scale=0.38]{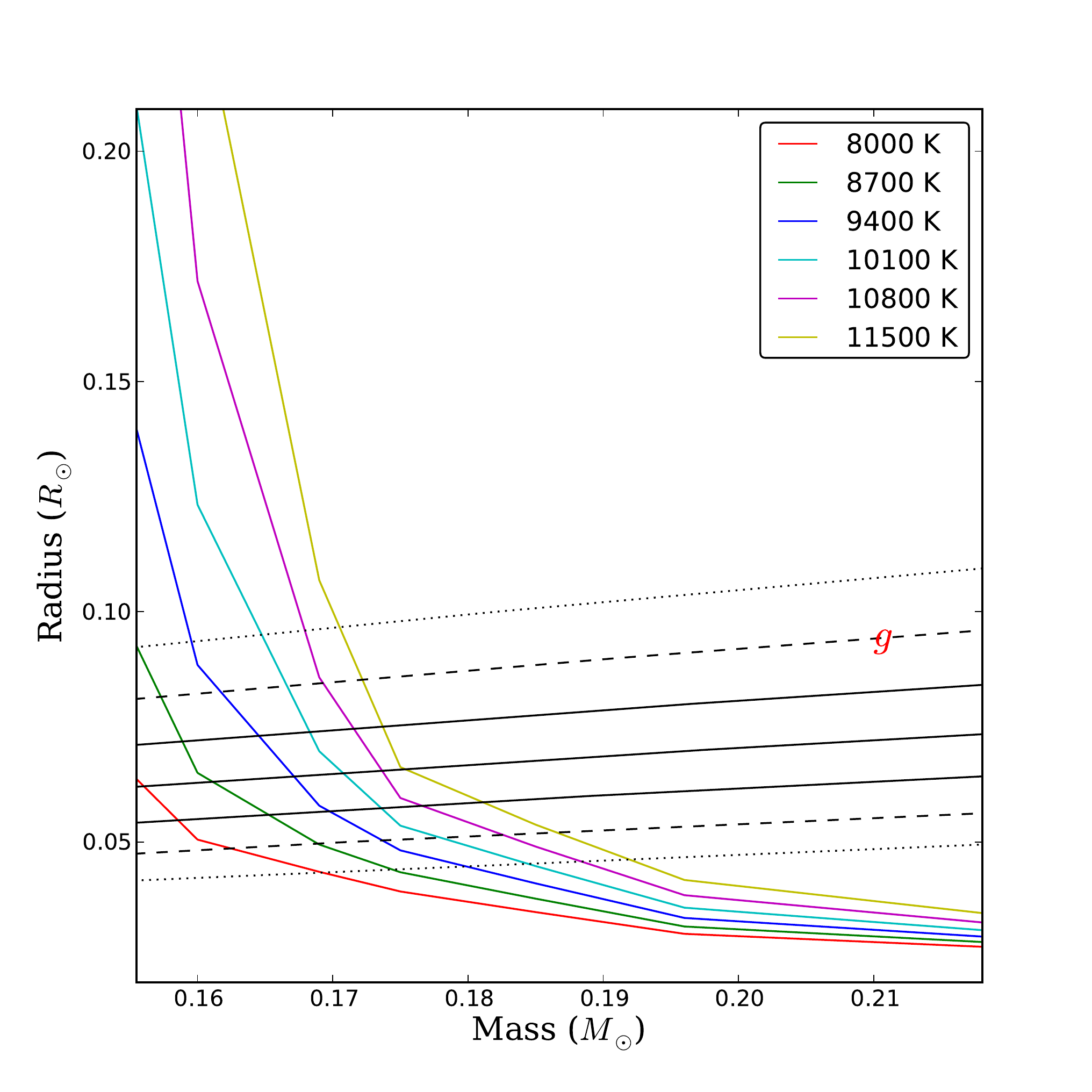}\\

   \end{array}$

\newpage

Figure~3. \\{\bf System Masses and Orbital-Inclination Constraints}\\ 
Constraints on system masses and orbital inclination from
     radio and optical measurements of  \psr\ and its WD companion.
     Each triplet of curves corresponds to the most likely value and standard 
     deviations (68.27\% confidence) of the respective parameters. Of these, two
     (the mass ratio $q$ and the companion mass $M_{\rm WD}$) are 
     independent of specific gravity theories (in black). The contours 
     contain the 68.27 and 95.45\% of the two-dimensional probability distribution. 
     The 
     constraints from the 
     measured intrinsic orbital decay
     ($\dot{P}_{\rm b}^{\rm int}$, in orange) are calculated {\em assuming} that
     GR is the correct theory of gravity. All curves intersect in the same
     region, meaning that GR passes this radiative test \cite{som}.  {\bf Left}: $\cos i$--$M_{\rm WD}$
     plane. The gray region is excluded by the condition $M_{\rm PSR} >$ 0. 
     {\bf Right}: $M_{\rm PSR}$--$M_{\rm WD}$ plane. The gray region is 
     excluded by the condition $\sin i \leq$ 1. The lateral panels depict the 
     one-dimensional  probability-distribution function for the WD mass (right), 
     pulsar mass (upper right) and inclination (upper left) based on the mass function,
      $M_{\rm WD}$ and $q$.
   \begin{center}
   \includegraphics[scale=0.65]{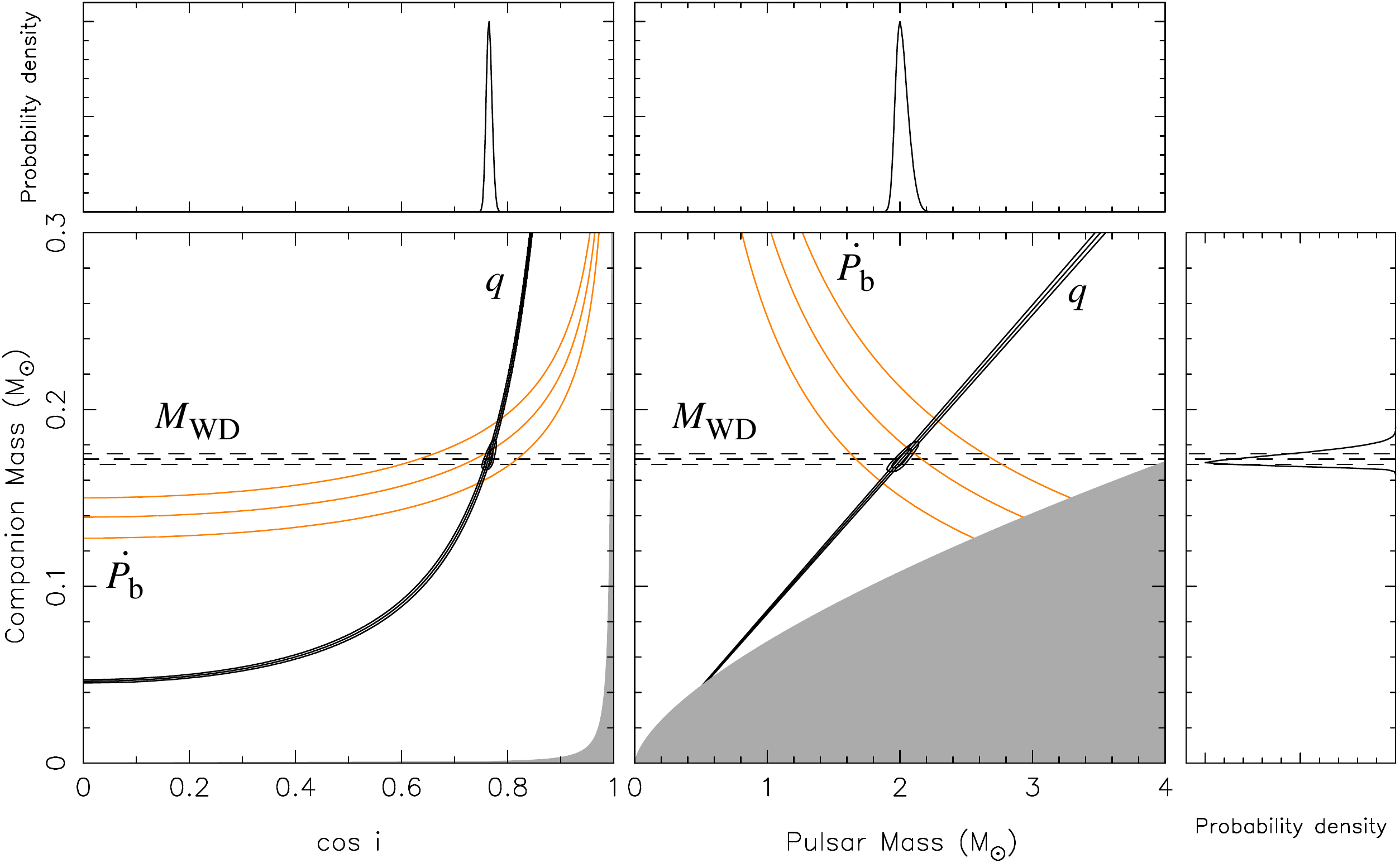}\\

   \end{center}

\newpage
Figure~4. \\{\bf Probing Strong Field Gravity with \psr}\\
\textbf{Left:} Fractional gravitational binding energy as a function of the 
inertial mass of a NS in GR (blue curve). The dots indicate the NSs of 
relativistic NS-NS (in green) and NS-WD (in red) binary-pulsar systems 
currently used for precision gravity tests \cite{som}.
\textbf{Right:}
Effective scalar coupling as a function of the NS mass, in the ``quadratic'' scalar-tensor
theory of \cite{de96}. For the linear coupling of matter to the scalar field we 
have chosen $\alpha_0 = 10^{-4}$, a value well below the sensitivity of any 
near-future Solar System experiment [e.g.\ GAIA \cite{hhl+10}]. The solid curves 
correspond to stable NS configurations for different values of the quadratic 
coupling $\beta_0$: $-5$ to $-4$ (top to bottom) in steps of $0.1$. The yellow 
area indicates the  parameter space allowed by the  best current limit on 
$|\alpha_{\rm PSR} - \alpha_0|$ \cite{fwe+12}, while only the green area is in 
agreement with the limit presented here. \psr\ probes deeper into the non-linear 
strong-field regime due to its high mass. 
 $    \begin{array}{cc}

   \includegraphics[scale=0.35]{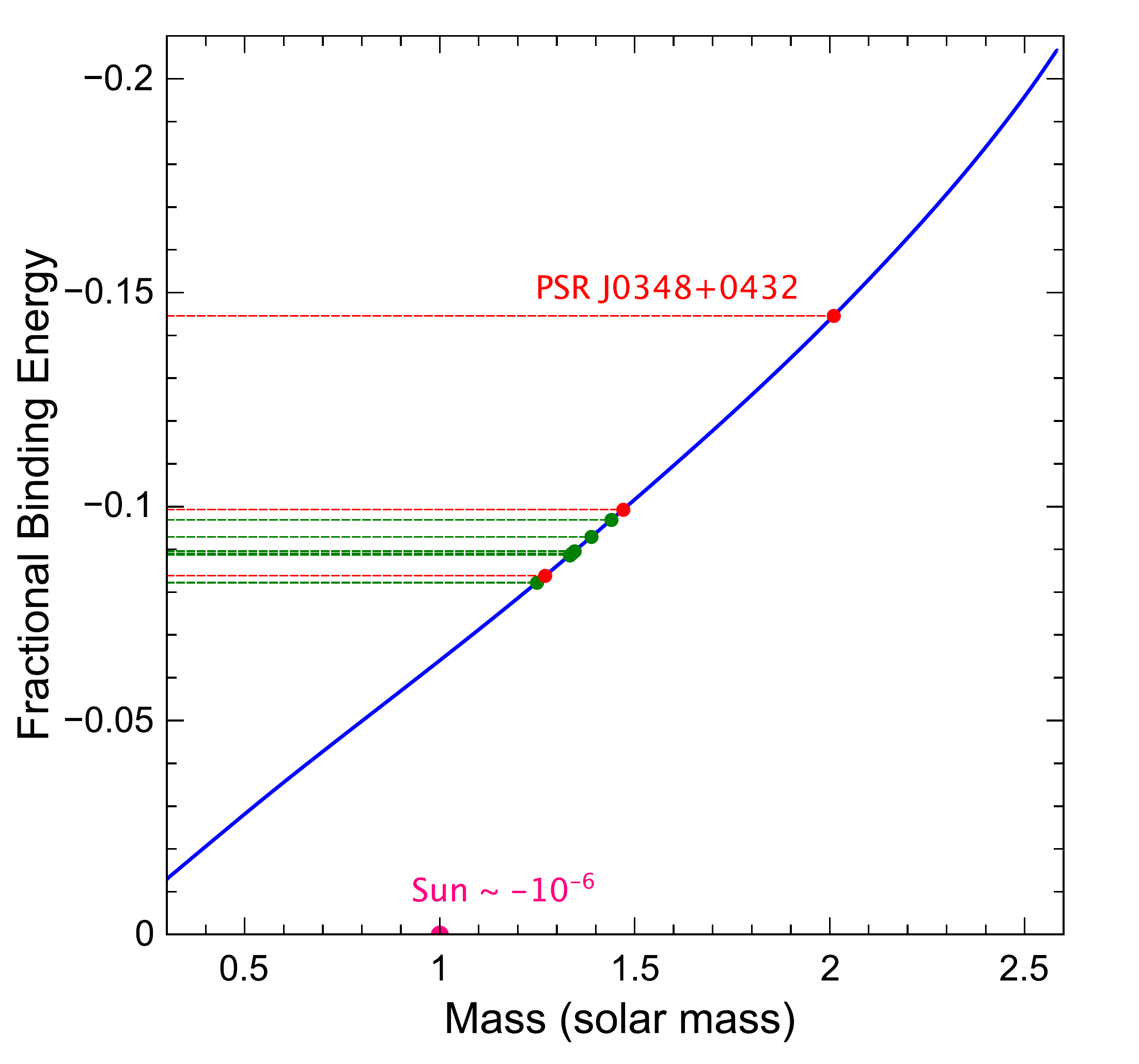} &
   \includegraphics[scale=0.35]{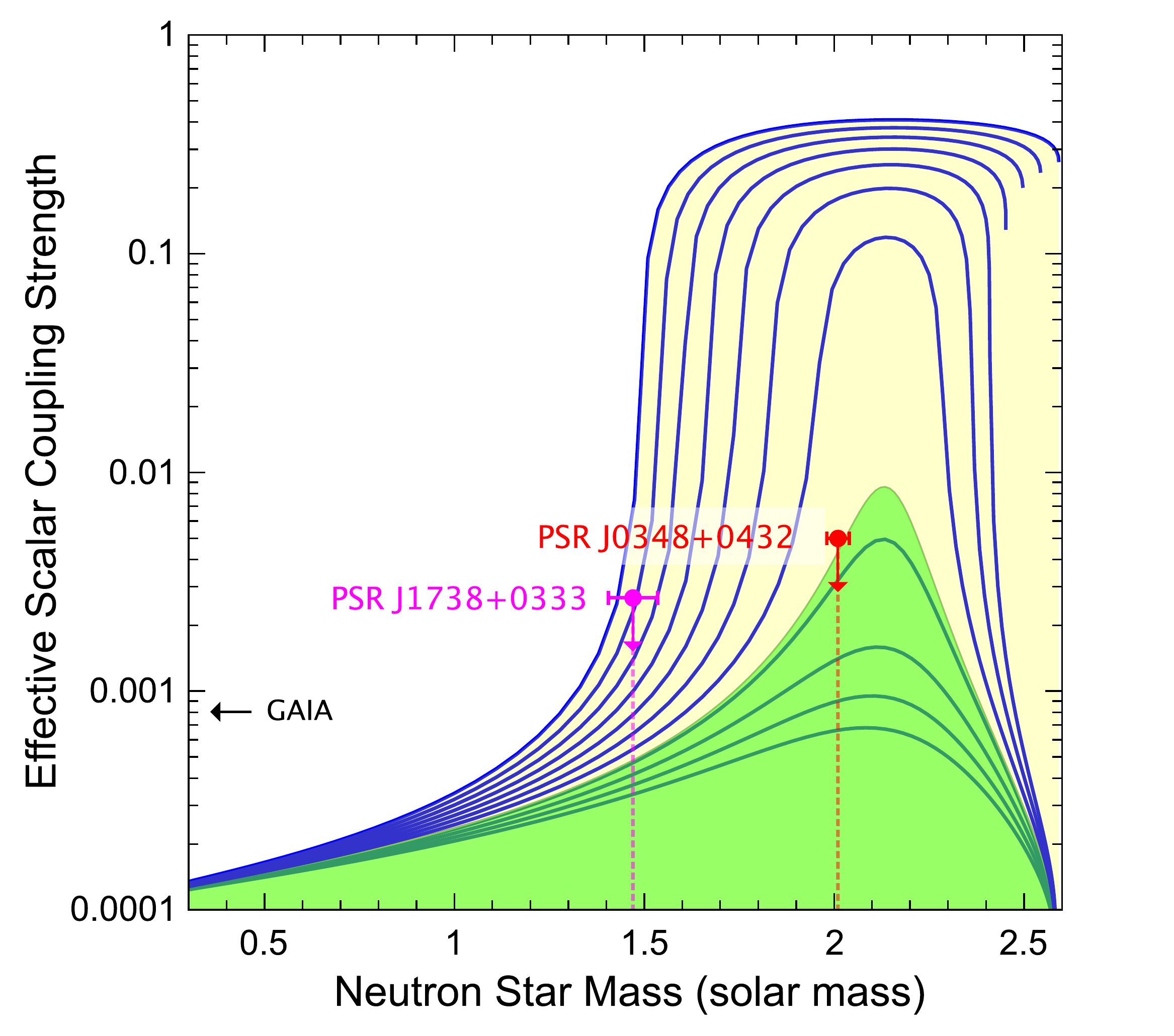}\\

   \end{array}$

\newpage

Figure~5. \\{\bf Constraints on the Phase Offset in Gravitational Wave Cycles in the 
LIGO/VIRGO bands}\\
Maximum offset in GW cycles in the LIGO/VIRGO band (20\,Hz to a few 
kHz) between the GR template and the true phase evolution of the in-spiral in 
the presence of dipolar radiation, as a function of the effective coupling of 
the massive NS for two different system configurations: a 2\,M$_\odot$ NS with 
a 1.25\,M$_\odot$ NS (NS-NS), and a merger of a 2\,M$_\odot$ NS with a 
10\,M$_\odot$ BH (NS-BH). In the NS-NS case, the green line is for $\alpha_B = 
\alpha_0$, and the gray dotted line represents the most conservative, rather 
unphysical, assumption $\alpha_0$ = 0.004 and $\alpha_B$ = 0 \cite{som}. 
In the NS-BH case, 
$\alpha_B$ is set to zero (from the assumption that the no-hair theorem holds). 
The blue line is for $\alpha_0$ = 0.004 
(Solar System limit for scalar-tensor theories), 
and the purple line represents $\alpha_0$ = 0. The gray area to the right of the 
red line is excluded by \psr. In this plot there is no assumption concerning the EOS.
   \begin{center}
   \includegraphics[scale=0.65]{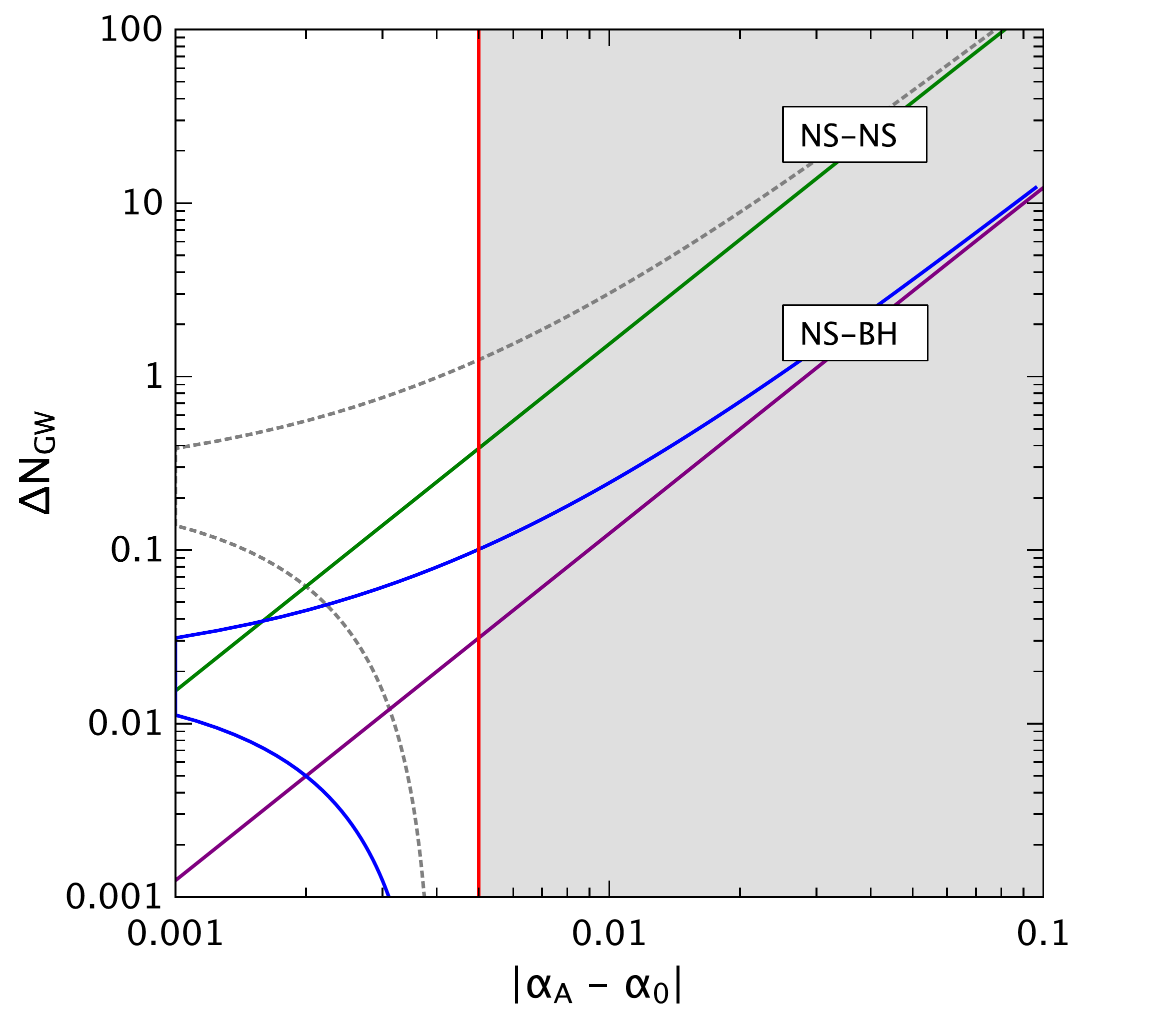}\\

   \end{center}

\newpage 
Figure~6. \\{\bf  Past and Future Orbital Evolution of \psr}\\
Formation of PSR~J0348+0432 from our converging LMXB model calculation. The plot 
shows orbital period as a function of time (calibrated to present day). The 
progenitor detached from its Roche lobe about 2\,Gyr ago (according to the 
estimated cooling age of the WD) when $P_{\rm b}$ $\simeq$ 5\,hr, 
and since then GW damping reduced the orbital period to its present value 
of 2.46\,hr (marked with a star).
   \begin{center}
   \includegraphics[scale=0.65]{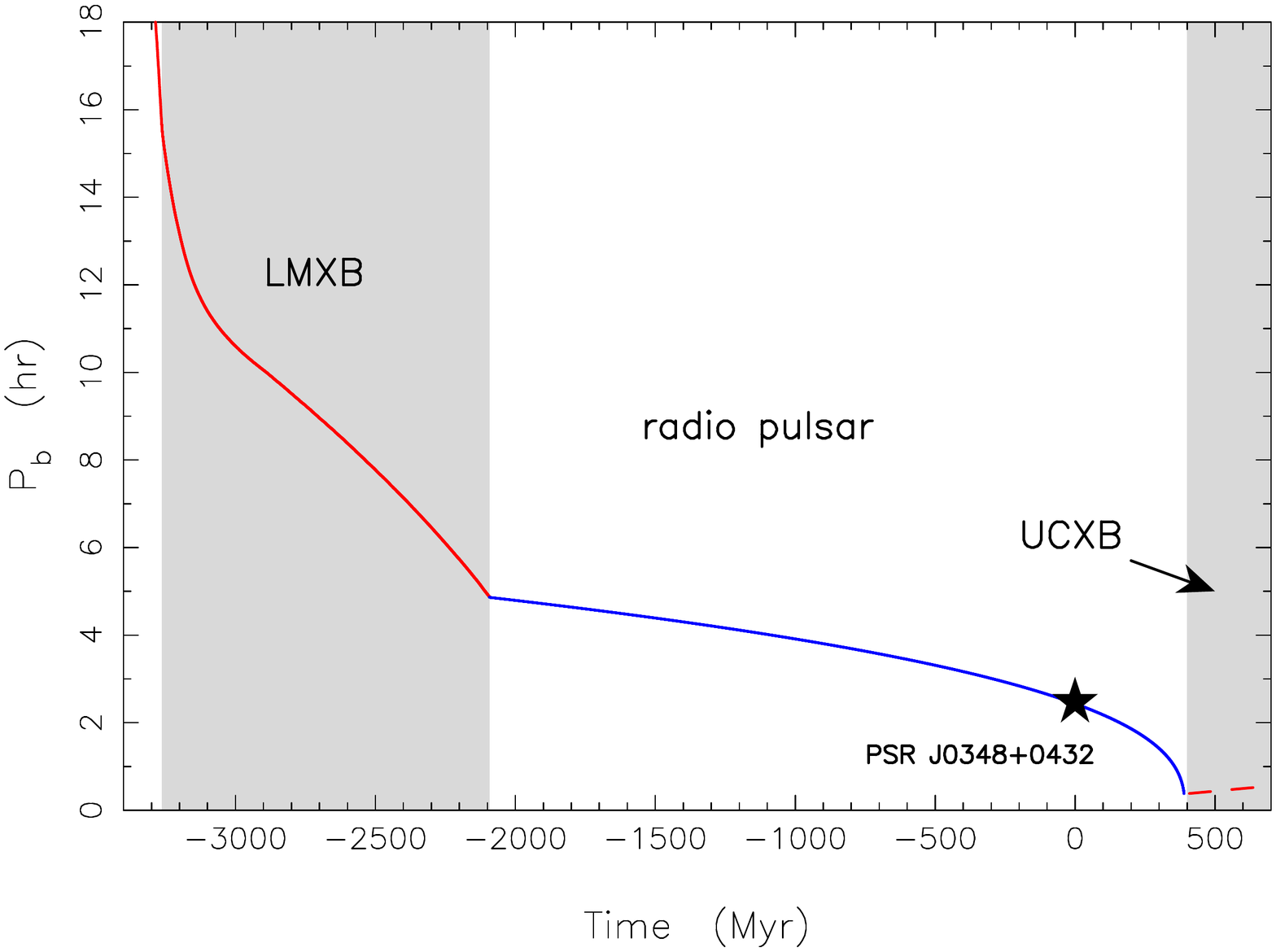}\\

   \end{center}

\newpage
Figure~7.  \\{\bf Possible Formation Channels and Final Fate of \psr}\\ 
 An illustration of the formation and evolution of PSR~J0348+0432.
The zero-age main sequence (ZAMS) mass of the NS progenitor is likely to be 
20 -- 25\,M$_\odot$, 
whereas the WD progenitor had a mass of 1.0 -- 1.6\,M$_{\odot}$ (LMXB) 
or 2.2 -- 5\,M$_\odot$ (common envelope, CE), depending on its formation 
channel. In $\sim$ 400\,Myr (when $P_{\rm b}$ $\simeq$ 23\,min)
the WD will fill its Roche~lobe and the system becomes an ultra-compact 
X-ray binary (UCXB) leading to the formation of a BH or a pulsar with a planet. 
   \begin{center}
   \includegraphics[scale=0.10]{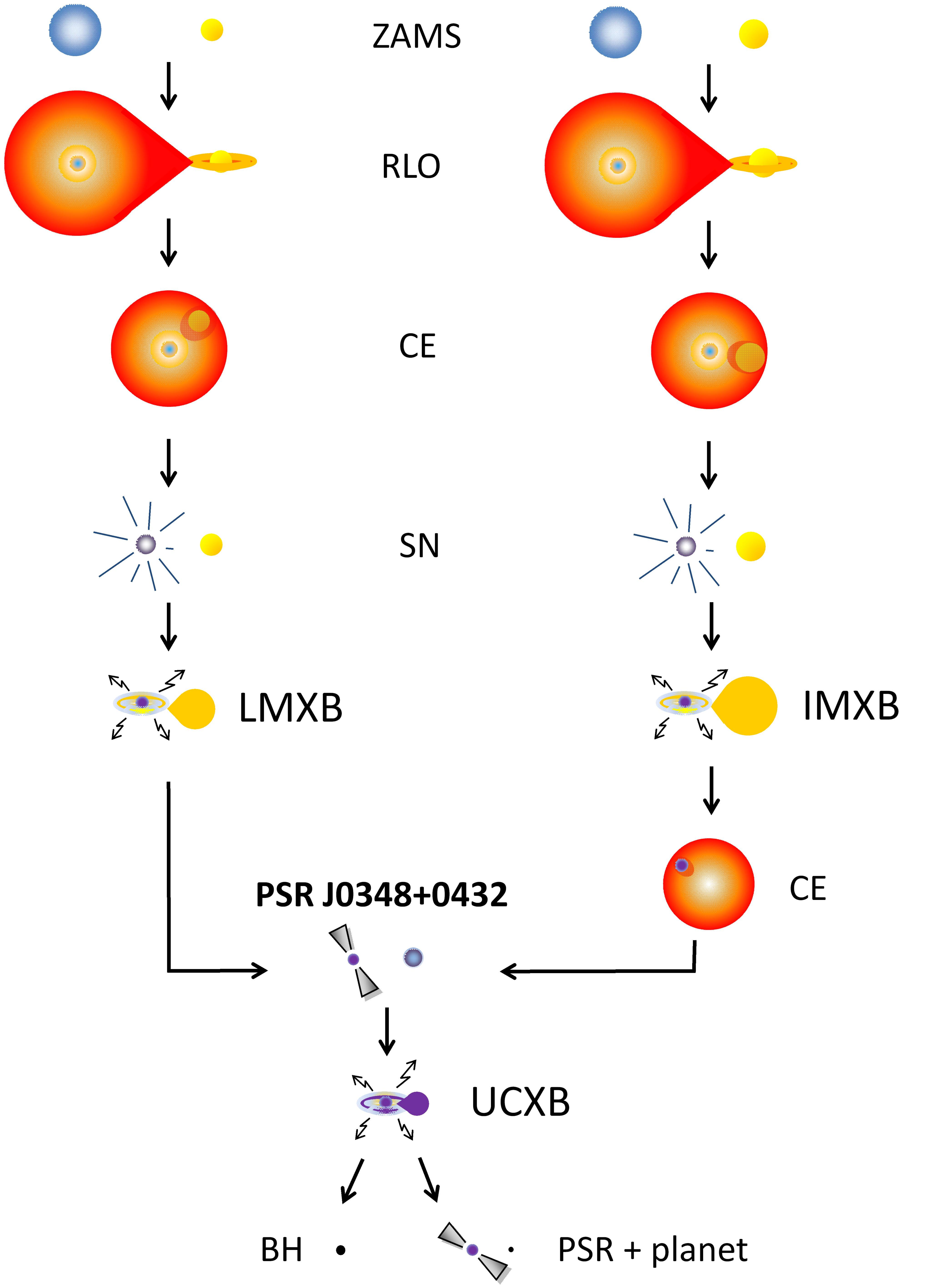}\\

   \end{center}

\section*{Supporting Online Material}
\paragraph*{VLT spectral observations and analysis}
We observed the companion of \psr\ 
during December 19 \& 20, 2011 with the FORS2 \cite{aff+98} instrument on
Unite-Telescope\,1 (Ant\'{u}), using its blue
sensitive E2V CCD detectors and the G1200B grism. This setup
delivers a resolution of 0.36\,$\AA$\,per\,binned-by-two-pixel along 
dispersion and 0\farcs25\,per\,binned-by-two-pixel along the spatial 
direction. Because of the short orbital period of the binary, we chose 
a relatively wide 1\arcsec\ slit to 
avoid severe radial velocity smearing (by reducing the exposure time) 
and minimize possible dispersion losses not corrected
by the dispersion corrector of the instrument. 
However, this choice may potentially result in
 systematic offsets in radial velocity measurements due to
non-uniform illumination of the slit.   
To monitor these effects we rotated the slit by $134\fdg8$ 
(north-through east)  
with respect to the parallactic angle
to include a bright nearby star for local flux and velocity
calibration (Fig.~S1).
Our setup covers the spectral range from $\sim 3700$ to 5200\,$\AA$
with a resolution ranging from $\sim2$ to 3\,$\AA$ depending on the seeing. 
During the first night the conditions were good to photometric
and the seeing varied between $\sim$ 0\farcs7 and 1\farcs2. The second 
night was sporadically plagued with thin cirrus and the seeing ranged from 
$\sim$ 0\farcs9 to 1\farcs7.
Bias, flat and Mercury-Cadmium (HgCd) 
frames for wavelength calibration were collected during
day-time after each run.
We collected a total of 34 spectra of  the white dwarf companion to \psr\  
and the nearby comparison star. 
Of these, 22 had 800-s
exposures and were taken with the slit rotated 
by the angle mentioned above, 
4 were taken with 850-s exposures during bad weather instances and
8 with the slit rotated by a slightly different angle during experimental
stages. In addition, we collected 2 spectra of the comparison through a wider, 
2\farcs5 slit and spectra of several
flux standards at the beginning and the end of each run through both 1\arcsec\
and 2\farcs5 slits. 

We reduced the data using
routines inside the Munich Image and Data Analysis System (\textsc{midas}).
Our analysis, from cosmetic corrections to extraction of spectra,
is identical to that followed for the white dwarf companion to PSR\,J1738+0333
and is described in detail elsewhere \cite{avk+12}.
The dispersion solution has root-mean-square (rms) residuals of 
$\sim 0.03$\,$\AA$ for 18 lines. 
Flux calibration was performed separately  for each night by comparing 
flux-standard observations (Table~S1) with high S/N templates
or appropriate white dwarf model spectra \cite{koe08}. 
Overall, the response curves from each standard are consistent with each other, 
with the largest differences (up to 10\%)  observed at short 
wavelengths ($\lambda \le 4000$\,$\AA$); 
we use their average for flux calibration.
Prior to the latter, 
we corrected the narrow-slit spectra for wavelength dependent slit losses
using the wide slit spectra of the comparison and accounted for atmospheric 
extinction using the average extinction curve for La Silla.

\paragraph*{Radial velocities}
We extracted the radial velocities of the white dwarf companion and the nearby comparison 
star following the procedure described in \cite{avk+12}.
First, we identified the nearby comparison star as being a type G1V star 
(with an uncertainty of about 2 subtypes) and used a high-resolution
spectrum of the similar star HD\,20807\cite{bjl+03} as a template. 
For the \psr\ white-dwarf companion we fitted a high S/N spectrum with a grid 
of DA model atmospheres \cite{koe08}, used the best-fit 
template to measure radial velocities, averaged the zero-velocity 
spectra and finally re-fitted the average spectrum 
to determine the final template. 
We scanned a grid of velocities from $-800$ to $+800$\,km\,s$^{-1}$ with
a step-size of 5\,km\,s$^{-1}$. The best fits had $\chi_{\rm red,min}^2 =1-1.5$ 
and $\chi_{\rm red,min}^2 = 1.1-3.0$ for the comparison star. As described in 
the main paper, we scaled the errors to account for the fact that $\chi_{\rm red,min}^2$ was not equal to unity. 

The velocities of the comparison star show a peak-to-peak variation of 
$\sim40$\,km\,s$^{-1}$, much higher than the typical 
0.8\,km\,s$^{-1}$ measurement error.
While we find no evidence for binarity, the measurements
form 5 distinct groups,  each of which  display a variability
only marginally higher than the formal errors. These coincide
with blocks of observations interrupted for target repositioning. 
The scatter of velocities is therefore clearly related to the instrument and most
probably associated with positioning uncertainties. 
For this reason we chose to use velocities relative to the comparison. 

The best-fit solution using all available (barycentred) data gave 
$K_{\rm WD}=346\pm6$\,km\,s$^{-1}$ and a systemic velocity of 
$\delta\gamma = +8\pm4$\,km\,s$^{-1}$ relative 
to the comparison with $\chi^2_{\rm red, min} = 2.78$ for 32 degrees of 
freedom (dof). However, 8 of the observations used here were taken with
the slit at a different angle and the white dwarf's 
velocity relative to the comparison 
is thus most likely contaminated with an extra systematic shift
due to slit rotation.
For this reason we neglect these data. Using the homogeneous 
set of observations only and 
further rejecting one outlier with spuriously shaped
continuum (no.\ 15 in Table~S1) we obtain 
$K_{\rm WD}=345\pm4$\,km\,s$^{-1}$ and
$\delta \gamma= +23 \pm 5$\,km\,s$^{-1}$ respectively
 with $\chi^2_{\rm red, min} = 0.99$ for 23 degrees of freedom (Fig.~1a). 

After correcting for the small effect of orbital smearing 
($\sin(\pi \langle t_{\rm exp}\rangle/P_{\rm b})/
(\pi \langle t_{\rm exp} \rangle/P_{\rm b})=0.98636$) 
we find a semi-amplitude of $K_{\rm WD} = 351\pm4$\,km\,s$^{-1}$. 
The best-fit systemic velocity of \psr\  using the raw white dwarf
velocity measurements is $\gamma = -1 \pm6$\,km\,s$^{-1}$. 
Given the large scatter of the comparison's velocity we adopt
$\gamma = -1\pm20$\,km\,s$^{-1}$ with the uncertainty being a conservative
estimate based on the scatter of the data. 

\paragraph*{Average spectrum and atmospheric parameters}
We scanned a grid of models covering effective temperatures from 
$T_{\rm eff}=8000$ to 25000\,K with a step-size of 250\,K and 
surface gravities ranging from $\log g = 5.00$ to $\log g = 8.00$ 
with a step-size of 0.25\,dex. 
At each point of the grid we fitted for the 
normalization using a polynomial function of the wavelength to account
for non-perfect flux calibration. Analysis of statistical  errors is again identical to
that followed for the white dwarf companion of PSR\,J1738+0333 \cite{avk+12}. 
We achieved the best fit to higher Balmer lines when excluding the 
continuum regions between $4000-4050$, $4180-4270$ 
and $4400-4790$\,$\AA$, which had small irregularities due to leftover detector
imperfections: $T_{\rm eff} = 10120 \pm 35$\,K
and $\log g = 6.042 \pm 0.032$ ($1\sigma$) with $\chi^2_{\rm red, min} = 1.02$.
To estimate the influence of systematics we varied the 
degree of the polynomial used for normalization (1st to 5th degree), 
the spectral regions used for the fit (lines-only to whole spectrum) 
and the assumed spectral resolution (by steps of $\sim 5\%$).
We also searched for velocity smearing 
by checking the consistency of the solution in  an average of 
spectra taken close to orbital conjunction 
 and an average of spectra taken close to the nodes.
Finally, we fitted each line (from H$\beta$ to H12) separately to
verify the consistency of the fit over the spectrum and examined
the influence of our flux calibration by fitting 
the average uncalibrated spectrum.
Overall, all  tests gave fits consistent within statistical errors
with only few exceptions that had (higher) central values that differed by 
120\,K and 0.11\,dex  compared to the numbers above.
The good agreement is probably due to the high S/N of the spectrum. 
The values adopted in the main paper are based on the solution using a 
third degree polynomial and the systematic error is a conservative estimate 
based on the scatter of the different fits mentioned above.

\paragraph*{Spectroscopic modeling and the ``high $\log g$'' problem}

Spectroscopic modeling of the Balmer lines in \textit{higher} mass white dwarfs 
shows a spurious increase in surface gravity for stars with temperatures between 
$\sim$ 8000 and 11000\,K. This well-known problem is linked to the incomplete 
treatment of convection in 1-D atmospheric models  and disappears with the use 
of 3-D model atmospheres \cite{tls+11,tls+13}. However, our modeling below shows 
that for the parameter space relevant to the \psr\ companion, the atmosphere is  
not yet convective (e.g.\ convection sets in at $T_{\rm eff }\le 9300$\,K for 
$M_{\rm WD} = 0.17$\,M$_\odot$). Therefore this problem is very unlikely to be 
relevant for the mass determination presented here. 

\paragraph*{Initial white-dwarf models}
To construct the white dwarf models presented in the main paper, 
we evolved solar composition stars (metal mass-fraction of $Z=0.02$) with masses
between 1.0 and 1.5\,M$_\odot$ and applied a large 
mass-loss wind at various points on the Red Giant Branch (RGB).
To constrain the upper limit of the envelope mass expected from natural 
binary evolution, we removed the mass before the star enters the 
asymptotic RGB, letting the star evolve and contract naturally 
to become a white dwarf. 
Our upper limits agree well with the results of previous studies
 \cite{dsb+98,sar+01,pac07}.
Finally, to fully control the envelope mass of the white dwarf at the final 
stages of evolution we neglected hydrogen
fusion through the CNO bi-cycle that is responsible for the 
hydrogen shell flashes\footnote{For a white dwarf at the final cooling branch, 
CNO luminosity accounts for less than 5\% of the total energy budget.
Hence, it is safe to neglect it without influencing the macroscopic 
characteristics of the models \cite{sba10}.}. 

In Fig.~S2 we show the post-contraction white dwarf cooling age when 
$T_{\rm eff}=10000$\,K, as a function of the total hydrogen mass 
(after cessation of the mass transfer), 
for  masses ranging from 0.155 to 0.185\,M$_\odot$.  
For low envelope masses, hydrogen burning cannot 
be initiated and  the white dwarf quickly radiates the latent thermal energy of the core and cools in a few Myr.  
The thick-envelope modes presented in the main text were constructed as above. 
 
 \paragraph*{Metallicity}

The metallicity of the white dwarf plays an important role in both regulating 
the CNO luminosity and changing the chemical profile of the stellar envelope. 
Qualitatively, our main models described above are in good agreement 
with the $Z$ = 0.001 models of \cite{sar+02} for the parameter space relevant
to the white dwarf companion to \psr. 
Specifically, their 0.172\,M$_\odot$ track has a thick envelope and 
predicts a surface gravity of $\log g = 6.13$ for $T_{\rm eff} = 10000$\,K 
which is reached at a cooling age of $\tau_{\rm cool} = 2.85$\,Gyr. This 
agreement is not surprising given that CNO burning is neglected in our analysis, 
 convective mixing has not yet set in at $T=10000$\,K and consequently metals 
are absent from surface layers due to gravitational settling. Therefore we 
consider that any uncertainties due to metallicity are small and anyway included 
in our adopted errors. 


\paragraph*{Input physics of the stellar evolution models}

Stellar models used in our analysis were constructed using the 1-D stellar
evolution code ``\texttt{star}" provided with the 
Modules for Experiments in Stellar Astrophysics (\textsc{mesa}) \cite{pbd+11}.  
\texttt{star} solves for the equations of hydrostatic equilibrium, 
nuclear energy generation, convection and  time-dependent 
element diffusion using a self-adaptive 
non-Lagrangian  mesh and analytic Jacobians. 
We used default options for the equation-of-state, radiative and
neutrino opacities, thermonuclear and weak reaction rates described in 
\cite{pbd+11} and references therein. 
We implemented the mixing length theory of convection from \cite{hvb65} that
takes into account radiative losses near the outer layers of the star.
Diffusion was taken into account using the method and coefficients from \cite{tbl94}
and transport of material was calculated using the method described in \cite{im85}
after grouping the elements in ``classes"  in terms of atomic mass ranges.
Finally, boundary atmospheric conditions were calculated using the 
gray-atmosphere approach of \cite{edd26}.

\paragraph*{Photometry}

A photometric campaign on the white dwarf companion to \psr\ was carried out during February 1, 2012  
using the ULTRACAM instrument \cite{ultracam} on the 4.2-m William-Herschel 
Telescope at La Palma, Spain. 
The data were reduced using the standard ULTRACAM pipeline (Fig.~S3).

The lightcurves have an rms scatter of $\sim$0.53, 0.07 and 0.08\,mag in 
$u'$, $g'$ and $r'$ respectively and show no evidence for variability 
over the course of the observations. 
The phase-folded light-curve shows no variability either. 
Additionally, our calibrated magnitudes are consistent with the SDSS catalogue 
magnitudes implying no significant variability 
at the $\sim 5$\,yr time-scale.

Qualitatively, this result supports
the use of the \psr\ system as a gravitational laboratory (see timing analysis below). 
In what follows, we discuss the limits on various parameters in more detail.
Three effects that cause phase-dependent variability are: deformation of
the white dwarf by tides raised by the neutron star, irradiation by the
pulsar wind, and Doppler boosting caused by the white dwarf's orbital motion.
For a circular orbit, the combined modulation in photon rate $n_{\gamma}$
is given by 
\begin{equation}
\Delta n_{\gamma}/n_{\gamma,0}\simeq 
 f_{\rm ell} \left(\frac{R_{\rm WD}}{a}\right)^3 q \sin^2i \cos(4\pi\phi) 
+ f_{\rm db}\frac{K_{\rm WD}}{c} \sin i \cos(2\pi\phi) 
- f_{\rm irr}\frac{T_{\rm irr}^4}{32T_{\rm eff}^4} \sin i \sin(2\pi\phi),
\end{equation}
with $f_{\rm ell}$, $f_{\rm db}$, and $f_{\rm irr}$ factors of order
unity describing the observability in a given filter, $0\leq\phi\leq1$ the
orbital phase, and $T_{\rm irr}=L_{\rm psr}/4\pi a^2\sigma\simeq2400\,$K
the effective temperature corresponding to the pulsar flux incident on
the white dwarf.  We find that all terms should be small.  For the tidal
deformation, $f_{\rm ell}=-3(15+u_1)(1+\tau_1)/20(3-u_1)=1.75$, 
where we use linear approximations for limb and gravity darkening,
with coefficients $u_1=0.36$ \cite{hkb+12} and $\tau_1=1$ (appropriate 
for a radiative atmosphere).  Thus, the expected modulation is $1.5\times10^{-3}$.  For
the Doppler boosting, approximating the white dwarf as a black-body
emitter, $f_{\rm db}\simeq
\alpha\exp\alpha/(\exp\alpha-1)\simeq 2.6$, where $\alpha=hc/\lambda
kT_{\rm eff}\simeq 2.8$ \cite{vkr+10}, with $\lambda\simeq550\,$nm the typical
observing wavelength.  Hence, the expected amplitude is
$\sim\!3\times10^{-3}$.  Finally, for the irradiation, $f_{\rm
irr}=(1-A)f_{\rm db}\leq2$, where the maximum is for albedo
$A\simeq0$.  Thus, irradiation could cause a modulation of up to
$\sim\!1.2\times10^{-4}$.

Fitting the observed lightcurves with a function of the form
$\Delta n_\gamma/\langle n_{\gamma}\rangle=1+a_{\rm ell}\cos(4\pi\phi)+a_{\rm
db}\cos(2\pi\phi)-a_{irr}\sin(2\pi\phi)$, we find good fits
($\chi^2_{\rm red}\simeq1$) but no significant detections, with averaged
amplitudes of the higher S/N $r$ and $g$ band lightcurves of
$a_{\rm ell}=0.003\pm0.003$, $a_{\rm db}=0.003\pm0.003$ and $a_{\rm
irr}=0.006\pm0.004$.  The marginal irradiation signal would correspond
to a temperature difference between the irradiated and non-irradiated
side of $\sim\!100\,$K, which is substantially larger than the expected
difference of~$2\,$K.  Even if confirmed, however, this would not
 affect our inferred radial velocity amplitude or
white dwarf parameters. 

Finally, another possible source of variability is quadrupole moment variations 
of the white dwarf \cite{sba10}: these typically change the star's luminosity 
by a few per-cent (e.g. $\sim 20\%$ for the only three known 
cases of pulsating low-mass white dwarfs \cite{hmw+12,hmw+12b})  
and result in changes of the orbital period $P_{\rm b}$ through classical 
spin-orbit coupling \cite{app92}. 
To our knowledge, all possible mechanisms for such variations would result 
in modulations much higher than the precision of our lightcurve. Therefore we
can neglect this effect and  assume that the star is in equilibrium. Our
assumption is further supported by the lack of second or higher-order 
derivatives 
in the measured orbital period (see below) and recent theoretical findings 
\cite{cra+12} 
that locate the instability  strip for g-mode oscillations outside the parameter
space relevant for the white dwarf companion to \psr.

The SDSS photometry places a constraint on
the distance to the system. Adopting the model of \cite{sfd98} for the interstellar 
reddening and the 0.169\,M$_\odot$ cooling track of \cite{sar+01}, we find 
that the luminosities (Fig.~S3) are consistent with a distance of $d\simeq2.1$\,kpc 
(and a reddening of $A_{\rm V} \sim 0.7$).
Given the uncertainties in the models the error is difficult to estimate but it should
be better than $\sim 10\%$. 
Our estimate is also consistent with the 
distance of $d_{\rm DM} \sim 2$\,kpc implied by
the dispersion measure (DM) of the pulsar and the {\sc NE2001} 
model for the Galactic free electron density \cite{cl02}.

\paragraph*{Radio observations}

The observing setup for the Arecibo telescope is identical to the well-tested 
setup described in
\cite{fwe+12}, with the exception of one WAPP now
being centered at 1610 MHz instead of 1310 MHz; the
former band is cleaner and its use improves the precision
of our DM measurements. Also, as in the former case,
data are taken in search mode and processed off-line.
This allows for iterative improvement of the pulsar ephemeris
which is important at the early stages when the timing
parameters are not yet very precise. With each improved
ephemeris, we de-disperse and re-fold the data,
obtaining pulse profiles with higher S/N 
that yield more accurate pulse times-of-arrival (TOAs). 
This helps to avoid orbital-phase dependent smearing and 
 timing artefacts, which may corrupt the
determination of orbital parameters, particularly
the orbital phase and orbital period variation \cite{nsk+08}.

We de-disperse and fold the
radio spectra following the procedure described in  \cite{fwe+12}. 
TOAs are derived every 4 minutes to
preserve the orbital information in the signal.
The pulse profile template, resulting from more than 1 
hour of data, is displayed in Fig.~\ref{fig:profile}. 
Although the pulse profile changes significantly from
350 to 2200 MHz \cite{discovery_paper}, the changes within the
band of the L-wide receiver used for timing
(1100-1660 MHz, also displayed in Fig.~\ref{fig:profile})
are small enough for us to consider this single average
profile taken at 1410 MHz as a good template for all the
data. The latter is cross-correlated with every
4-minute/25\,MHz-wide pulse profile in the Fourier
domain \cite{lk05,tay92} and the phase offset that yields the 
best match is used to derive the topocentric TOA 
of a reference sub-pulse (normally that
closest to the start of each sub-integration).
The results described below are obtained using 7773 TOAs 
with stated rms uncertainty smaller than 10\,$\mu$s.

In order to verify the Arecibo data we have been timing
\psr\, with the 100-m radio telescope
in Effelsberg, Germany, which has a very different
observing system. The polarization characterization of
the radio emission of \psr, displayed in the top plot of
Fig.~\ref{fig:profile}, was made with this telescope.
Overlaid on the polarization data is a theoretical
Rotating Vector Model (RVM). It is generally difficult to fit a
RVM model to polarization data from recycled pulsars, but
for \psr\, this model works surprisingly well. 
For instance, as explained in \cite{lk05}, the covariance between 
the angle between the spin and magnetic axis, $\alpha$, and the angle between 
the spin axis and the line of sight $\zeta$, allows for a wide range of possible solutions (Fig.\,S5). However, 
if we assume that during the
accretion episode that recycled the pulsar the
spin axis of the pulsar was aligned with the orbital angular
momentum (which has an angle $i = 40\fdg2 \pm 0\fdg6$
to the line of sight) then $\alpha \simeq 45^\circ$. The
minimum angle between the magnetic axis and the line of
sight is then given by $\beta = \zeta - \alpha = -5\fdg$.

Apart from the polarimetry, the Effelsberg data yielded a total of 179
high-quality TOAs. As can be seen in Fig.~\ref{fig:residuals},
these follow the Arecibo timing very closely, providing added
confidence in both.

\paragraph*{Timing analysis}
The combined timing dataset contains 8121 TOAs.
The TOA  residuals obtained with the best ephemeris (Table~1) 
are displayed as a function of time 
in the top panel and as a function of orbital phase in the
bottom plot of Fig.~\ref{fig:residuals}.
To derive the ephemeris in Table~1
(using {\sc tempo2}) we increased
the TOA uncertainties by factors of 1.3 for the GBT and Arecibo data and by
1.8 for the Effelsberg data. This results in the residuals of each
dataset having a normalized $\chi^2$ of 1. Using these slightly
increased (but more realistic) TOA uncertainties results in
more conservative (i.e. larger) uncertainties for the fitted timing parameters.
Globally, the residuals have a weighted rms of 4.6$\,\mu$s and  the reduced
$\chi^2$ is 1.019 for 8102 degrees of freedom. The TOA uncertainties
presented in Fig.~\ref{fig:residuals} are those used to derive the
timing solution.

The orbit of \psr\ has a very low eccentricity, therefore we use the
``ELL1'' orbital model \cite{lcw+01} to parametrize it.\footnote{The
ELL1 timing model as implemented in the {\sc tempo2} software package 
is a modification of the DD timing 
model \cite{dd85,dd86} adapted to low-eccentricity binary pulsars.
In terms of  
post-Keplerian observables, it contains all those which are numerically
relevant for systems with $e \ll 1$. The ``Einstein delay" term is
not relevant for such systems and is therefore not taken into account.}
This parametrization yields Keplerian and
post-Keplerian parameters very weakly correlated with each
other. In order to estimate the intrinsic (``real'') eccentricity of the
binary  (Table~1) we adopt $M_{\rm WD}\,=\,0.172$\,M$_\odot$ and 
$i\,=\,40\fdg 2$ obtained from the optical observations. 
This assumption is safe because GR is known to provide a
sufficiently accurate description of spacetime
around weakly self-gravitating objects \cite{bit03}.  
According to \cite{fw10}, the orthometric amplitude of the
Shapiro delay (which quantifies the time amplitude of
the {\em measurable} part of the Shapiro delay) is $h_3\,=\,42\,$ns.  
Fitting for this quantity
 we obtain $h_3\,=\,69\,\pm\, 53\,$ns.
This is 1-$\sigma$ consistent with the prediction but
the low relative precision of this measurement implies that
we cannot determine $M_{\rm WD}$ and $\sin i$ independently
from the existing timing data. A precise measurement of the
component masses of this system from Shapiro delay would
require an improvement in timing precision that
is much beyond our current capabilities.

\paragraph*{Intrinsic orbital decay}

As described in the main text, we detect an orbital decay
consistent with the prediction of General Relativity.
When we say that this decay is stable, we mean
that we detect no higher-order
variations of the orbital frequency $f_{\rm b} \equiv 1 / P_{\rm b}$
nor large variations in $x \equiv a_p \sin i / c$:
\begin{eqnarray}
\frac{d^2 f_{\rm b}}{dt^2} & = &-4.5 \pm 4.4 \times 10^{-23}\rm \, Hz\, s^{-2}, \\
\frac{d^3 f_{\rm b}}{dt^3} & = &+4.1 \pm 2.5 \times 10^{-36}\rm \, Hz\, s^{-3}, \\
\frac{d x}{dt} & = & +7.4 \pm 4.4 \times 10^{-15} \rm \, s\, s^{-1},
\end{eqnarray}
where the values and 1-$\sigma$ uncertainties
were obtained using the {\sc tempo}
implementation of the BTX orbital model.
In systems where the quadrupole moment of the white dwarf changes,
we should expect such timing effects \cite{lvt+11}
plus significant photometric variations with orbital phase
(discussed above). Since none are observed,
the companion to \psr\, is very likely to
have a stable quadrupole moment.

The constraints on the total proper motion $\mu$ combined 
with the optically derived distance $d= 2.1\pm0.2$\,kpc allow us to
calculate the two kinematic corrections to the observed $\dot{P}_{\rm b}$.
The more important one is the Shklovskii effect \cite{shk70}:
\begin{equation}
\dot{P}_{\rm b}^{\rm Shk} = P_{\rm b} \frac{\mu^2 d}{c} =
0.0129^{+0.0025}_{-0.0021} \times 10^{-13} \,\rm s\,s^{-1},
\end{equation}
where we have adopted the 10\% error-estimate on the distance. 
The second correction is caused by the difference of
Galactic accelerations between the binary and the Solar System.
Using the detailed procedure outlined in \cite{fwe+12}, we
obtain:
\begin{equation}
\dot{P}_{\rm b}^{\rm Acc} = P_{\rm b} \frac{a_c}{c} =
0.0037^{+0.0006}_{-0.0005} \times 10^{-13} \,\rm s\,s^{-1}.
\end{equation}
A third correction could arise from a possible variation of the gravitational 
constant $\dot{G}$. Conservative limits are given by 
\cite{dgt88,dt91,nor90a,nor90b}:
\begin{equation}
  \dot{P}_{\rm b}^{\dot{G}} 
  = - 2 P_{\rm b} \frac{\dot{G}}{G} 
  = (0.0003\pm 0.0018) \times 10^{-13} \,{\rm s\,s}^{-1},
\end{equation}
where we used the latest limit on $\dot{G}$ from Lunar Laser Ranging \cite{hmb10}.

Adding these corrections, we obtain a total of
$\sim (+1.6\pm 0.3) \times 10^{-15}\,\rm s\,s^{-1}$, or about 0.006
of the measured value. This is much smaller than
the current measurement uncertainty and therefore
we can conclude that, at the current precision limit, 
the observed value is intrinsic to the system.
Its magnitude is entirely consistent with the GR 
prediction for the orbital decay caused by
emission of gravitational waves:
$\dot{P}_{\rm b} / \dot{P}_{\rm b}^{\rm \, GR} = 1.05 \pm 0.18$. This
agreement is depicted graphically in a $\cos i - M_{\rm WD}$ and $M_{\rm PSR} - 
M_{\rm WD}$ diagram (Fig.~3); the
consequences  are discussed in the main paper and
in detail further below. 

\paragraph*{Mass loss contribution to $\dot{P}_{\rm b}$}

If the system is losing mass, that should cause a change
in the orbital period \cite{dt91}:
\begin{equation}\label{eq:massloss}
  \dot{P}_{b}^{\dot{M}} = 2 \frac{\dot{M}_{\rm T}}{M_{\rm T}} P_{\rm b},
  \end{equation}
where $\dot{M}_{\rm T} = \dot{M}_{\rm PSR} + \dot{M}_{\rm WD}$ is the
change of mass of both components.

We now estimate both mass loss terms. The pulsar is losing rotational energy at
a rate given by $\dot{E} = 4 \pi I_{\rm PSR} \dot{P} P^{-3}\,=\,1.6\,\times\,10^{32}$ erg s$^{-1}$,
where $I_{\rm PSR}$ is the pulsar's moment of inertia, normally
assumed to be $10^{45} \rm g \, cm^{2}$. This dominates
the mass loss for the pulsar \cite{dt91}:
\begin{equation}
  \frac{\dot{M}_{\rm PSR}}{M_{\rm T}}
    = \frac{\dot{E}}{M_{\rm T} c^2}
    = 4.1 \times 10^{-23} \, {\rm s}^{-1}.
\end{equation}
Most of this energy is emitted as a wind of relativistic particles, which
we assume to be isotropic to first order.
A fraction of this energy $F = R_{\rm WD}^2 / 4 a^2 = 0.00074$
(where $a = x c (q + 1) / \sin i = 8.32 \times 10^8\,$m is the separation between components)
strikes the surface of the white dwarf. This is the energy available to
power mass loss from the white dwarf. Conservation of energy
requires that
\begin{equation}
\dot{E}  F = \frac{1}{2} \dot{M}_{\rm WD} v^2,
\end{equation}
where $v$ is the velocity of the escaping particles. This equation
shows that $\dot{M}$ increases as $v$ decreases, however $v$
must be at least equal to the escape velocity for the star to lose mass,
i.e., $v^2 / 2 > G M_{\rm WD} / R_{\rm WD}$. Putting all the constraints
together, we obtain:
\begin{equation}
\frac{\dot{M}_{\rm WD}}{M_{\rm T}} < 5.4 \times 10^{-21} \rm \, s^{-1}.
\end{equation}
Therefore, $\dot{M}_{\rm T} \simeq \dot{M}_{\rm WD}$. Evaluating
eq.~\ref{eq:massloss}, we obtain
$\dot{P}_{b}^{\dot{M}} < 0.4 \times 10^{-16}$, which is
$\sim 5 \times 10^2$ times smaller than the current uncertainty in the
measurement of $\dot{P}_{\rm b}$.


\paragraph*{Tidal contribution to $\dot{P}_{\rm b}$}

We now calculate the orbital decay caused by tides. If these change the
angular velocity of the white dwarf $\dot{\Omega}_{\rm WD}$,
this will be compensated by a change in the orbital period of the system
$\dot{P}_{\rm b}^{\rm T}$. We can relate the two because of
conservation of angular
momentum:
\begin{equation}
 \dot{P}_{\rm b}^{\rm T} = \frac{3 k \Omega_{\rm WD}}{2 \pi q (q +1)}
 \left( \frac{R_{\rm WD} P_{\rm b} \sin i}{x c} \right)^2 \frac{1}{\tau_s},
\end{equation}
where $\tau_s = - \Omega_{\rm WD} / \dot{\Omega}_{\rm WD}$ is the
synchronization timescale and $k \equiv I_{\rm WD} / (M_{\rm WD} R_{\rm WD}^2)$,
where $I_{\rm WD}$ is the white dwarf moment of inertia.
For idealized white dwarfs
(particularly those with a mass much below the Chandrasekhar limit)
sustained solely by degeneracy pressure of non-relativistic electrons,
a polytropic sphere with $n = 1.5$ provides a good approximation.
For such stars, we have $k = 0.2$ \cite{motz52}. However, for this
light white dwarf only the core is degenerate, and is surrounded by a deep
non-degenerate layer that accounts for only about 5\% of the mass of the
star. Therefore, the mass distribution is much more centrally condensed
than for an $n = 1.5$ polytrope and the moment of inertia is much
smaller. We therefore use the output of our white dwarf model calculations (see Fig~2. and above)
to estimate that factor. For the model closer to the mean of the white dwarf mass distribution, 
with $M_{\rm WD} = 0.169\,\rm M_{\odot}$, $R_{\rm WD}=0.069$\,R$_\odot$ and $T_{\rm eff}=9950$\,K 
we obtain $k = 0.0267$. We adopt this value in subsequent calculations.

The only unknown parameters in this expression are
$\Omega_{\rm WD}$ and $\tau_s$. If $\tau_s$ were much smaller than
the characteristic age of the pulsar $\tau_c = 2.6\,$Gyr
(which is similar to the cooling age of the white dwarf, i.e., this
number is likely to be a good approximation to the true age of
the system), then the white dwarf rotation would already be synchronized
with the orbit ($\Omega_{\rm WD} = 2\pi/P_{\rm b}$).
In this case the orbital decay would be
slightly affected because, as the orbital period decreases, the
white dwarf spin period would decrease at exactly the same rate in order to
preserve tidal locking.
The resulting exchange of angular momentum would change the orbital
decay by a factor $\Delta \dot{P}_{\rm b}$ given by
the ratio of the moment of inertia of the white dwarf and the binary:
\begin{equation}
\frac{\Delta \dot{P}_{\rm b}}{\dot{P}_{\rm b}^{\rm GR}} \simeq
\frac{I_{\rm WD}}{I_{\rm b}}
= \frac{k}{q (q + 1)} \left( \frac{R_{\rm WD} \sin i}{x c} \right)^2 =
1.2 \times 10^{-4}.
\end{equation}
This means that, were the system synchronized, $\Delta \dot{P}_{\rm b}$
would be an insignificant correction given our current
measurement precision.

If the white dwarf is not yet
synchronized, then $\tau_s > \tau_c$.
In this case $\Omega_{\rm WD}$ can be much larger than $2\pi/P_{\rm b}$,
but it must still be smaller than the break-up
angular velocity, i.e.,
$\Omega_{\rm WD} < (G M_{\rm WD} / R_{\rm WD}^3)^{1/2} = 0.0142\, \rm
rad\, s^{-1}$.
These conditions for $\Omega_{\rm WD}$ and $\tau_s$ yield
$\dot{P}_{\rm b}^{\rm T} < 4.2 \, \times \, 10^{-16}\,\rm s\,s^{-1}$.
Thus, even if the white dwarf were rotating near break-up velocity,
$\dot{P}_{\rm b}^{\rm T}$ would still be two orders of
magnitude smaller than the uncertainty in the measurement of
$\dot{P}_{\rm b}$. We note, however, that the progenitor of the white dwarf was
very likely synchronized with the orbit at formation, which
had a period of $\sim 5$ hours (see below). When the
white dwarf formed, fall-back of material within the Roche lobe into it
would have spun it up, but not by more than 1 order of magnitude
(e.g., Appendix B2.2 of \cite{bkkv06}).
Therefore, at formation $\Omega_{\rm WD}$ was of the order
of $3.5 \times 10^{-3} \, \rm rad \, s^{-1}$; this would yield
$\dot{P}_{\rm b}^{\rm T} < 1.0 \, \times \, 10^{-16}\,\rm s\,s^{-1}$.

\paragraph*{Constraints on dipolar radiation and Scalar-Tensor gravity}
In scalar-tensor gravity, like for most other alternatives to GR, the dominant 
contribution to the GW damping of the orbital motion of a binary system would 
come from the scalar dipolar waves, proportional to $(\alpha_A - \alpha_B)^2$, 
where $\alpha_A$ and $\alpha_B$ denote the effective scalar-coupling constants 
of the two masses $m_A$ and $m_B$, respectively, of the binary system. Such 
deviations should then become apparent as a modification in the orbital period 
decay observed in binary pulsars. In GR the emission of quadrupolar tensor waves 
enters the orbital dynamics at the 2.5 post-Newtonian (pN) level, which 
corresponds to corrections of order $(v/c)^5$ in the equations of motion, $v$ being 
a typical orbital velocity. A contribution from dipolar GWs enters already at 
the 1.5pN level, i.e.\ terms of order $(v/c)^3$. As an example, in scalar-tensor 
gravity the change in angular orbital frequency $n_{\rm b} \equiv 2\pi/P_{\rm b}$ for a 
circular orbit caused by gravitational wave damping up to 2.5pN order is given 
by \cite{de92,de98} 
\begin{equation}\label{eq:ndot}
\frac{\dot{n}_{\rm b}}{n_{\rm b}^2} = 
  \frac{X_AX_B}{1 + \alpha_A \alpha_B} \, 
  \left[ \frac{96}{5} \,\kappa \, 
  \left( \frac{v}{c} \right)^5 + (\alpha_A - \alpha_B)^2 \, 
  \left( \frac{v}{c} \right)^3 \right] \;,
\end{equation}
where
\begin{equation}
  v \equiv [G_\ast(1 + \alpha_A\alpha_B)(m_A+m_B)n_{\rm b}]^{1/3} \;,
\end{equation}
with $G_\ast$ denoting the bare gravitational constant, and
$X_A \equiv m_A/(m_A+m_B)$ and $X_B \equiv m_B/(m_A+m_B)$. The quantity 
$\kappa$, where $\kappa=1$ in GR, holds terms arising from the emission of 
scalar quadrupolar waves and higher order terms of the scalar dipolar emission 
\cite{de98}:
\begin{equation}
  \kappa = 1 + \frac{1}{6}\,(\alpha_A X_B + \alpha_B X_A)^2 
             + d_1\,(\alpha_A - \alpha_B) +  d_2\,(\alpha_A - \alpha_B)^2 \end{equation}
and\footnote{To our knowledge, the quantity $d_2$ has been calculated 
here for the first time.}
\begin{eqnarray}
  d_1 &=& \frac{1}{6}(\alpha_A X_A + \alpha_B X_B) (X_A - X_B) +
          \frac{5}{48} \, \frac{\beta_B \alpha_A X_A - \beta_A \alpha_B X_B}
          {1 + \alpha_A \alpha_B} \;,\\
  d_2 &=& \frac{5}{64} + \frac{253}{576}X_A X_B - 
          \frac{39 + 49\,\alpha_A \alpha_B}{144 (1 + \alpha_A \alpha_B)} -
          \frac{5 (X_B \alpha_B^2 \beta_A + X_A \alpha_A^2 \beta_B)}
               {72 (1 + \alpha_A \alpha_B)^2} \;.
\end{eqnarray}
GR is recovered for $G_\ast = G$ and $\alpha_A = \alpha_B = 0$. Equation 
(\ref{eq:ndot}) can directly be confronted with the results compiled in 
Table~1, in combination with the p.d.f.\ of the white dwarf mass in 
Fig.~3, 
where we use index $A$ for the pulsar and index $B$ for the white dwarf
companion.\footnote{Strictly speaking, when using the masses of Table~1 in
equation (\ref{eq:ndot}) one has to keep in mind the difference between
the bare gravitational constant $G_\ast$ and Newton's gravitational constant 
$G = G_\ast(1 + \alpha_0)^2$ as measured in a Cavendish-type experiment. 
However, since $\alpha_0^2 < 10^{-5}$, we can ignore this difference in our
calculations.}
One finds from the mass ratio $q$ that $X_A = q/(q + 1) = 0.9213\pm0.0008$ and 
$X_B = 1/(q + 1) = 0.0787\pm0.0008$. Furthermore, since $G_\ast (m_A + m_B) 
\simeq G m_B (q + 1)$, one has $v/c = (0.001970\pm 0.000016) \times 
(1 + \alpha_A \alpha_B)^{1/3}$. With the observed change in the 
orbital frequency $\dot{n}_{\rm b} = -2 \pi \dot{P}_{\rm b} / P_{\rm b}^2 = 
(2.23 \pm0.36) \times 10^{-20}$, which agrees with GR, one can infer the 
following constraint on $(\alpha_A - \alpha_0)$ using equation (\ref{eq:ndot}):
\begin{equation}\label{eq:aAma0limit}
  |\alpha_A - \alpha_0| < 0.005 \quad \mbox{(95\% C.L.)} \;.
\end{equation}
Our detailed calculations show that this limit is solely enforced by the 
dominant 1.5pN term of equation (\ref{eq:ndot}) (see also equation (1) in
the main text), and is practically insensitive to the values assumed by 
$\beta_A$ and $\beta_B$. Consequently, as in \cite{lwj+09,fwe+12} 
the limit (\ref{eq:aAma0limit}) can be seen as a generic limit, that is 
independent of the EOS. 

To illustrate how \psr\ probes a new gravity regime, we present detailed 
calculations based on a specific EOS and a specific class of alternative gravity 
theories. As an EOS we use the rather stiff EOS ``.20'' of \cite{hkp81}, which 
supports (in GR) neutron stars of up to $2.6$\,M$_\odot$. Concerning the alternative 
gravity theories, we use the class of ``quadratic'' mono-scalar-tensor theories 
used in \cite{de93,de96}, where the (field-dependent) coupling strength 
$\alpha(\varphi)$ between the scalar field and matter contains two parameters: 
$\alpha(\varphi) = \alpha_0 + \beta_0 \varphi$. Every pair $(\alpha_0,\beta_0)$
represents a specific scalar-tensor theory of gravity. As discovered
in \cite{de93}, for certain values of $\beta_0$, neutron stars can develop a significant 
scalarization, even for vanishingly small $\alpha_0$, if their mass exceeds a
critical ($\beta_0$-dependent) value. For this reason, this class of gravity 
theories is particularly well suited to demonstrate how  the limit 
(\ref{eq:aAma0limit}) probes a new gravity regime that has not been tested 
before (see Fig.~2b). The specific parameters and EOS in Fig.~2b have been 
chosen for demonstration purposes. A change in the EOS, for instance, would lead 
to a modification in the details of the functional shape of $\alpha_A$, but 
would not change the overall picture.

\paragraph*{Constraints on the phase evolution of neutron-star mergers}
So far, the best constraints on dipolar gravitational wave damping in compact 
binaries come from the observations of the millisecond pulsar PSR~J1738+0333, 
a $1.47_{-0.06}^{+0.07}$\,M$_\odot$ neutron star in a tight orbit ($P_{\rm b} \approx
8.5$\,h) with a spectroscopically resolved white-dwarf companion 
\cite{avk+12,fwe+12}. However, as discussed in detail above, such timing 
experiments are insensitive to strong-field effects that might only become 
relevant in the strong gravitational fields of high-mass neutron stars. Consequently, the 
dynamics of a merger of a 2\,M$_\odot$ neutron star with a ``canonical'' neutron star or a black 
hole (BH) might have a significant contribution from dipolar GW damping, leading 
to a modification of the orbital dynamics that is incompatible with the 
sophisticated GR templates used to search for GWs with ground-based 
GW detectors, like LIGO and VIRGO, \cite{ss09}. With the results on 
\psr, in particular with the limit given in eq. (\ref{eq:aAma0limit}), this 
question can finally be addressed in some details. For this purpose, we 
decompose equation (\ref{eq:ndot}) into the 2.5pN contribution, that is matched 
by an appropriate GR template, and the 1.5pN contribution, that drives the phase 
evolution away from the 2.5pN dynamics. Following \cite{will94,de98}, we 
introduce the dimensionless orbital angular velocity 
\begin{equation}
  u \equiv {\cal M}n_{\rm b} = \pi {\cal M} f_{\rm GW} \;,
\end{equation}
where $f_{\rm GW}$ denotes the frequency of the GW and
\begin{equation}
  {\cal M} \equiv \frac{G_\ast M}{c^3}
                  \frac{(X_A X_B\kappa)^{3/5}}
                       {(1 + \alpha_A \alpha_B)^{2/5}} \;.
\end{equation}
To leading order, one then finds
\begin{equation}\label{eq:udot}
  {\cal M} \dot{u} = \frac{96}{5} (u^{11/3} + {\cal B} u^3) \;,
\end{equation}
where 
\begin{equation}
  {\cal B} \equiv \frac{5}{96}
                  \left(\frac{X_A X_B}{1 + \alpha_A \alpha_B}\right)^{2/5} 
                  \frac{(\alpha_A - \alpha_B)^2}{\kappa^{3/5}} \;.
\end{equation}
The observed GW cycles in a frequency band $[f_{\rm in},f_{\rm out}]$ can be computed as follows:
\begin{equation}
  N_{\rm GW} = \int_{t_{\rm in}}^{t_{\rm out}} f \, dt 
             = \int_{f_{\rm in}}^{f_{\rm out}} (f/\dot{f}) \, df 
             = \frac{1}{\pi}\int_{u_{\rm in}}^{u_{\rm out}}  
               \frac{u}{{\cal M}\dot{u}} \, du \;.
\end{equation}
Consequently, the difference between the 2.5pN dynamics and the 2.5pN + 1.5pN
dynamics is given by
\begin{equation}
  \Delta N_{\rm GW} = \frac{5}{32\pi} 
    \left.\left(
    \frac{1}{5 u^{5/3}}
    - \frac{1}{3u {\cal B} } +
    \frac{1}{u^{1/3} {\cal B}^2} 
    + \frac{\arctan\left(u^{1/3}/{\cal B}^{1/2}\right)}
         {{\cal B}^{5/2}}\right)\right|_{u_{\rm in}}^{u_{\rm out}} \;,
\end{equation}
where we made no assumption about the size of the value for ${\cal B}$. 
For the LIGO/VIRGO 
band $u_{\rm in} \ll u_{\rm out}$.\footnote{For a detector that is sensitive up 
to a few kHz, the frequency $f_{\rm GW}^{\rm out}$ is determined by the 
innermost circular orbit, which is $\sim$ 1350\,Hz for a 
2/1.25\,M$_\odot$ system and $\sim$ 370\,Hz for a 2/10\,M$_\odot$ system (see 
\cite{bla06}).} Fig.~5 gives $\Delta N_{\rm GW}$ for the LIGO/VIRGO detectors,
for which a typical bandwidth of 20\,Hz to a few kHz was assumed, as a function 
of $|\alpha_A - \alpha_0|$ for two different systems, a
2/1.25\,M$_\odot$ NS-NS system and a 2/10\,M$_\odot$ NS-BH system. 
Concerning the NS-BH systems, we considered the class of 
alternative gravity theories where BHs are practically identical to GR, 
and consequently used $\alpha_B = 0$. For instance, this is the case in 
scalar-tensor gravity theories with negligible time dependence of the 
asymptotic scalar field \cite{de98}. 
For the NS-NS system an extreme case is represented by the assumption that
only the massive neutron star has a significant scalar coupling strength $\alpha_A$,
while the lighter companion behaves like a weakly self-gravitating body, meaning
$\alpha_B = \alpha_0$. Besides this, for the NS-NS system we have also performed 
calculations using a hypothetical most conservative (maximal $\Delta 
N_{\rm GW}$) value for the effective  coupling strength of the companion $B$, 
which is $\alpha_B = 0$. However, such an assumption seems unphysical for a 
non-zero $\alpha_0$, where $\alpha_B$ is expected to approach $\alpha_0$ (and
not 0) for less massive stars. With the limit obtained 
from \psr\ we find a conservative upper limit for the dipolar phase offset
of $\sim$ 0.5 (NS-NS) and 0.04 (NS-BH) cycles, an amount that would not 
jeopardize the detection of the gravitational wave signal in the LIGO/VIRGO 
band \cite{mag08}.\\

\paragraph*{Formation via a common envelope and spiral-in phase}
Common-Envelope (CE) evolution \cite{pac76,il93} in X-ray binaries is initiated by
dynamically unstable mass transfer, often as the result of a high 
mass-transfer rate and a large initial donor/accretor mass 
ratio, $q_{\rm i}\equiv M_2/M_{\rm PSR} >1$. If the CE is initiated while the 
donor star is still early in its main sequence stage (i.e.\ if $P_{\rm b}<1\,
{\rm day}$), the outcome is expected to be a merger \cite{tas00}.
It is generally believed that a binary can only survive the CE evolution, and 
thereby successfully eject the envelope of the donor star, if the binding energy 
of the envelope, $E_{\rm bind}$, is less than the released orbital energy from 
the in-spiral process, $\Delta E_{\rm orb}$ \cite{web84}. The orbital energy of 
\psr\ is: $|E_{\rm orb}|=GM_{\rm PSR}M_{\rm WD}/2a \simeq 
5.5\times 10^{47}\,{\rm erg}$. Hence, even if assuming in-spiral from infinity 
to the current orbital separation, the amount of liberated orbital 
energy from the CE~phase cannot exceed this value. 
From calculations of $E_{\rm bind}$ of intermediate-mass stars (Table~S3), 
we find that $E_{\rm bind} \gg \Delta E_{\rm orb}$ during most of their
evolutionary stages. Only if the donor star (i.e. the white dwarf progenitor)
is an evolved giant is it possible to eject the 
envelope. However, in this case the core mass of such an evolved star, $M_{\rm core}$, 
is more massive than the observed white dwarf companion by at least a 
factor of $2-3$. (As argued in the main text, 
a reduction in white dwarf mass via evaporation from the pulsar wind seems to be ruled out
for \psr\ and therefore cannot help circumvent this discrepancy.) 

From Fig.~\ref{0348.Mcore} we see that
only a low-mass donor star with mass $M_2\le2.2\,{\rm M}_{\odot}$, and not evolved 
beyond the terminal age main-sequence (TAMS), 
would leave behind $M_{\rm core} = M_{\rm WD} \simeq 
0.17\,{\rm M}_{\odot}$. In this case, it is clear
that energy sources other than $\Delta E_{\rm orb}$ must contribute to expel 
the envelope (since in this case $E_{\rm bind} \gg \Delta E_{\rm orb}$). 
Such an energy source could be the release of gravitational potential 
energy from material which accretes onto the neutron star 
during the CE. The amount of released energy per accreted unit mass  
is roughly $\Delta U/m\sim GM/R\sim 2\times 10^{20}\,{\rm erg}\,{\rm g}^{-1}$. 
Hence, assuming full absorption and 100\% energy conversion of 
this released energy to eject the envelope, this would require accretion of 
$\sim 4\times 10^{-5}\,{\rm M}_{\odot}$; a value which is not 
unrealistic given a timescale of the CE event of $\sim 10^3\,{\rm yr}$ 
with Eddington limited accretion (a few 
$10^{-8}\,{\rm M}_{\odot}\,{\rm yr}^{-1}$). 

As a consequence of this relatively short CE~phase, 
the currently observed mass of $M_{\rm PSR}=2.01\,{\rm M}_{\odot}$ 
should be close to the original mass of the neutron star 
after its formation in a type~Ib/c supernova. 
According to recent studies by \cite{ujma12}, 
neutron star birth masses of $2.0\,{\rm M}_{\odot}$ are indeed possible.
As mentioned in the main text, however, 
having an initially massive neutron star would be a more serious problem 
for formation via a CE event with a $\le 2.2\,{\rm M}_{\odot}$ donor star. 
Such a high value of $M_{\rm PSR}$ would lead to a value of $q_{\rm i}$ close to 
unity, in which case the Roche-lobe overflow (RLO) is expected to be dynamically {\em stable},
thereby avoiding the formation of a CE \cite{ts99,prp02}. The only solution to 
this problem would be that the neutron star was originally born with a 
more typical mass of $\sim\!1.4\;M_{\odot}$, in which case $q_{\rm i}$ 
would be sufficiently high to ensure formation of a CE. However, in that case
one would have to accept the concept of hypercritical accretion 
\cite{che93,iva11}, allowing the neutron star to accrete a large amount of mass 
$\sim$ $0.5$ -- $0.7\,{\rm M}_{\odot}$
on a timescale of $\sim 10^3\,{\rm yr}$. 
One could argue that \psr\ would then be the best 
(and to our knowledge the only) candidate known
in which hypercritical accretion might have been at work.\\
To summarize, given the many issues discussed above we find that a CE formation
channel is less favorable to explain \psr\ and we now proceed with investigating
another solution, the LMXB formation channel.

\paragraph*{Formation via a converging low-mass {X}-ray binary}
As mentioned in the main text, a handful of binary pulsars exist with values of  
$P_{\rm b}\le 8\,{\rm hr}$ and $M_{\rm WD}\approx 0.14 - 0.18\,{\rm M}_\odot$, 
similar to those of \psr. These systems are tentatively thought to descend from
low-mass {X}-ray binaries (LMXBs) in which the binary suffered from loss of 
orbital angular momentum caused by magnetic braking \cite{ps89,prp02,vvp05}.
However, there remains a general problem for reproducing these pulsar binaries using current stellar evolution codes. 
A main issue is that converging LMXBs most often do not detach
but keep evolving with continuous mass transfer to more and more compact systems 
with $P_{\rm b} \le 1\,{\rm hr}$ and ultra-light donor masses 
$M_2 < 0.08\,{\rm M}_{\odot}$. 
In a few instances, where fine-tuning may lead to detachment and the right
values of $P_{\rm b}$ and $M_2$, the donor star is typically too hydrogen rich 
to settle and cool as a compact He~white dwarf [however, see sequence~$d$ in 
fig.~16 of \cite{prp02} for an exception]. 
Our numerical studies are no exception from this general picture. 

Using the Langer stellar evolution code [e.g.\ \cite{wl99,tlk11}] 
we have attempted to model the formation and evolution of the \psr\ system. 
Here we present a solution where we have 
forced the donor star to detach its Roche~lobe at $P_{\rm b}\sim5\,{\rm hr}$,  
such that the system subsequently shrinks in size to its present 
value of $P_{\rm orb}\simeq 2.46\,{\rm hr}$ due to GW radiation 
within the estimated cooling age of the white dwarf
($t_{\rm WD}\simeq 2\,{\rm Gyr}$, 
depending on cooling models and assumed metallicity).
To be more precise, the estimated $t_{\rm WD}$ is actually a lower 
limit on the timescale during which the detached system
evolved via GW radiation since it takes $10^8 - 10^9\,{\rm yr}$ for 
the detached pre-white dwarf to settle on the final cooling track. 
This can be compensated for by choosing a slightly larger $P_{\rm b}$ at the ZAMS, 
which causes the system to detach from the LMXB in a somewhat wider orbit.

In Fig.~\ref{0348.LMXB_2} (and see also Fig.~6 in the main text) 
we show an example of our LMXB calculations.
The model binary shown here consisted initially of a $1.75\,{\rm M}_{\odot}$ 
neutron star and a $1.1\,{\rm M}_{\odot}$ donor star with metallicity $Z=0.02$, 
mixing length parameter, $\alpha=2.0$ and ZAMS orbital period, 
$P_{\rm b}=2.55\,{\rm days}$. The initial $P_{\rm b}$
depends on the modeling of magnetic braking. Here the value corresponds to
onset of RLO at $P_{\rm b}\simeq 0.65\,{\rm days}$, shortly after the donor star ceased central hydrogen burning. 
Our high value of the initial neutron star mass is motivated from studies which show that the accretion efficiency in LMXBs must be
rather small --- even for systems which are expected to have accreted at sub-Eddington levels \cite{avk+12,jhb+05,ts99}.
Hence, by adopting an accretion efficiency of only 30\% we need an initial high-mass neutron star in order to reach the present mass of
PSR~J0348+0432. 
Note, that some neutron stars are indeed expected to have been born massive [for a discussion, see \cite{tlk11} and references therein].
The outcome of our calculations would possibly have been somewhat similar by assuming an accretion efficiency close to 
100\% and starting with $M_{\rm PSR}=1.3\,{\rm M}_{\odot}$. 
To model the loss of orbital angular momentum due to mass loss from the system, 
we adopted the isotropic re-emission model \cite{bvh91}.

Based on its proper motion and radial velocity measurements, \psr\ has an estimated 3D space velocity
of $56 \pm 8\;{\rm km}\,{\rm s}^{-1}$ with respect to the Solar System. 
From Monte Carlo simulations
of its past motion through our Galaxy [following the method described in \cite{avk+12,fbw+11}], 
we find that this velocity corresponds to a peculiar velocity with respect to the local standard of rest 
at every transition of the Galactic plane  of $75\pm 6\;{\rm km}\,{\rm s}^{-1}$.
This result is rather independent of the applied Galactic model.
From subsequent simulations of the dynamical effects of the supernova explosion,
we find that a relatively small kick magnitude of $w<150\;{\rm km}\,{\rm s}^{-1}$ was imparted to the
newborn neutron star, by probing a broad range of values of the pre-supernova orbital period and the masses of the collapsing naked He-core and its companion star (the white dwarf progenitor).

\paragraph*{Spin evolution of \psr}
A peculiarity of \psr, compared to other recycled pulsars with similar $P_{\rm b}$
and $M_{\rm WD}$, is its slow spin period, 
$P=39\,{\rm ms}$ and its high value of the spin period 
derivative, $\dot{P} = 2.41 \times 10^{-19}\,{\rm s\,s}^{-1}$, 
cf. the unusual location of \psr\ in both the 
$P\dot{P}$-diagram and the Corbet-diagram (Figs.\,S9, S10). 
In particular, the Corbet~diagram clearly displays the 
unique characteristics of \psr\ with
a small $P_{\rm b}$ and a large value of $P$. 

During the LMXB phase, a pulsar is generally expected to accrete much more mass and 
angular momentum than needed to be spun-up to a few milliseconds \cite{tlk12}. In the same 
process, its B-field should have decayed significantly --- typically to values 
$\le 10^8\,{\rm G}$. However, for some reason the B-field ($\sim 2\times 10^9\;{\rm G}$)  
remained relatively high in \psr. 
In contrast, the other known binary radio pulsars with similar values of
$P_{\rm b}\le 8\,{\rm hr}$ and $M_{\rm WD}\approx 0.14-0.18\,{\rm M}_{\odot}$ 
(e.g. PSRs~J0751+1807 and J1738+0333), 
besides from the many black-widow-like systems, 
have low B-fields and spin periods of a few milliseconds, 
as expected from current theories of LMXB evolution.

In Fig.~\ref{0348_spin} we have plotted the past and the future evolution of 
PSR~J0348+0432. In the upper panel is seen the evolution of $P_{\rm b}$. 
In the lower panel is seen the spin evolution of the pulsar 
assuming different values of a constant braking index $2\le n\le 5$.
If the estimated cooling age of $\sim 2\,{\rm Gyr}$ is correct 
(and adding to this value a pre-white dwarf contraction phase 
between RLO detachment and settling on the final cooling track, yielding an
assumed total age of about $2-2.5\,{\rm Gyr}$) we can estimate that
PSR~J0348+0432 was recycled with an initial spin period of about 
$10-20\,{\rm ms}$. 
This relatively slow spin could be (partly) caused by enhanced braking 
of the spin rate, due to the high B-field of the pulsar, 
during the Roche-lobe decoupling phase
when the progenitor of the white dwarf ceased its mass transfer \cite{tau12}. 
If the total post-LMXB 
age is $\sim 2.6\,{\rm Gyr}$ then the pulsar could, at first sight, have been
recycled with an initial spin period of 1~ms for $n\ge3$. However, calculations 
of the pulsar spin-up line \cite{tlk12} do not predict such a rapid spin for 
pulsars with high B-fields and which accreted with typical mass-accretion rates 
of $\dot{M} < 10^{-2}\,\dot{M}_{\rm Edd}$ (evident from both theoretical modeling
of the LMXB RLO and observations of LMXB luminosities \cite{lv06}).

\newpage
 
 \begin{figure}[htp]
\begin{center}
  \includegraphics[scale=0.4]{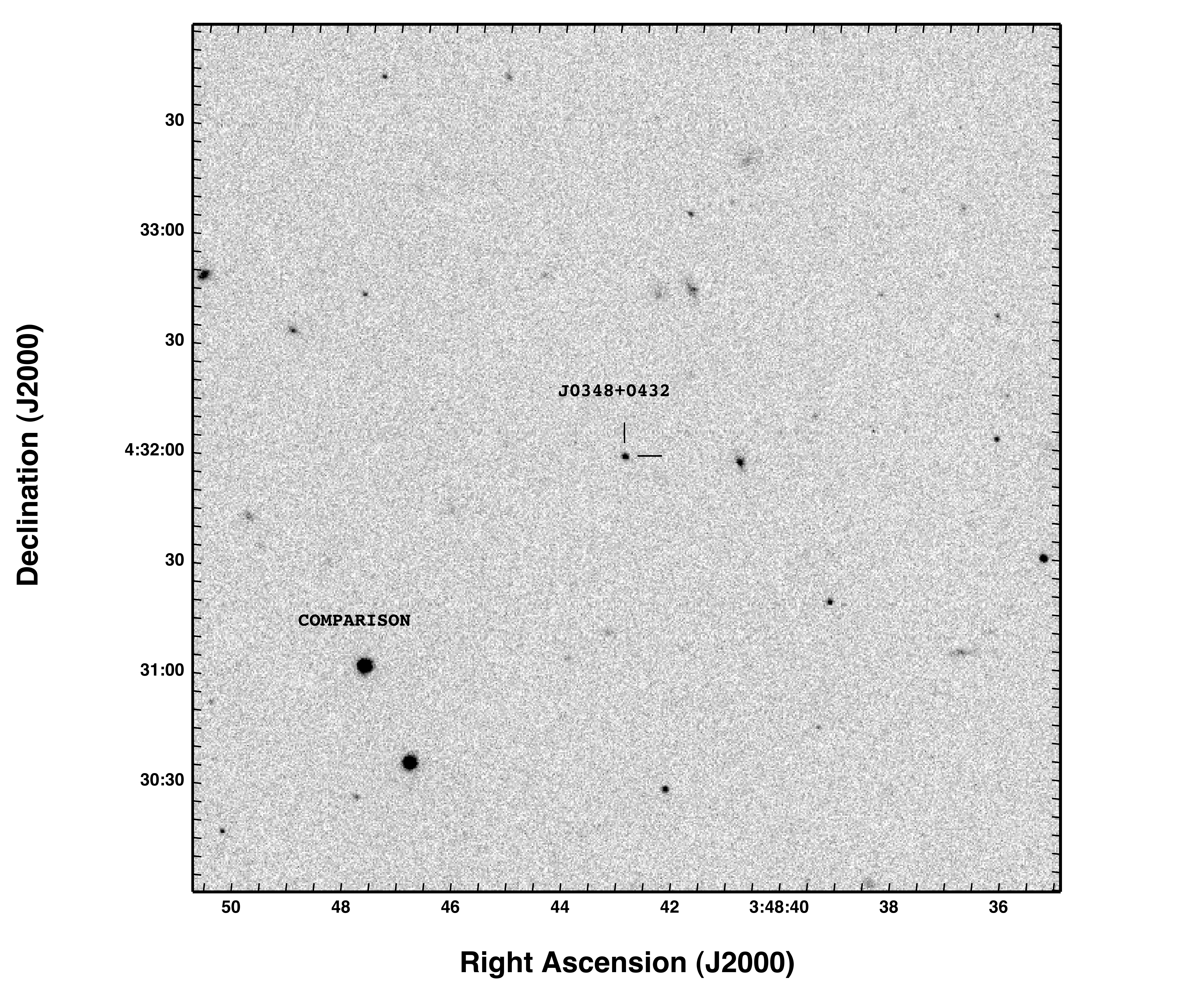}
  \caption{Finding chart for the \psr\ system and the comparison star used in our analysis (see text), created from the archived SDSS 
  $g'$ image.
  \label{fig:sdss}}
  \end{center}
\end{figure}

\begin{figure}[h]
\begin{center}
$
\begin{array}{cc}
\resizebox{8.5cm}{!}{\includegraphics{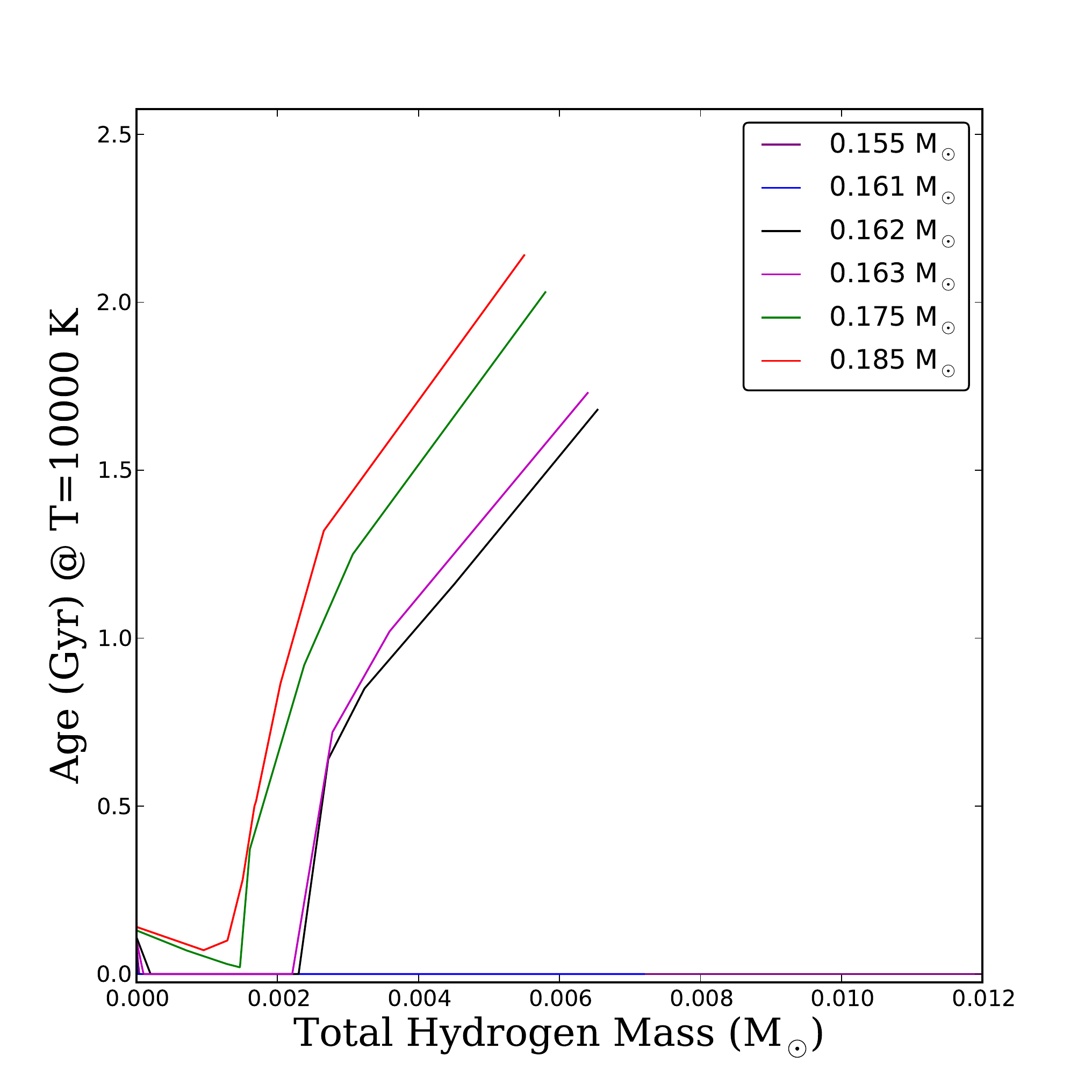}} &
\resizebox{8.5cm}{!}{\includegraphics{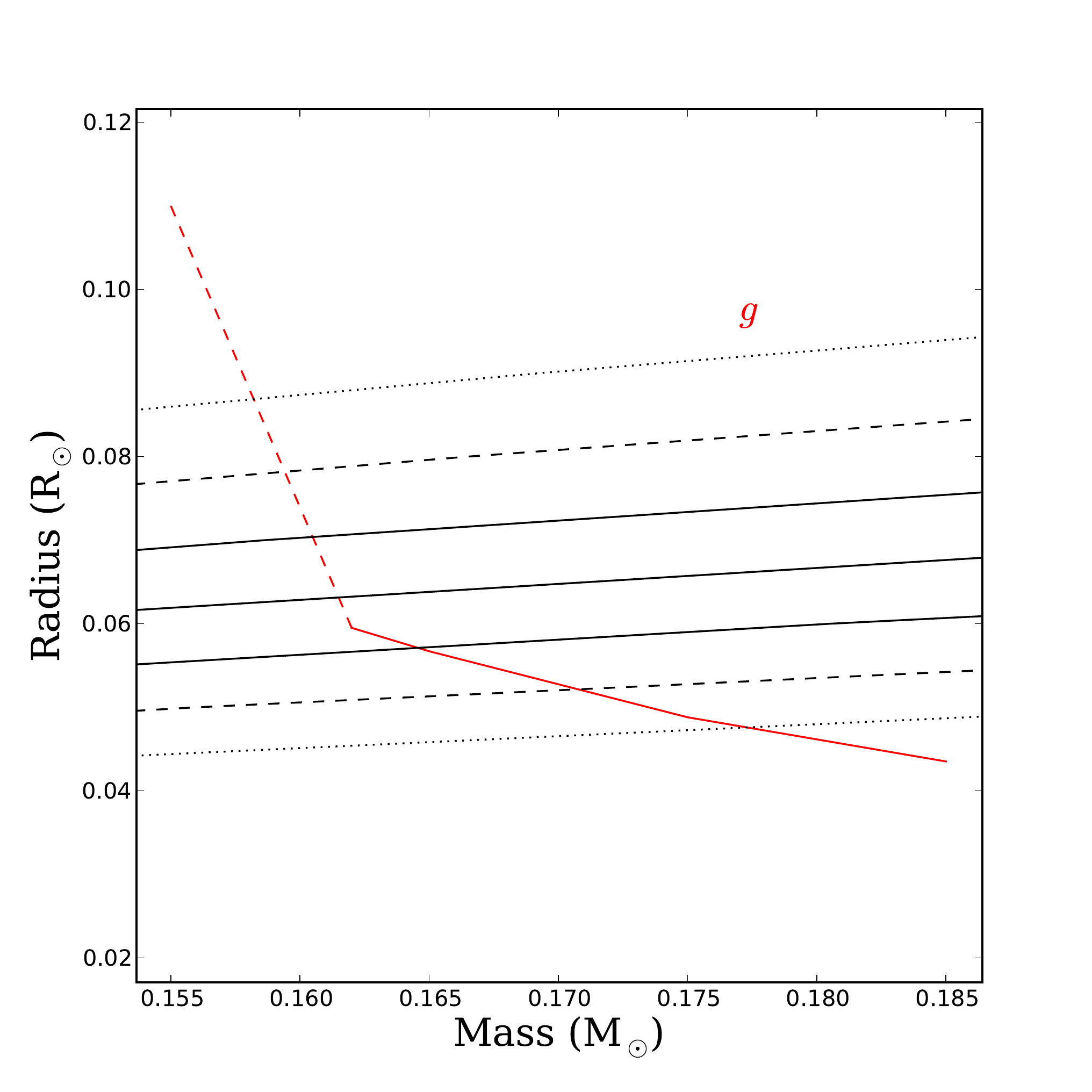}} 
\end{array} 
$
\caption{
\textbf{Left:} White dwarf cooling age (measured from the onset of the core 
contraction) when the temperature reaches $T_{\rm eff}$ = 10000\,K as a function of 
the total hydrogen mass of the star. Each line depicts a different total mass 
(from 0.155 to 0.185\,M$_{\odot}$). For each model, hydrogen burning through the 
pp-chain at the bottom of the stellar envelope cannot be initiated below a 
critical envelope mass limit. As a result the white dwarf cools in a few Myr.
For models below $\sim$ 0.162\,M$_\odot$ a temperature of 10000\,K cannot be
reached regardless of the envelope size.
\textbf{Right:} Finite-temperature mass-radius relation (for 10000\,K) for 
models that have the minimum envelope mass required for hydrogen burning (red 
line). Over-plotted are the most-likely value and 1, 2 and 3$\sigma$ constraints on the surface 
gravity for \psr\ (solid, dashed and dotted black lines respectively). For masses 
below 0.162\,M$_\odot$ the radius is an extrapolation from lower 
temperatures (in dashed red).}
\end{center}
\end{figure}

\begin{figure}[htp]
\begin{center}
  \includegraphics[scale=0.43]{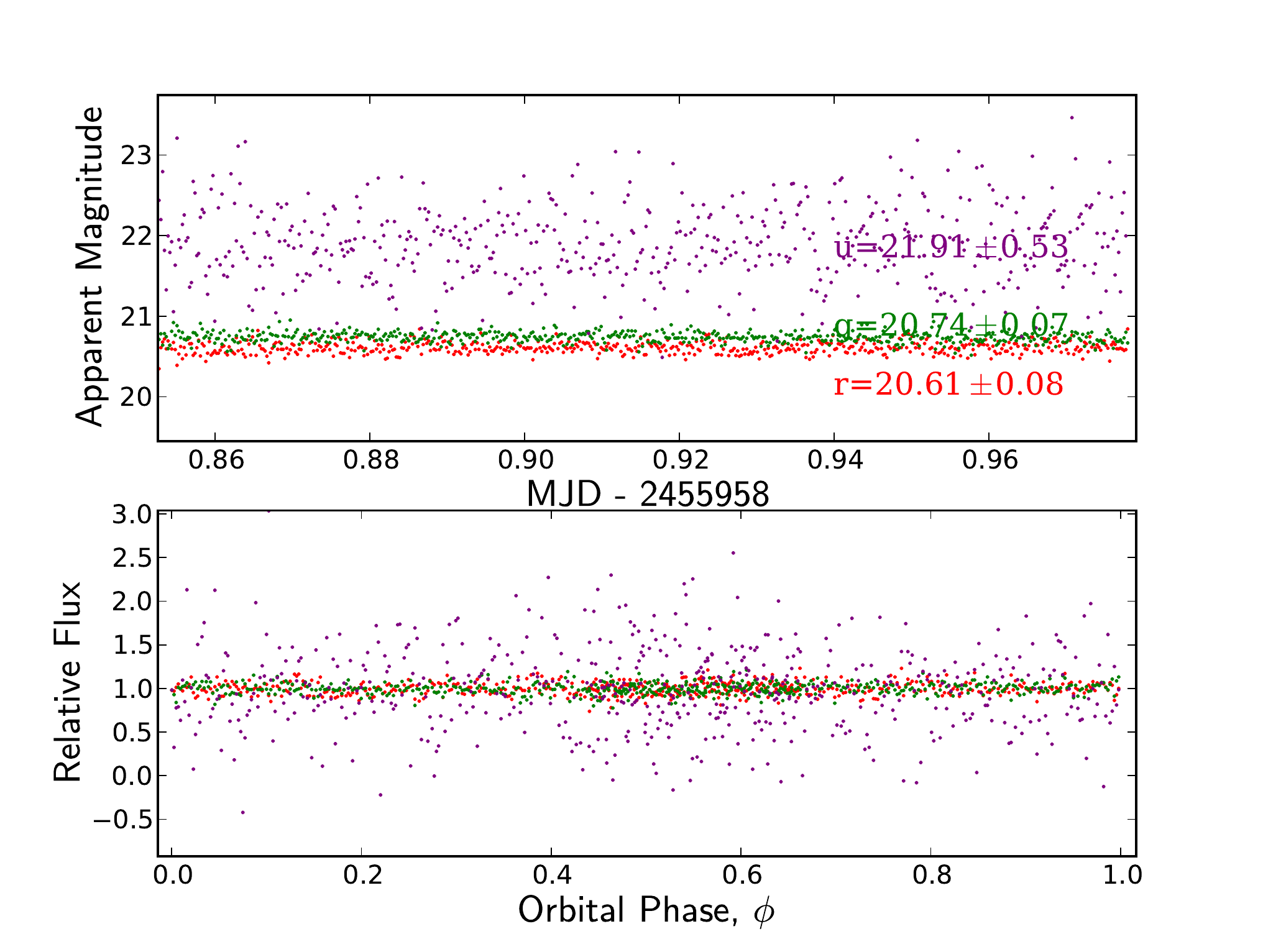}
  \caption{Photometric  (upper) and phase-folded (lower) light-curve  
  of the white dwarf companion to \psr\ in $u'$, $g'$ and $r'$.
}
\end{center}
\end{figure}

\begin{figure}[htp]
\begin{center}
  \includegraphics[height=8cm]{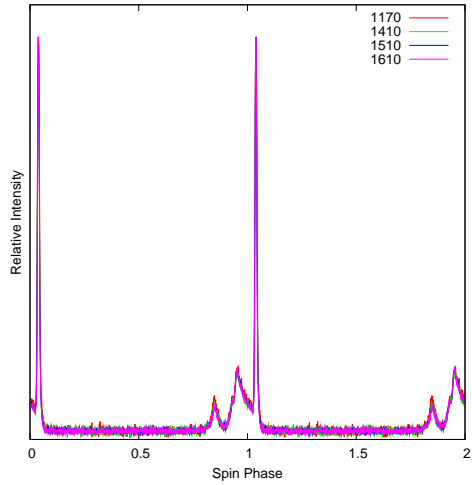}
  \caption{
     Pulse profiles for \psr\, obtained with the WAPP
    spectrometers at frequencies of 1170, 1410, 1510 and
    1610 MHz. Two full cycles are displayed for clarity.
    Their (almost) perfect overlap indicates that there
    is little pulse profile evolution between 1170 and 1610 MHz.
    The 1410 MHz pulse profile is the template used to derive
    all TOAs. The TOAs correspond to integer phases in this
    plot, which mark the maximum of the fundamental harmonic.
  \label{fig:profile}}
\end{center}
\end{figure}
\begin{figure*}[htp]
\begin{center}
\includegraphics[height=10cm]{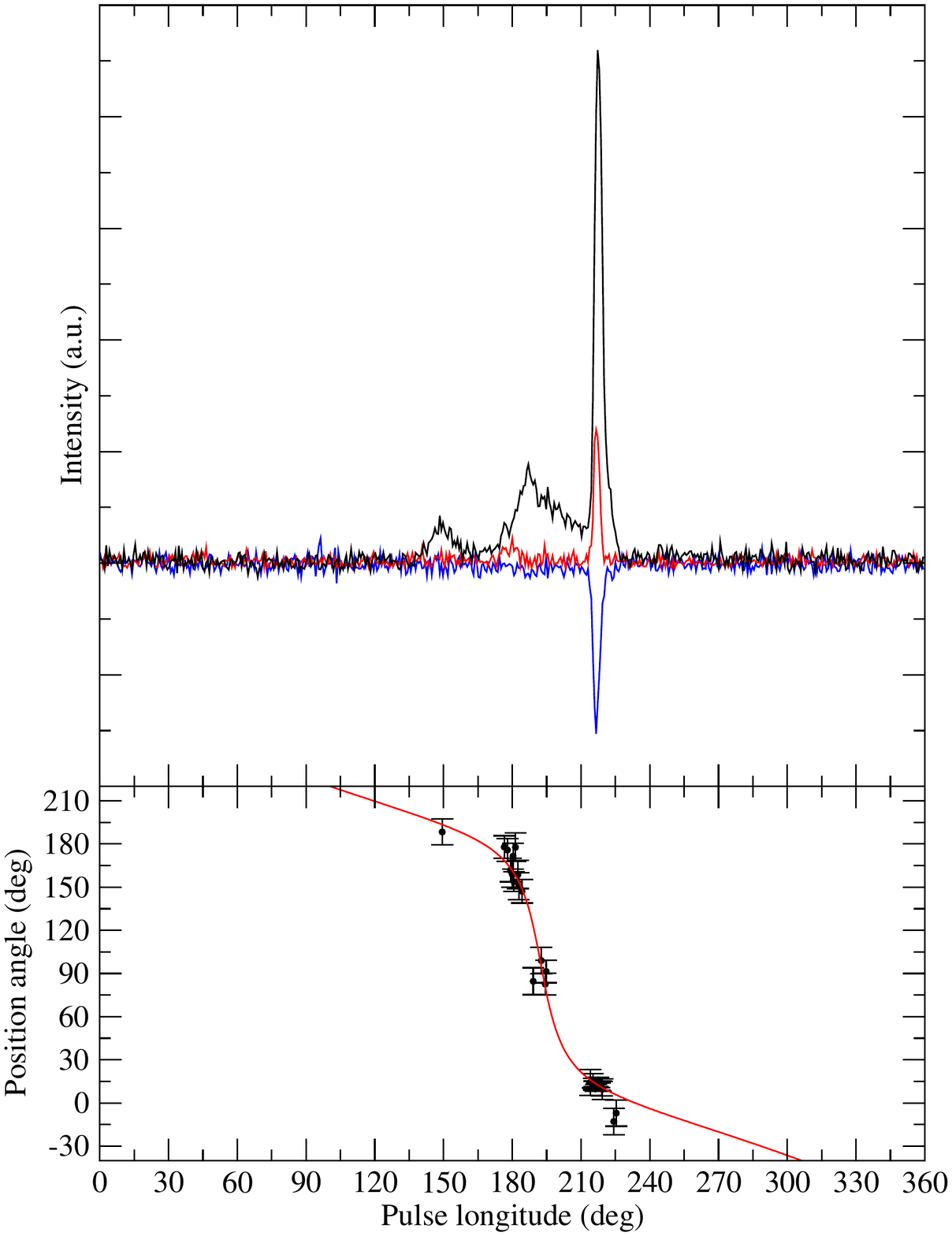} \\
\includegraphics[height=7.5cm,angle=-90]{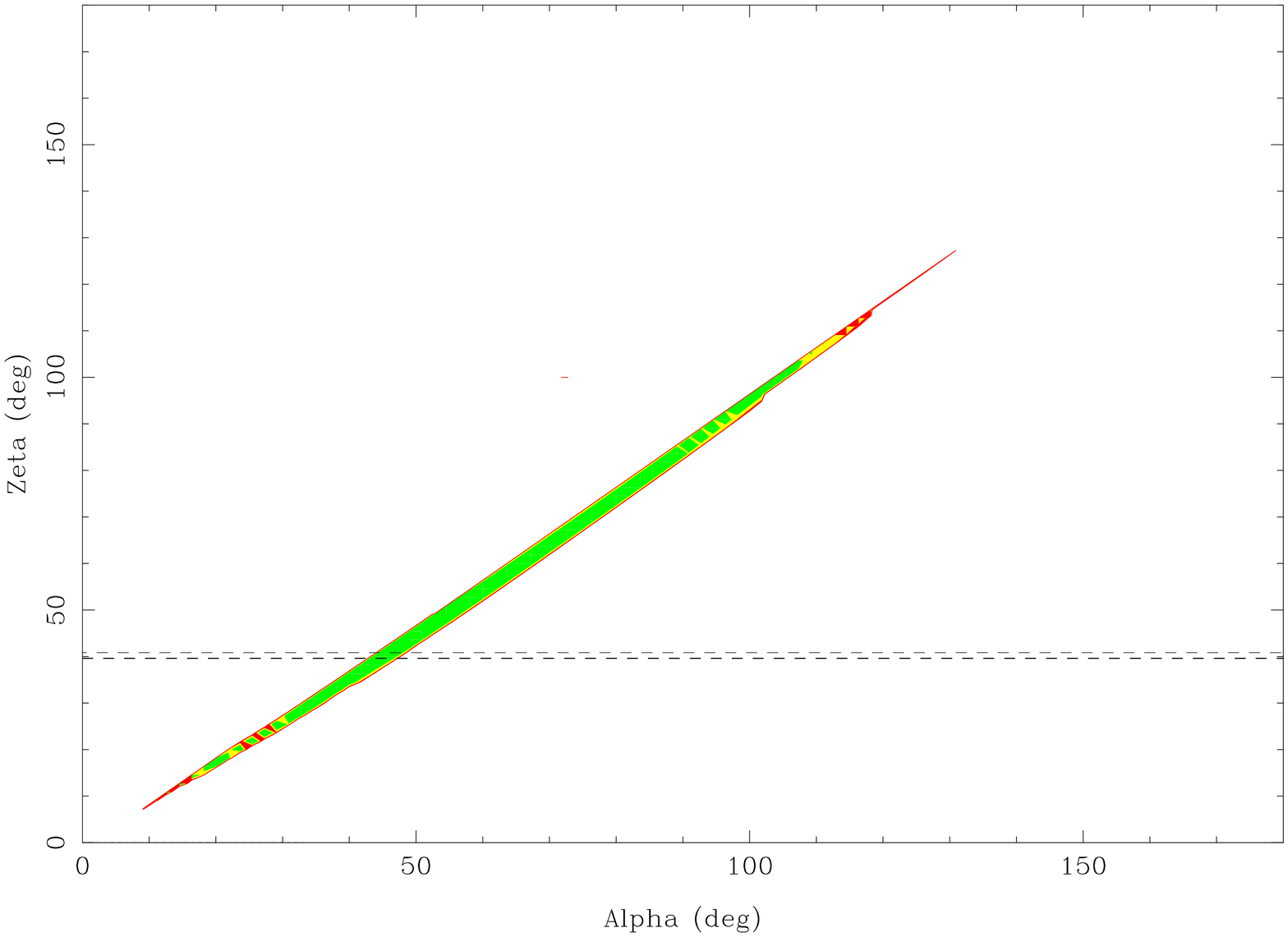}
\caption{{\bf Top:} Polarization profile of \psr\/ obtained with the Effelsberg Telescope. 
The upper panel shows total intensity ($I$, black), the linearly polarized
intensity ($L$, red) and the circularly polarized intensity ($V$, blue). The lower panel shows the position angle
of $L$ measured at pulse longitudes where $L$ exceeds $2\sigma$ measured from an off-pulse RMS. The red
line shows the resulting fit of a Rotating Vector Model (RVM), which indicates an ``outer line-of-sight'' (see \cite{lk05} for details). 
{\bf Bottom:} Map of the RVM parameters $\alpha$ (the angle between the spin axis and magnetic axis)
 and $\zeta$ (the angle between the line of sight and the spin axis).
The green region corresponds to combinations of $\alpha,\zeta$ for which the RVM provides a good description of
the polarimetry of \psr . Based on the polarimetry alone we would have a large uncertainty regarding $\alpha$
and $\zeta$. However, if we assume that during the accretion episode that recycled the pulsar the
spin axis of the pulsar was aligned with the orbital angular momentum
(which has an angle $i = 40\fdg2 \pm 0\fdg6$ to the line of sight) then $\alpha \simeq 45^\circ$. The
minimum angle between the magnetic axis and the line of sight is then given by $\beta = \zeta - \alpha = -4\fdg8$.
  \label{fig:pol}}
\end{center}
\end{figure*}

\begin{figure*}[htp]
  \includegraphics[width=16cm]{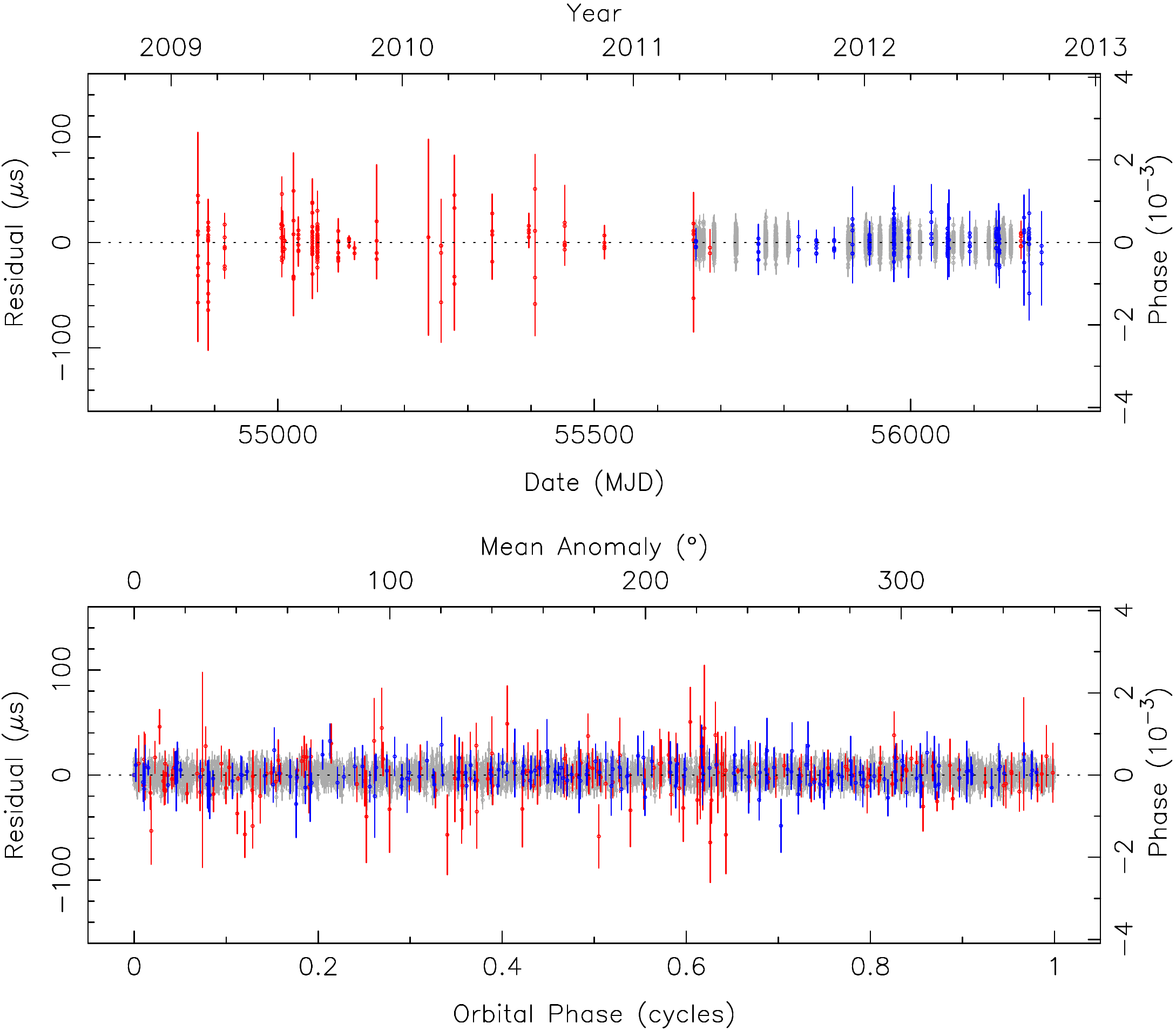}
  \caption{
    Post-fit residuals from the GBT (red), Arecibo (gray) and Effelsberg (blue)
    TOAs, obtained with the timing model presented in Table~1. 
    {\bf Top:} Residuals versus time. No significant un-modeled trends can be 
    found in the TOA residuals.
    {\bf Bottom:} Post-fit residuals versus orbital phase, which for this very 
    low-eccentricity system is measured from the ascending node (i.e., the mean 
    anomaly is equal to the orbital longitude). No significant trends can be 
    identified in the residuals; indicating that the orbital model can describe 
    the orbital modulation of the TOAs correctly. No dispersive delays or 
    unaccounted Shapiro delay signatures are detectable near orbital phase 0.25 
    (superior conjunction), nor artifacts caused by incorrect de-dispersion or 
    folding of the data.    
  \label{fig:residuals} }
\end{figure*}

\begin{figure}[htp]
  \centering
  \includegraphics[height=130mm,angle=0]{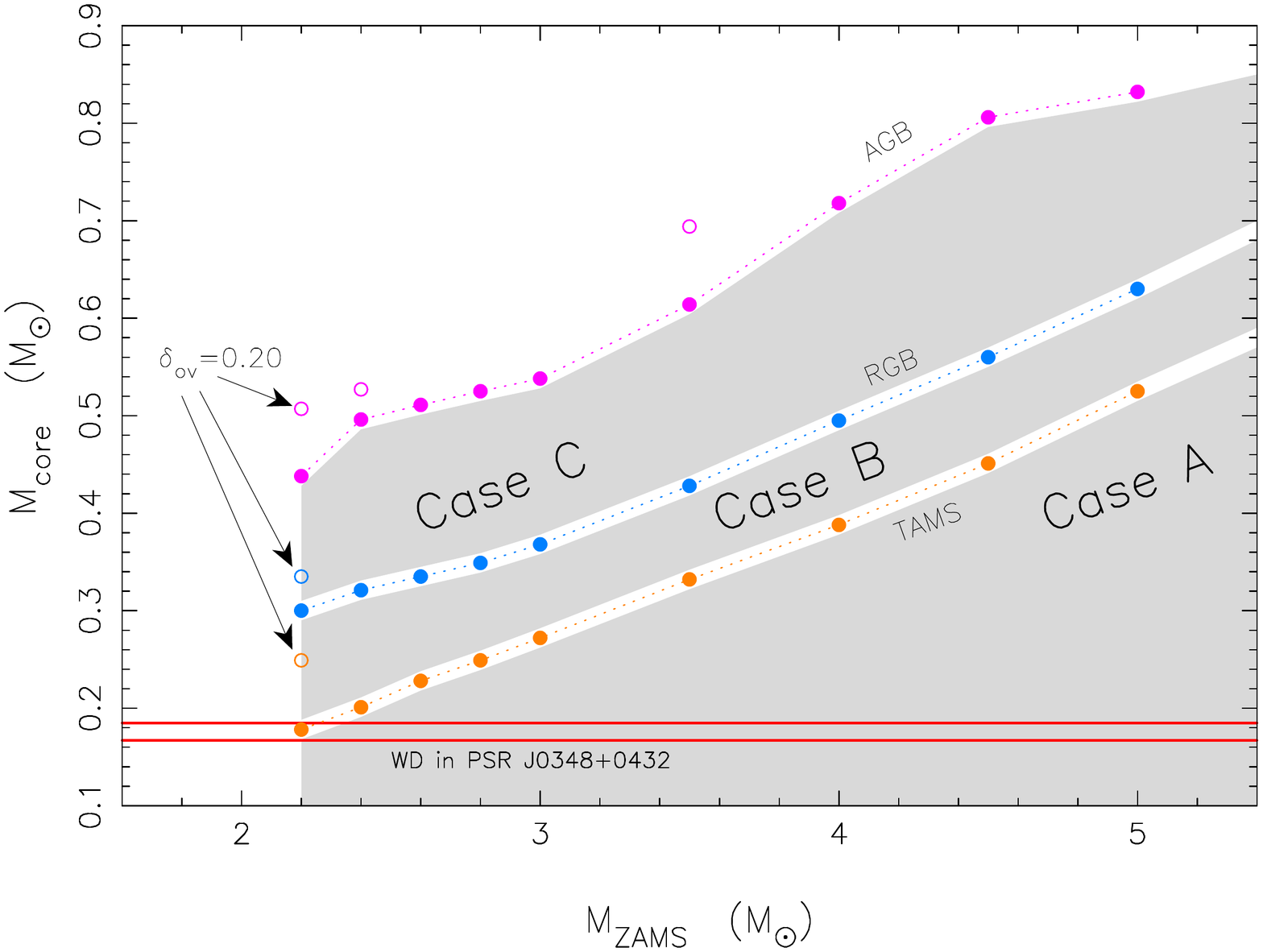}
  \caption{Stellar core mass at different evolutionary epochs as a function of 
    zero-age main sequence (ZAMS) mass. Assuming $M_{\rm core}$ = $M_{\rm WD}$
    $\simeq$ 0.17\,${\rm M}_\odot$ (as observed in PSR~J0348+0432) constrains 
    the progenitor star ZAMS mass to be $\le$ 2.2\,${\rm M}_{\odot}$ and that 
    its envelope was lost near the terminal-age main sequence (TAMS). All 
    calculations were performed without convective core overshooting. Including 
    this effect (for example, using $\delta_{\rm OV}$ = 0.20) would lower the 
    required donor mass even more. [Figure adapted from \cite{tlk11}].
}
  \label{0348.Mcore}
\end{figure}
\begin{figure}[htp]
  \centering
  \includegraphics[height=120mm,angle=0]{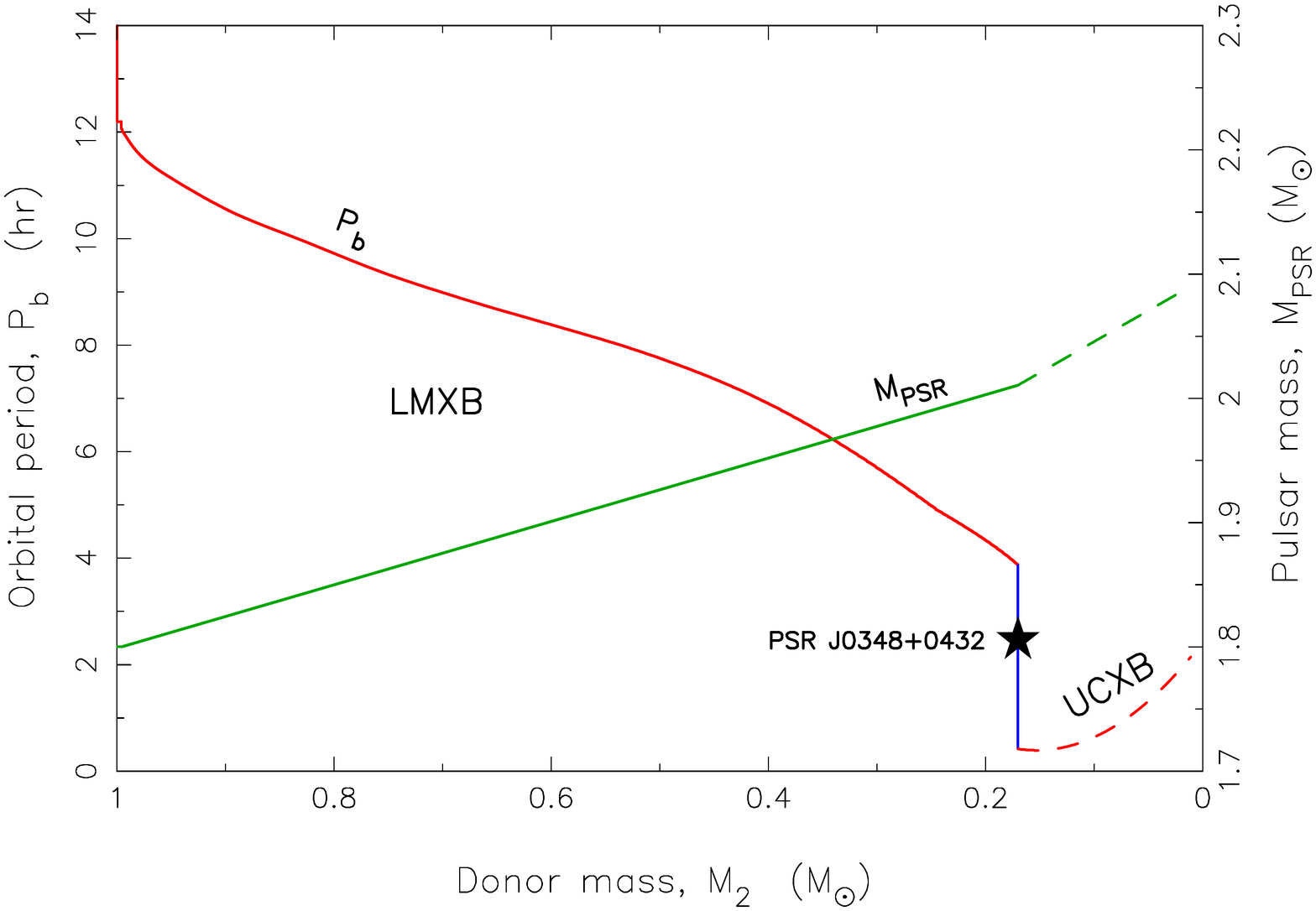}
  \caption{Formation of PSR~J0348+0432 from a converging LMXB for the same
           model as shown in Fig.~6 in the main text. 
           The plot shows how $P_{\rm b}$ (red line) and the mass of the 
           accreting neutron star (green line) 
           evolved as a function of decreasing donor star mass (here assumed to 
           be 1.1\,M$_{\odot}$ on the ZAMS).
           The RLO was initiated when
           $P_{\rm b}$ $\simeq$ 16\,hr and detached when $P_{\rm b}
           \simeq$ 4.9\,hr.
           In this model the initial mass of the neutron star was assumed to be 
           1.75\,M$_\odot$, although it may have been significantly
           lower if the neutron star accreted with 
           an efficiency close to 100\%.
           The present location of PSR~J0348+0432 is marked with a star.
  }
  \label{0348.LMXB_2}
\end{figure}

\begin{figure}[htp]
  \centering
  \includegraphics[height=130mm,angle=-90]{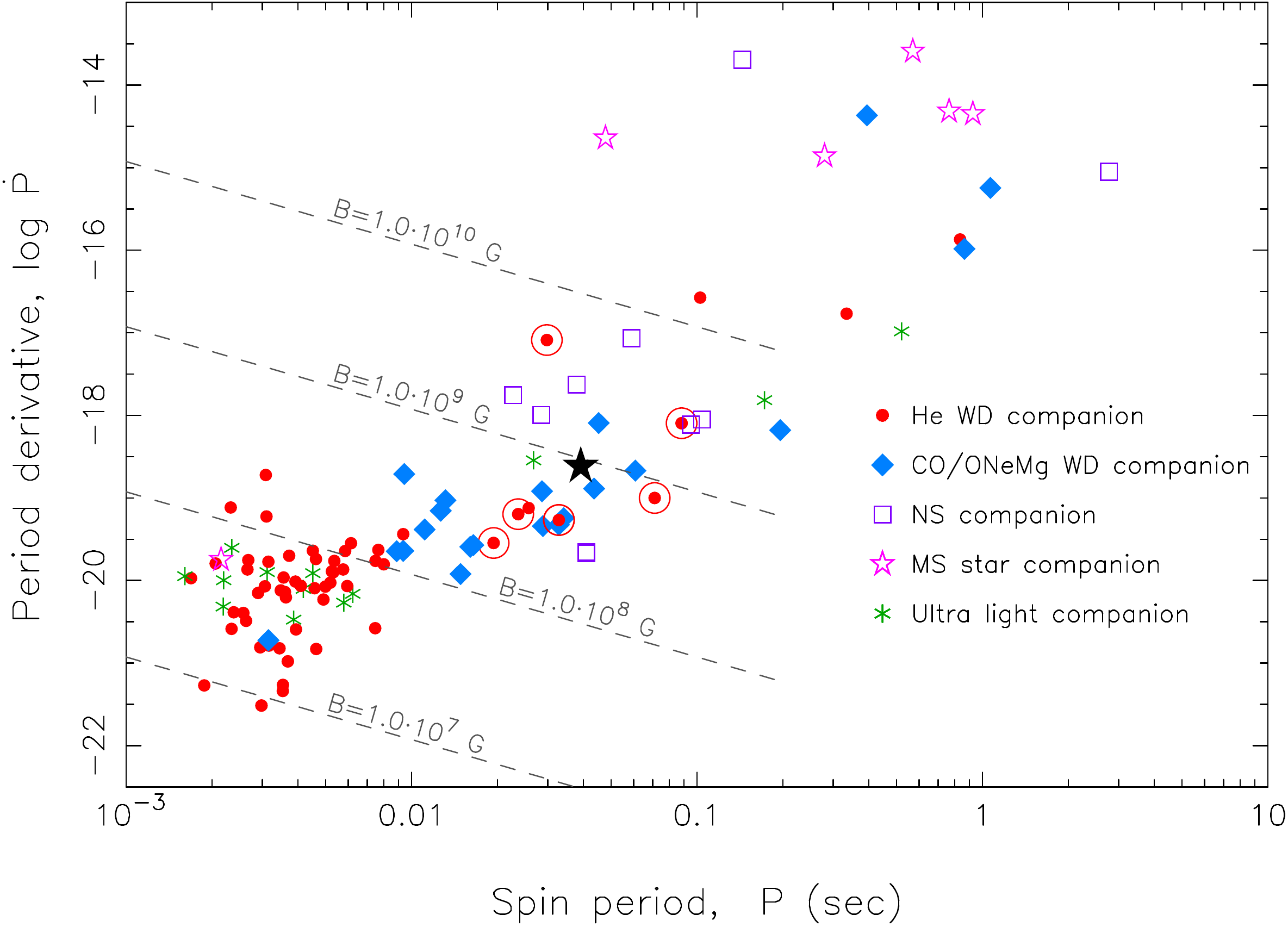}
  \caption{A $P\dot{P}$--diagram of the 111 known binary radio pulsars in the 
    Galactic disk. The location of PSR~J0348+0432 is marked with a black star in 
    a region which is mainly dominated by slow spin and high B-field pulsars 
    with massive white dwarf companions (marked with blue diamonds). 
    The dashed lines of constant B-fields were calculated 
    following \cite{tlk12} and assuming for simplicity $M_{\rm PSR}$ = 1.4 M$_\odot$ and 
    $\sin\alpha = \phi = \omega_{\rm c}$ = 1. All $\dot{P}$ values in this 
    plot are intrinsic values obtained from kinematic corrections to the 
    observed values. Data taken from the ATNF Pulsar Catalogue \cite{mhth05}, 
    in Oct.~2012.}
  \label{PPdot}
\end{figure}
\begin{figure}[htp]
  \centering
  \includegraphics[height=130mm,angle=0]{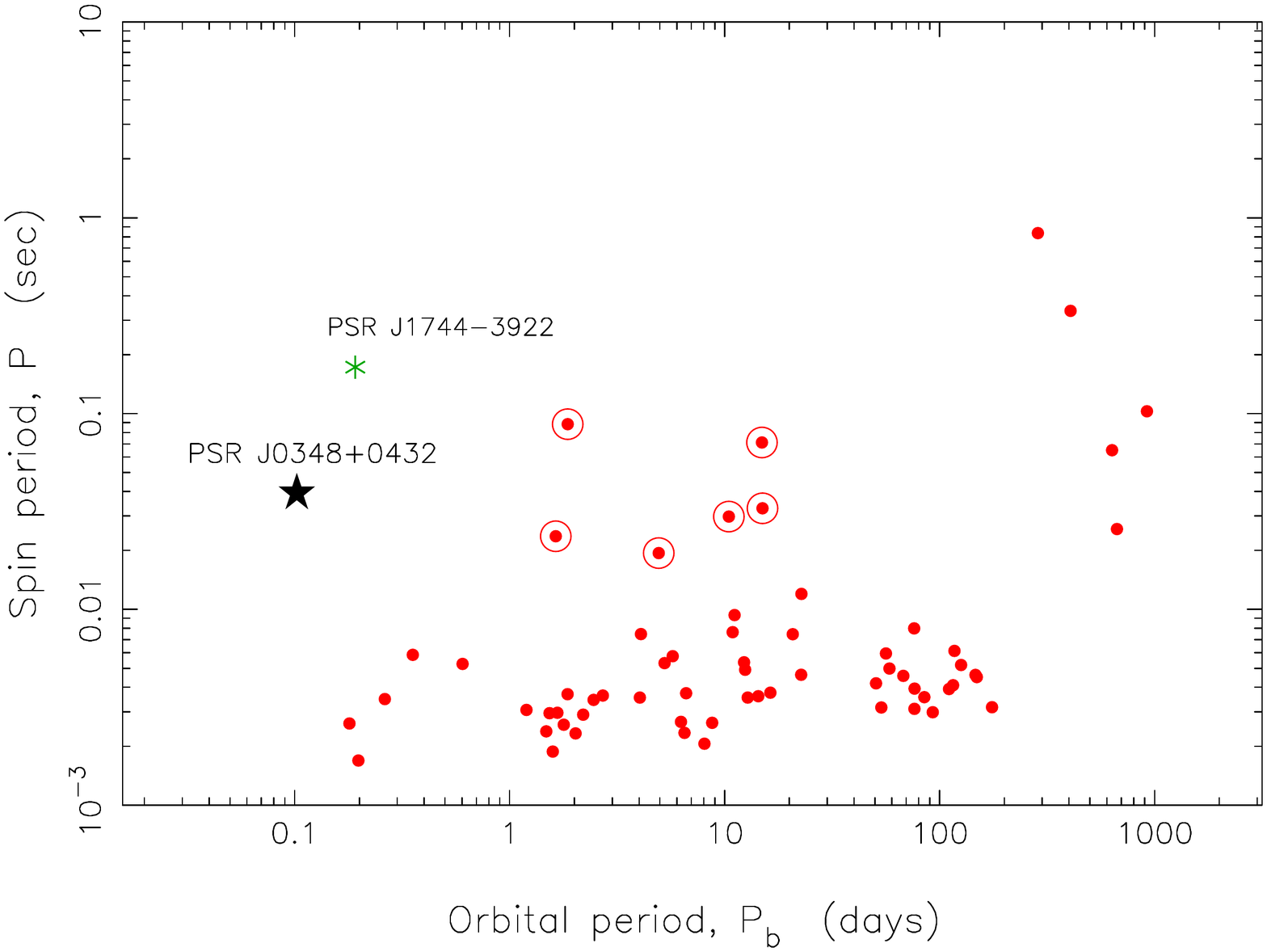}
  \caption{A $P_{\rm b}-P$ (Corbet) diagram of the 63 known Galactic binary pulsars with a 
    He~white dwarf companion of mass $M_{\rm WD} > 0.14\,{\rm M}_{\odot}$. The unique 
    location of PSR~J0348+0432 is shown with a star. 
    Another puzzling pulsar, PSR~J1744$-$3922 \cite{brr+07} marked with a green asterisk, 
    is included in this plot. These two pulsars seem to share high B-field properties  
    with the 6~pulsars in circles. Data taken from 
    the ATNF Pulsar Catalogue \cite{mhth05}, in Sep.~2012.}
  \label{Corbet}
\end{figure}
\begin{figure}[htp]
  \centering
  \includegraphics[height=180mm,angle=0]{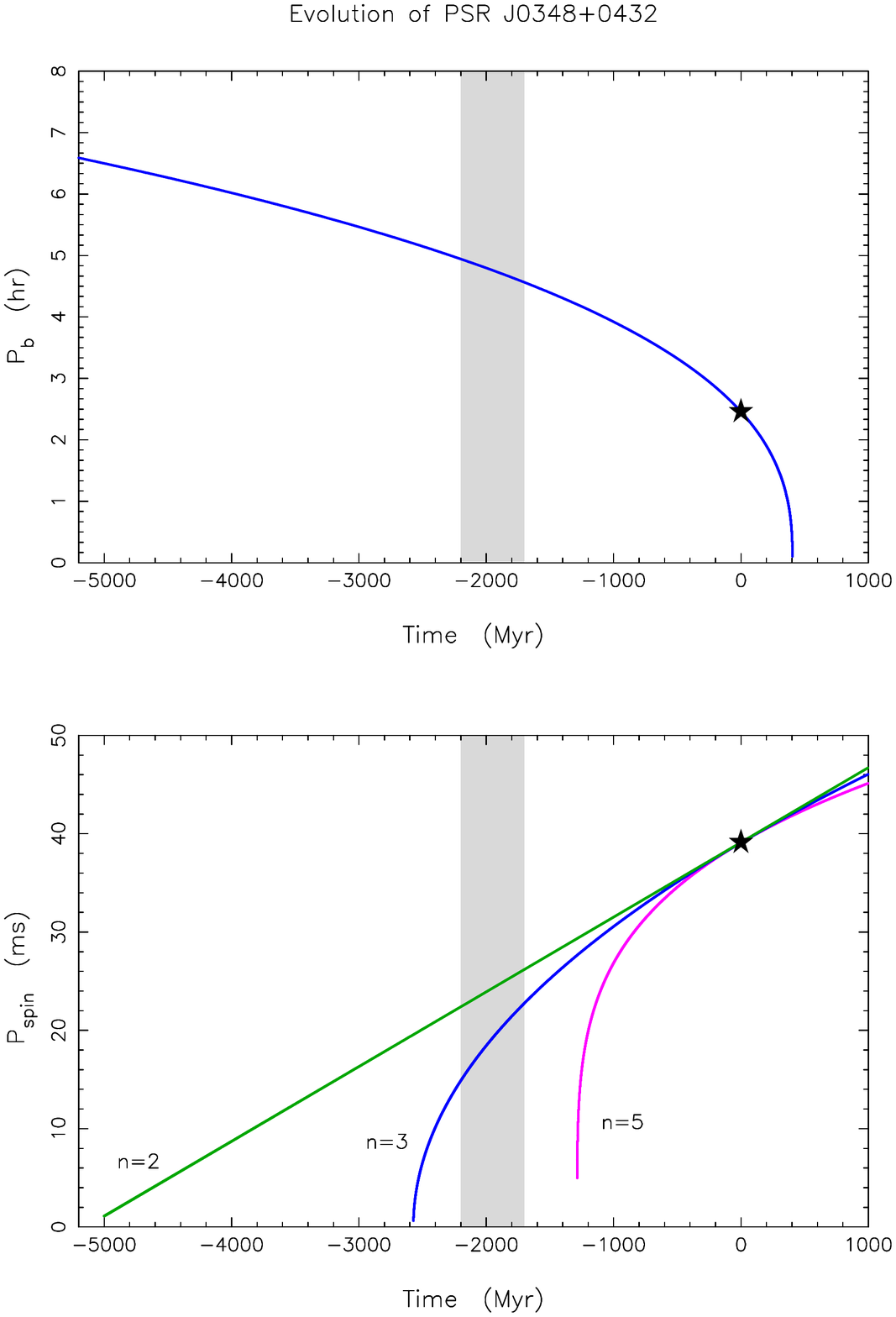}
  \caption{Orbital evolution (top) and spin evolution (bottom) of PSR~J0348+0432 
    in the past and in the future. Different evolution tracks are plotted for 
    different assumed values of a (constant) braking index, $n$. The grey shaded 
    region marks the estimated white dwarf cooling age. The past is for a 
    negative value of time, future is for a positive value of time. See text for 
    a discussion.}
  \label{0348_spin}
\end{figure}

\captionsetup{width=18cm}
\begin{landscape}
\footnotesize
\begin{center}
\begin{longtable}{lcccccccccc}
\caption{Observations log and radial velocity measurements\\
Notes: (1) refers to the barycentric mid-exposure 
time. (2) is the orbital phase, $\phi$, using the ephemeris in Table\,1.
(3) is the comparison's velocity in respect to the Solar System Barycenter and 
(4) the raw barycentric velocities of the white dwarf companion to 
\psr.} \\
\hline
Target & No. &MJD$_{\rm{bar}}^1$ & $\phi^2$ & slit &exposure & rotation &pos. angle &$v_{\rm R}^3$& $v_{\rm{WD}}^4$& \\
&     &     &                              &            &       s        & $\deg$  &   $\deg$    & (km\,s$^{-1}$) & (km\,s$^{-1}$) & \\
\hline
\endfirsthead

\multicolumn{3}{c}{{\tablename} \thetable{} -- Continued} \\[0.5ex]
\hline
Target & No. &MJD$_{\rm{bar}}$ & $\phi$ & slit &exposure & rotation &pos. angle &$v_{\rm R}$& $v_{\rm{WD}}$& \\
 &    &     &                              &            &       s        & $\deg$  &   $\deg$    & (km\,s$^{-1}$) & (km\,s$^{-1}$) & \\
\hline
\endhead

\multicolumn{3}{l}{{Continued on Next Page\ldots}} \\
\endfoot

\\[-1.8ex] 
\endlastfoot

\\
\psr& $1$ & $55915.070159$ & $0.9742$ & 1\arcsec & $799.96$ & $-135.0$ & $-150.0$ & $-18.24\pm0.41$ & $-348.25\pm7.17$ \\
& $2$ & $55915.080011$ & $0.0704$ & 1\arcsec & $799.96$ & $-135.0$ & $-155.1$ & $-22.46\pm0.39$ & $-345.54\pm7.55$ \\
& $3$ & $55915.099359$ & $0.2593$ & 1\arcsec & $799.95$ & $-135.0$ & $-166.5$ & $-29.37\pm0.40$ & $-12.36\pm9.71$ \\
& $4$ & $55915.109036$ & $0.3538$ & 1\arcsec & $799.96$ & $-135.0$ & $-172.8$ & $-25.21\pm0.39$ & $+159.76\pm9.54$ \\
& $5$ & $55915.120877$ & $0.4694$ & 1\arcsec & $799.96$ & $-135.0$ & $+179.2$ & $-19.26\pm0.38$ & $+332.63\pm8.37$ \\
& $6$ & $55915.130692$ & $0.5652$ & 1\arcsec & $799.98$ & $-135.0$ & $+172.7$ & $-20.79\pm0.38$ & $+295.18\pm8.08$ \\
& $7$ & $55915.140364$ & $0.6596$ & 1\arcsec & $799.97$ & $-135.0$ & $+166.4$ & $-19.54\pm0.38$ & $+149.72\pm8.77$ \\
& $8$ & $55915.150194$ & $0.7556$ & 1\arcsec & $799.98$ & $-135.0$ & $+160.4$ & $-19.41\pm0.38$ & $-45.85\pm8.76$ \\
& $9$ & $55915.217387$ & $0.4116$ & 1\arcsec & $799.96$ & $-134.8$ & $+132.1$ & $-8.28\pm0.44$ & $+311.34\pm11.94$ \\
& $10$ & $55915.227349$ & $0.5089$ & 1\arcsec & $849.96$ & $-134.8$ & $+129.7$ & $-9.74\pm0.43$ & $+360.61\pm11.87$ \\
& $11$ & $55915.237602$ & $0.6090$ & 1\arcsec & $849.97$ & $-134.8$ & $+127.3$ & $-5.13\pm0.44$ & $+278.11\pm11.03$ \\
& $12$ & $55915.247859$ & $0.7092$ & 1\arcsec & $849.95$ & $-134.8$ & $+125.4$ & $-4.85\pm0.46$ & $+86.48\pm13.56$ \\
& $13$ & $55915.261379$ & $0.8412$ & 1\arcsec & $799.96$ & $-134.8$ & $+123.1$ & $-3.10\pm0.42$ & $-143.98\pm13.42$ \\
& $14$ & $55915.271051$ & $0.9356$ & 1\arcsec & $799.97$ & $-134.8$ & $+121.7$ & $-2.20\pm0.46$ & $-297.87\pm14.87$ \\
& $15$ & $55915.280729$ & $0.0301$ & 1\arcsec & $799.96$ & $-134.8$ & $+120.4$ & $-0.34\pm0.52$ & $-272.12\pm18.87$ \\
& $16$ & $55915.290404$ & $0.1245$ & 1\arcsec & $799.97$ & $-134.8$ & $+119.5$ & $+1.56\pm0.61$ & $-188.66\pm33.02$ \\
& $17$ & $55916.060700$ & $0.6452$ & 1\arcsec & $799.98$ & $-134.8$ & $-146.8$ & $-49.62\pm0.44$ & $+197.14\pm15.88$ \\
& $18$ & $55916.070535$ & $0.7412$ & 1\arcsec & $799.97$ & $-134.8$ & $-151.6$ & $-52.40\pm0.50$ & $-10.28\pm21.58$ \\
& $19$ & $55916.080364$ & $0.8372$ & 1\arcsec & $799.96$ & $-134.8$ & $-156.8$ & $-50.10\pm0.51$ & $-209.03\pm20.37$ \\
& $20$ & $55916.091598$ & $0.9469$ & 1\arcsec & $799.97$ & $-134.8$ & $-163.4$ & $-48.75\pm0.49$ & $-339.79\pm20.05$ \\
& $21$ & $55916.101421$ & $0.0428$ & 1\arcsec & $799.96$ & $-134.8$ & $-169.6$ & $-43.97\pm0.44$ & $-354.42\pm15.82$ \\
& $22$ & $55916.111221$ & $0.1384$ & 1\arcsec & $799.97$ & $-134.8$ & $-176.1$ & $-41.06\pm0.42$ & $-255.63\pm13.45$ \\
& $23$ & $55916.122672$ & $0.2502$ & 1\arcsec & $799.96$ & $-134.8$ & $+176.6$ & $-44.69\pm0.46$ & $-15.94\pm18.30$ \\
& $24$ & $55916.132782$ & $0.3490$ & 1\arcsec & $849.96$ & $-134.8$ & $+169.6$ & $-46.23\pm0.46$ & $+153.76\pm19.31$ \\
& $25$ & $55916.142895$ & $0.4477$ & 1\arcsec & $799.96$ & $-134.8$ & $+163.1$ & $-43.02\pm0.46$ & $+332.62\pm18.44$ \\
& $26$ & $55916.154536$ & $0.5613$ & 1\arcsec & $799.96$ & $-134.8$ & $+156.3$ & $-41.91\pm0.49$ & $+285.43\pm20.94$ \\
& $27$ & $55916.164373$ & $0.6574$ & 1\arcsec & $799.97$ & $-134.8$ & $+151.1$ & $-40.10\pm0.48$ & $+165.98\pm18.92$ \\
& $28$ & $55916.174210$ & $0.7534$ & 1\arcsec & $799.97$ & $-134.8$ & $+146.4$ & $-39.74\pm0.48$ & $-54.07\pm20.13$ \\
& $29$ & $55916.229170$ & $0.2900$ & 1\arcsec & $799.96$ & $-134.8$ & $+128.5$ & $-12.85\pm0.44$ & $+89.37\pm15.28$ \\
& $30$ & $55916.238967$ & $0.3857$ & 1\arcsec & $799.97$ & $-134.8$ & $+126.5$ & $-9.24\pm0.43$ & $+300.61\pm17.14$ \\
& $31$ & $55916.250488$ & $0.4982$ & 1\arcsec & $799.96$ & $-134.8$ & $+124.4$ & $-9.52\pm0.43$ & $+350.95\pm18.00$ \\
& $32$ & $55916.260274$ & $0.5937$ & 1\arcsec & $799.97$ & $-134.8$ & $+122.8$ & $-15.17\pm0.55$ & $+331.23\pm22.73$ \\
& $33$ & $55916.270071$ & $0.6894$ & 1\arcsec & $799.97$ & $-134.8$ & $+121.4$ & $-18.58\pm0.62$ & $+89.53\pm33.20$ \\
& $34$ & $55916.279869$ & $0.7850$ & 1\arcsec & $799.97$ & $-134.8$ & $+120.2$ & $-19.10\pm0.76$ & $-110.32\pm50.99$ \\
& $35$ & $55915.181590$ & $0.0641$ & 2\farcs5 & $799.97$ &$-134.8$& $+144.0$ & & \\ 
& $36$ & $55916.190536$ & $0.9148$ & 2\farcs5 & $799.97$ &$-134.8$& $+139.4$ & & \\
EG\,21& $37$ & $55915.031970$ &  & 1\arcsec & $21.99$ & $0.0$ &$-25.8$ & & \\
 & $38$ & $55915.034856$ &  & 2\farcs5 & $21.99$ & $0.0$ & $-24.5$& & \\
HD\,49798 & $39$ & $55915.350264$ &  & 1\arcsec & $2.00$ & $0.0$ & $+72.6$& & \\
 & $40$ & $55916.343776$ & & 1\arcsec & $2.00$ & $0.0$ &$+71.3$ & & \\
& $41$ & $55916.346515$ &  & 2\farcs5 & $2.01$ & $0.0$ &$+72.3$ & & \\
LTT\,3218& $42$ & $55916.352238$ &  & 2\farcs5 & $22.01$ & $0.0$ &$+58.8$ & & \\
& $43$ & $55916.357062$ & & 2\farcs5 & $22.01$ & $0.0$ &$+62.4$ & & \\
& $44$ & $55916.369131$ & & 1\arcsec & $35.00$ & $0.0$ &$+64.5$ & & \\
GD\,108 & $45$ & $55916.360948$ &  & 2\farcs5 & $22.00$ & $0.0$ &$-176.4$ & & \\
& $46$ & $55916.365961$ &  & 1\arcsec & $35.00$ & $0.0$ &$+179.9$ & & \\

\hline
\end{longtable}
\end{center}

\end{landscape}

\begin{table}[ht]
\caption{\small Fractional binding energies of neutron stars (Data for Fig~4a).}

\smallskip
\centering
\begin{tabular}{llcr}
\hline\hline           
Neutron Star & Mass (M$_\odot$) & Reference & Fractional Binding Energy 
\\[0.5ex] 
\hline   
\multicolumn{4}{c}{pulsars with white dwarf companions} \\
\hline
PSR J0348$+$0432  & 2.01 & (this paper)  & $-$0.1446 \\
PSR J1141$-$6545  & 1.27 & \cite{bbv08}  & $-$0.0838 \\
PSR J1738$+$0333  & 1.47 & \cite{avk+12} & $-$0.0993 \\
\hline
\multicolumn{4}{c}{pulsars with neutron star companions} \\
\hline
PSR J0737$-$3039A & 1.338 & \cite{ksm+06} & $-$0.0890 \\
PSR J0737$-$3039B & 1.249 & \cite{ksm+06} & $-$0.0822 \\
PSR B1534$+$12    & 1.333 & \cite{sttw02} & $-$0.0887 \\
\dots companion   & 1.345 & \cite{sttw02} & $-$0.0896 \\
PSR B1913$+$16    & 1.440 & \cite{wnt10}  & $-$0.0969 \\
\dots companion   & 1.389 & \cite{wnt10}  & $-$0.0929 \\[0.2ex]
\hline 
\end{tabular}
\begin{flushleft}
{\footnotesize 
Fractional binding energies of neutron stars in relativistic binaries, 
which are currently used in precision tests for gravity, and where the neutron 
star masses are determined with good ($<$ few \%) precision. The masses are 
taken from the given references. The specific numbers for the fractional 
binding energy are based on the equation-of-state ``.20'' of \cite{hkp81}. 
A different equation-of-state gives different numbers, but does not change the 
fact that \psr\ significantly exceeds the tested binding energy range.
}
\end{flushleft}
\label{tab:egrav} 
\end{table}

\begin{table}[htp]
\caption{\small Stellar envelope binding energies, $E_{\rm bind}$, for given donors and evolutionary stages.
} 
\smallskip
\centering
\begin{tabular}{l c c l}
\hline\hline           
Stage      & $E_{\rm bind}$ ($2.2\,{\rm M}_\odot$) & $E_{\rm bind}$ ($4.0\,{\rm M}_\odot$) & CE outcome\\ [0.5ex] 
\hline   
$X_c$ = 0.40 & $1.8\times 10^{49}\,{\rm erg}$ & $3.8\times 10^{49}\,{\rm erg}$  & merger$^{*}$\\ 
$X_c$ = 0.20 & $1.6\times 10^{49}\,{\rm erg}$ & $3.4\times 10^{49}\,{\rm erg}$  & merger$^{*}$\\
$X_c$ = 0.02 & $1.5\times 10^{49}\,{\rm erg}$ & $3.2\times 10^{49}\,{\rm erg}$  & merger$^{*}$\\ [0.2ex]
\hline \noindent
TAMS & $1.6\times 10^{49}\,{\rm erg}$ & $3.6\times 10^{49}\,{\rm erg}$ & may survive if $L_{\rm acc}$ can eject envelope\\ [0.2ex]
\hline 
RGB & $9.8\times 10^{47}\,{\rm erg}$ & $2.1\times 10^{48}\,{\rm erg}$ 
& survives with 0.30 $\le M_{\rm WD}/M_{\odot} \le$ 0.50\\ [0.2ex]      
\hline 
AGB & $2.0\times 10^{47}\,{\rm erg}$ & $1.9\times 10^{47}\,{\rm erg}$ 
& survives with 0.44 $\le M_{\rm WD}/M_{\odot} \le$ 0.72\\ [0.2ex]      
\hline 
\end{tabular}
\begin{flushleft}
  {\footnotesize Envelope binding energies of stars with a total mass of 
  2.2\,M$_{\odot}$ and 4.0\,M$_{\odot}$, respectively.
  In all cases $E_{\rm bind}$ was calculated assuming $M_{\rm core}$ = 
  $M_{\rm WD}$ = 0.17\,M$_\odot$, except for the cases where the
  RLO was initiated at the tip of the RGB/AGB with resulting values of 
  $M_{\rm WD}$ as listed in the table.}\\
  $^{*}$  {\footnotesize Note, that intermediate-mass donor stars on the main 
  sequence ($X_c >$ 0) with $P_{\rm b}$ $>$ 1\,day, or at the TAMS, may avoid 
  the onset of a CE altogether and evolve as a stable intermediate mass {X}-ray 
  binary, IMXB \cite{tvs00}.}
\end{flushleft}
\label{table:Ebind} 
\end{table}

\bibliographystyle{Science}

\begin{thebibliography}{10}
\bibitem{nor90b}
The effect of a variable gravitational constant on the orbital period decay
$\dot{P}_{\rm b}$ of \psr\ is actually mitigated by the change of the binding 
energy of the neutron star related to $\dot{G}$ \cite{nor90a}. Since this change
is both EOS and gravity theory dependent, we keep the more conservative 
expression of \cite{dgt88}, which is anyway two orders of magnitude below our
measurement precision of $\dot{P}_{\rm b}$.
\end{thebibliography}

\end{document}